\documentclass[a4paper,12pt]{article}
\pdfoutput=1
\usepackage{jheppub} 
\usepackage{amsmath,amssymb}
\usepackage{graphicx}
\usepackage{float}
\usepackage{soul}
\usepackage[export]{adjustbox}
\usepackage[utf8]{inputenc}
\usepackage{slashed}
\usepackage{bbold}
\usepackage{tikz}
\usetikzlibrary{arrows,shapes}
\usepackage{subfig}
\usepackage[utf8]{inputenc}

\usepackage{amssymb}
\usepackage{amsmath}
\usepackage{amsfonts}

\usepackage{array}

\usepackage{multirow}
\usepackage{psfrag}

\graphicspath{{figures/}}

\def\bea{\begin{eqnarray}}
\def\eea{\end{eqnarray}}

\def\nn{\nonumber}

\newcommand{\eqs}[2][0.3]{\includegraphics[width=#1\linewidth, valign=c]{#2}}

\newcommand\psla{\slashed p}
\newcommand\ksla{\slashed k}
\newcommand\lsla{\slashed l}

\def \psla{\rlap{$p$}{/}}
\def \ksla {\rlap{$k$}{/}}
\def \lsla {\rlap{l}{/}}

\def \rsla {\rlap{$r$}{/}}

\title{Locally finite two-loop QCD amplitudes from IR universality for electroweak production}

\author[a]{Charalampos Anastasiou}
\affiliation[a]{Institute for Theoretical Physics, ETH Zurich, 8093
  Z\"urich, Switzerland}
\emailAdd{babis@phys.ethz.ch}

\author[b]{George Sterman}
\affiliation[b]{ C.N.\ Yang Institute for Theoretical Physics and Department of Physics and Astronomy\\
Stony Brook University, Stony Brook NY, 11794-3840 USA}
 \emailAdd{george.sterman@stonybrook.edu}

\abstract{ 
We describe the implementation of infrared subtractions for two-loop QCD corrections to quark-antiquark annihilation
to electroweak final states.   The subtractions are given as form-factor integrands whose integrals are known.   The
resulting subtracted amplitudes are amenable to efficient numerical integration. 
Our procedure is based on the universality of infrared singularities and requires a relatively
limited set of subtractions, whose number grows as the number of two-loop diagrams, rather than with
the number of singular regions of integration.  
}

           
\begin{document}
\preprint{YITP-SB-2022-43}
\maketitle 

\section{Introduction}
\label{sec:introduction}

The statistical power of  measurements at the  LHC experiments 
makes possible the observation and study of rare processes.
A remarkable example is the observation and measurements of the Higgs boson in association with a pair of top quarks by ATLAS~\cite{ATLAS:2020ior,ATLAS:2018mme} and CMS~\cite{CMS:2020cga,CMS:2018uxb}. 
This Higgs production channel offers the possibility to analyze directly the nature
of the  Higgs and top-quark interaction and to extract its strength.   
A different recent example of rare processes becoming accessible experimentally is the observation by CMS ~\cite{CMS:2020hjs} and ATLAS~\cite{ATLAS:2022xnu,ATLAS:2016jeu} of triboson production (VVV with V=W,Z). Triple electroweak production processes probe the Standard Model in novel ways, as they are sensitive to quartic gauge-boson interactions, light-quark Yukawa interactions of the Higgs boson and to possible novel electroweak states predicted in extensions of the Standard Model~\cite{Degrande:2013rea,Falkowski:2020znk}.
Electroweak diboson production is less rare, and it is copiously associated with high transverse momentum jet radiation (VV+jet) at the LHC.  The corresponding cross sections are measured differentially with high precision~\cite{CMS:2020mxy,ATLAS:2021jgw}. Differential studies of electroweak diboson production probe stringently the electroweak sector of the Standard Model and its extensions (see, for example, \cite{Franceschini:2017xkh}).

Measurements of  hard scattering hadron collisions with $ n \geq 3 $ electroweak particles, heavy quarks or jets  in the final state, as in these 
examples,  are an important part of the LHC physics program. These processes are of a special interest for the high luminosity phase, during which a twenty-fold increase in the number of events is expected, as well as at future high energy colliders. Correspondingly, the calculation of perturbative cross sections for high multiplicity partonic cross sections is important as well.

In the last five years, groundbreaking results towards next-to-next-to-leading-order (NNLO)  cross-sections for $2 \to 3$ processes, which require   two-loop amplitudes,  were derived~\cite{Abreu:2020xvt,Canko:2020ylt,Papadopoulos:2019iam,Abreu:2017xsl,Badger:2017jhb,Chicherin:2018old,Chicherin:2020oor,Kallweit:2020gcp,Gehrmann:2015bfy,Chicherin:2018yne,Chawdhry:2019bji,Badger:2019djh,Chawdhry:2020for,Badger:2021imn,Abreu:2018zmy,Abreu:2021oya,Chawdhry:2021mkw,Chawdhry:2021hkp,Czakon:2021mjy,Agarwal:2021grm,Agarwal:2021vdh,Abreu:2021asb,Hartanto:2022qhh,Kardos:2022tpo,Badger:2022ncb,Chicherin:2021dyp}. This spectacular theoretical progress concerns, at the moment, two-loop processes with no internal massive particles and at most one massive external particle.

The techniques that 
made these advances possible
are  analytic or semianalytic.
Given the rapid pace of developments, the same methods have a potential to evolve further and become suited for more complicated final states.  However, it is noted that the introduction of further kinematic or mass scales in two-loop scattering amplitudes beyond what constitutes the state of the art will have a daunting computational cost. This observation motivates  
the development of
methods for computing two-loop amplitudes in $2 \to n (\geq 3)$ scattering 
that
are less steeply sensitive to the number of masses in internal or external particles.

An appealing idea is to compute multi-loop amplitudes numerically, with a direct integration over momentum space.   Per loop, this approach requires a fixed number of, at most, four integrations. The number of integrations does not depend on the number of kinematic scales.
The integrand remains  a
rational function whose size scales with the number of Feynman diagrams,
and that is in principle simple to evaluate. 
For most practical purposes in phenomenology, recent work shows~\cite{Pozzorini:2022ohr} that it can be evaluated efficiently with a well tolerated computational cost. 
 
Methods for a direct integration of one-loop amplitudes in momentum space were introduced in Refs.~\cite{Soper:1999xk,Nagy:2006xy,Gong:2008ww,Becker:2010ng,Assadsolimani:2009cz,Becker:2012aqa,Becker:2011vg,Becker:2012bi,Seth:2016hmv}.  These works treated two issues that one encounters in direct momentum space integration: (i) the local subtraction or local cancellation of soft, collinear and ultraviolet singularities at one loop (ii) the automated deformation of the contour of integration away from threshold singularities at one loop and, in a visionary publication~\cite{Becker:2012bi}, beyond. Recently, the prospects of direct numerical integration were widened with the introduction of a Four Dimensional Regularization approach~\cite{Pittau:2012zd,Pittau:2021jbs,Page:2018ljf,Pittau:2018frs,Zirke:2015spg,Page:2015zca,Pittau:2014tva,Fazio:2014xea,Donati:2013voa}  and Loop Tree Duality methods~\cite{Capatti:2022tit,Capatti:2020ytd,Capatti:2019edf,Capatti:2019ypt,Runkel:2019zbm,Kromin:2022txz,Baumeister:2019rmh,Runkel:2019yrs,Chachamis:2017yzl,Chachamis:2016olm,Sborlini:2016gbr,Buchta:2015wna,Buchta:2014dfa,Bierenbaum:2013nja,TorresBobadilla:2020ekr,Bierenbaum:2010cy,Catani:2008xa,Kermanschah:2021wbk}. The latter, in the same spirit as of Ref.\ \cite{Soper:1999xk}, integrate out elegantly the energy variable in loop momenta at generic loop orders.  In this framework, loop integrations and phase-space integrations over real particles become very similar.

Indeed, it was recently demonstrated at generic orders in perturbation theory~\cite{Capatti:2020xjc,Capatti:2022tit} that the integrands of real and virtual corrections can be combined under a single integration so that  all possible cancelations of infrared singularities 
that
are expected from unitarity 
occur locally \cite{Sterman:1978bj,Sterman:1995fz}. This procedure is consistent with a generic method for contour deformation to treat threshold singularities~\cite{Capatti:2019edf}. The framework of ``local unitarity''\footnote{Referred to as ``generalized unitarity" in Ref.\ \cite{Sterman:1995fz}, in a discussion based on time-ordered perturbation theory.} is suitable for differential cross sections where the combination of real and virtual contributions alone yields  well-defined finite cross-sections, as in numerous processes and observables  at $e^+ e^-$ colliders. For cross sections with  identified hadrons in the initial or final state, where partonic cross sections are folded with parton densities or  fragmentation functions, an extension of the method is required.
Several techniques have 
also
been developed for the subtraction of infrared divergences and the numerical evaluation of the process-dependent finite remainders of real radiation contributions to cross sections (including hadroproduction) at NNLO~\cite{Anastasiou:2003gr,Binoth:2004jv,Anastasiou:2010pw,Weinzierl:2003fx,Weinzierl:2003ra,Catani:2007vq,Campbell:1997hg,Gehrmann-DeRidder:2005btv,Daleo:2006xa,Daleo:2009yj,NigelGlover:2010kwr,Abelof:2011jv,Gehrmann:2011wi,Gehrmann-DeRidder:2012too,Czakon:2010td,Czakon:2011ve,Czakon:2014oma,Boughezal:2011jf,DelDuca:2015zqa,Chen:2022clm,Chen:2022ktf,Agarwal:2021ais,Asteriadis:2021gpd,Campbell:2021mlr,Magnea:2020trj,Asteriadis:2020sud,Ebert:2019zkb,Asteriadis:2019dte,Engel:2019nfw,Billis:2019vxg,Caola:2019pfz,Caola:2019nzf,Cox:2018wce,Melnikov:2018jxb,Behring:2018cvx,Cieri:2018oms,Ebert:2018lzn,Caola:2018pxp,Herzog:2018ily,Currie:2018fgr,Boughezal:2018mvf,Magnea:2018jsj,Caola:2017xuq,Campbell:2017hsw,Moult:2017jsg,Caola:2017dug,Boughezal:2016zws,Moult:2016fqy,DelDuca:2016ily,Boughezal:2016wmq,Czakon:2016ckf,Caola:2015wna,Gaunt:2015pea,Boughezal:2015aha,Boughezal:2015eha,Abelof:2014fza,Alioli:2013hqa,Bernreuther:2013uma,Currie:2013vh,Buhler:2012ytl,Duhr:2022cob,Ebert:2020qef,Ebert:2020unb,Ebert:2020lxs,Ebert:2020yqt,Grazzini:2017mhc}. 
%

In this paper, we continue a complementary development, of a flexible subtraction method for the numerical computation of two-loop amplitudes for the production of multiple massive particles, based on the universality properties of infrared singularities in gauge theories. 
%
We build upon Refs~\cite{Anastasiou:2018rib,Anastasiou:2020sdt} and formulate simple infrared and ultraviolet subtractions that remove locally all singularities from  a family of massless QCD two-loop amplitudes for generic electroweak production in quark scattering $q \bar q  \to V_1\ldots V_n, \quad V_i \in {\gamma^*, W, Z} $.  Our subtractions  permit the direct integration of the finite amplitude remainders with numerical methods in four spacetime dimensions.  To our understanding, this is the first time that a 
local
subtraction technique for virtual corrections is derived at two loops for a non-abelian gauge theory.

Our method is inspired by and founded on the factorization of infrared divergences in QCD amplitudes~\cite{Erdogan:2014gha,Ma:2019hjq,Sterman:2022gyf,Akhoury:1978vq,Sen:1982bt,Korchemsky:1994jb,Catani:1998bh,Sterman:2002qn,Bauer:2000yr,Aybat:2006wq,Aybat:2006mz,Ferroglia:2009ii,Becher:2019avh,Collins:1989bt,Sterman:1995fz,Collins:2020euz,Dixon:2008gr,Gardi:2009qi,Gardi:2009zv,Becher:2009cu,Feige:2014wja,Sterman:1978bi,Libby:1978qf,Collins:1989gx,Becher:2014oda}. Divergences originating from  integration of loop momenta over regions with soft and collinear singularities organize themselves into ``soft function'' factors (in color space) and ``jet function'' factors, and multiply a process-dependent hard scattering function that receives contributions from non-singular regions of loop integration. Soft and jet functions are universal and can be determined from calculations of scattering amplitudes of simple processes. After loop integrations for a scattering amplitude are carried out to a certain perturbative order, one extracts  the hard function by multiplying it with the inverse of the soft and jet functions.

We aim to derive a hard function for scattering amplitudes in the form of an integrand that is free of infrared and ultraviolet singularities locally, so that it can be amenable to numerical integration. 
Factorization theorems do not guarantee, a priori, the existence of a  locally finite representation. While proofs of factorization extract the singular behaviour of amplitudes in  all possible soft and/or collinear configurations of the loop momenta with intrinsically local methods, such as power-counting and subtractions, in the combination of these singularities into soft and jet function factors one assumes symmetries and properties from the fact that the integration over loop momenta can, at least in principle, be carried out. Our strategy is to remove reliance on the properties of subintegrals, and to make factorization manifestly local, by constructing an alternative form of the loop integrand, compared the one derived from the standard application of Feynman rules.    
In Ref.~\cite{Anastasiou:2020sdt} it was demonstrated that this approach can be succesful for a class of QED processes through two-loop order. In this article, we achieve a complete local factorization for an analogous class of processes in QCD, addressing for the first time issues that emerge in non-abelian gauge theories at two loops. 


After a review of the notation and the basic formalism in Sec.\ \ref{sec:framework}, the construction of our amplitude integrand starts in Sec.\ \ref{sec:momentumflow} with the expression derived from a conventional application of Feynman rules and a judicious assignment of loop momentum flows.  We review in Sec.\ \ref{sec:momentumflow} how, at one loop, virtual gluons collinear to external particles  are longitudinally polarized manifestly, and, with our momentum flow assignment, the corresponding collinear singularities are factorized, due to Ward identities governing the propagation of longitudinal degrees of freedom in loops.

At two loops, virtual collinear gluons acquire an arbitrary polarization, which in Ref.~\cite{Anastasiou:2020sdt}  was termed {\it loop polarization}. Loop polarizations do not contribute to amplitude divergences, as the corresponding regions integrate to a factorizable form. However, the corresponding terms in the integrand are singular and spoil factorization locally.  In Sec.\ \ref{sec:VandS}, we remedy this problem by adding terms to one-loop QCD vertex and self-energy subgraphs adjacent to external partons. 
On the one hand, our added terms remove contributions of the original integrand that are responsible for nonfactorizable loop polarizations  in collinear limits. On the other hand,
they integrate to zero, preserving the value of the scattering amplitude. After  making manifest locally that virtual gluons in collinear limits have purely a longitudinal polarization, we can rely, once again, on Ward identities to factorize collinear singularities.  
Once loop polarizations are removed, this factorization is automatic and local in the sum over diagrams in QED \cite{Anastasiou:2020sdt}, and does not require further counterterms.
For two-loop QCD amplitudes, however, Ward identity cancelations occur only after shifts of loop momenta, and are hence not yet local.  In Sec.\ \ref{sec:construction} we add to the amplitude further terms to solve this problem. Each term is proportional to the difference of the integrand of a planar diagram evaluated with two different momentum flows, chosen to remove non-local cancelations in the Ward identity. 

The interventions discussed in Secs. \ref{sec:momentumflow} - Sec.\ \ref{sec:construction} may be summarized as, (i) to assign appropriate momentum flows, (ii) to eliminate loop polarizations and (iii) to implement Ward identity cancelations locally, so that  singularities of the integrand factorize algebraically in all collinear limits. At this stage, we can realize these modifications with counterterms, as described in Secs.\ \ref{sec:VandS} and \ref{sec:construction}. Exploiting their universal structure within the class of  massive electroweak gauge boson production, we use the integrand of the simplest $2 \to 1$ electroweak form factor amplitude to provide infrared counterterms for amplitudes of a higher multiplicity of massive electroweak bosons in the final state. Finally, in Sec.\ \ref{sec:UV}, we proceed to remove ultraviolet singularities following the method of Ref.~\cite{Anastasiou:2020sdt}, with ultraviolet counterterms that respect Ward identities and preserve collinear factorization.   We verify the infrared and ultraviolet finiteness of our subtracted amplitudes in Sec.\ \ref{sec:check} and conclude with a brief summary and discussion of prospects for further development of our approach.


\section{Framework}
\label{sec:framework}

We aim to compute amplitudes through two loops in perturbative QCD
for quark initiated processes of the type
\begin{equation}
q(p_1) + \bar q (p_2) \to  \mbox{ massive color neutral particles,}
\end{equation}
where $p_1$ and $p_2$ are the momenta of the incoming partons.

The perturbative expansion reads
\begin{equation}
  {M}  = {\cal M}_0
  +  \int \frac{d^Dk}{\left(2 \pi\right)^D} {\cal M}_1\left(k \right)
  +  \int \frac{d^Dk}{\left(2 \pi\right)^D}\, \int \frac{d^Dl}{\left(2 \pi\right)^D}\, {\cal M}_2\left(k
    ,l\right) \, ,
    \label{eq:cal-M-normalization}
\end{equation}
where here, as below, script letters refer to integrands.

Our goal is to extend the analysis of Ref.\ \cite{Anastasiou:2020sdt} from QED to QCD.  We want to show how to construct two-loop amplitude integrands for the production of massive color-neutral final state bosons from annihilating quarks, which are both locally integrable and manifestly convergent at infinity.   Such integrands can then be evaluated numerically.

The basic construction, in the notation of Ref.\ \cite{Anastasiou:2020sdt}, is summarised by the pattern 
\bea
{\cal M}^{(1)}_{\rm finite}\ &=&\ {\cal M}^{(1)}_{\rm UV\; finite}\ -\ {\cal F}^{(1)}_{\rm UV\; finite}\, \left[ {\bf P}_1\, \widetilde {\cal M}^{(0)} {\bf P}_1 \right ]\, ,
\nn\\[2mm]
{\cal M}^{(2)}_{\rm finite}\ &=&\ {\cal M}^{(2)}_{\rm UV\; finite}\ -\ {\cal F}^{(1)}_{\rm UV\; finite}\, \left[ {\bf P}_1\, \widetilde{\cal M}^{(1)}_{\rm finite} {\bf P}_1 \right ] 
\nn\\
&\ & \hspace{5mm} \ -\ {\cal F}^{(2)}_{\rm UV\; finite}\, \left[ {\bf P}_1\,\widetilde{\cal M}^{(0)} {\bf P}_1 \right ] \, ,
\label{eq:cal-M-finite-12}
\eea
where $\widetilde{\cal M}^{(a)}$ is the integrand of the $a$-loop amplitude with external spinors removed.   In this expression, 
the subscripts ``UV finite" indicate that we also regularize UV divergences by the introduction of local counterterms that make all integrals manifestly convergent in the ultraviolet.
The factor ${\bf P}_1$ in Eq.\ (\ref{eq:cal-M-finite-12}) is a Dirac projector that isolates terms with singularities at the next loop order \cite{Anastasiou:2020sdt},
\bea
\label{eq:projector}
  \mathbf P_1 \ \equiv\ \frac{\slashed p_1 \slashed p_2} {2p_1 \cdot p_2}\, .
\quad 
\eea
In Eq.\ (\ref{eq:cal-M-finite-12}), infrared singularities, both collinear and soft, are eliminated from amplitudes ${\cal M}^{(a)}$ iteratively, and on a local basis, by the subtractions shown, in terms of specific form factor integrands, ${\cal F}^{(a)}_{\rm UV finite}$, which we will specify below, multiplying lower-order IR regulated amplitudes.   We note that there are actually fewer IR subtraction terms than diagrams that contribute to the original amplitude.   This is because collinear singularities of the full amplitude for production of massive electroweak particles factorize into universal factors that are the same in the full amplitude as in the form factors.   

The basis of the procedure we have just described is the universality of infrared singularities associated with annihilation, and more generally wide-angle scattering amplitudes.   In the QCD annihilation amplitudes that we study in this paper, this universality can be expressed in terms of two factorized ``jet"  functions, a ``soft" function and a perturbative short-distance function \cite{Dixon:2008gr,Ma:2019hjq}.   That is, if we denote by $M_{ew}$ the amplitude for the process $q+\bar q \to ew$ for some explicit electroweak final state ``$ew"$,  
\bea
M_{ew}\ =\ J_q \times J_{\bar q} \times S_{q\bar q} \times H_{ew}\, ,
\label{eq:M_F-fact}
\eea
where all information on final state $ew$ is in the function $H_{ew}$.   
The jet functions and the soft function are exactly the same for a form factor as for any hard annihilation process, and all process-dependence is contained in the hard function.  This feature enables us to use the form factors to organize infrared subtractions, rather than the jet and soft functions themselves, and this is the approach we take below.   Although straightforward in principle, when we implement this procedure at the local level, we will find it necessary look closely at the factorization properties of the form factor amplitudes as well as those of the production amplitudes in question, at the local level.   

As we will review below, the factorization of all collinear singularities into jet functions  $J_i$ is a result of the gauge invariance of the theory, enforced in perturbation theory by its Ward identities.  The jet functions include all collinear singularities, including all double poles in dimensional regularization.    The function $S_{q\bar q}$ organizes any soft singularities of the amplitude that are not absorbed into the jet functions, which are themselves ambiguous up to soft contributions.   Indeed, for this set of amplitudes, it is often convenient to absorb all such contributions into the jet functions, so that $S$ is unity \cite{Sterman:2002qn}.   We shall not need to make choices like this in our discussion below, because they are shared between the simple form factors (in which state ``ew" is a virtual photon, say) and the much larger class of electroweak final states involving multiple weak bosons and Higgs particles.  
To use this universality, we turn to the specific structures of the diagrams that contribute to the processes at hand.

\section{Momentum flows and leading regions}
\label{sec:momentumflow}

We now turn to the assignment of momentum flows, and go on to identify relevant regions in the corresponding integration volume.

\subsection{Assigning momenta}
\label{subsec:momenta}
As a first step, we generate the Feynman diagrams of ${\cal M}_0$,
${\cal M}_1(k)$ and  $ {\cal M}_2(k,l)$ with a direct application of Feynman
rules in the Feynman gauge. In this article, we adopt the Feynman rule conventions of Refs.~\cite{Ellis:1996mzs,Sterman:1993hfp}. Diagrams with self-energy corrections to the gluon propagator and fermion loops are treated as in Ref.~\cite{Anastasiou:2020sdt} and will not be discussed further. In this article, we concentrate on the remaining diagrams. 

\begin{figure}[h]
\begin{center}
  \includegraphics[width=6.5cm]{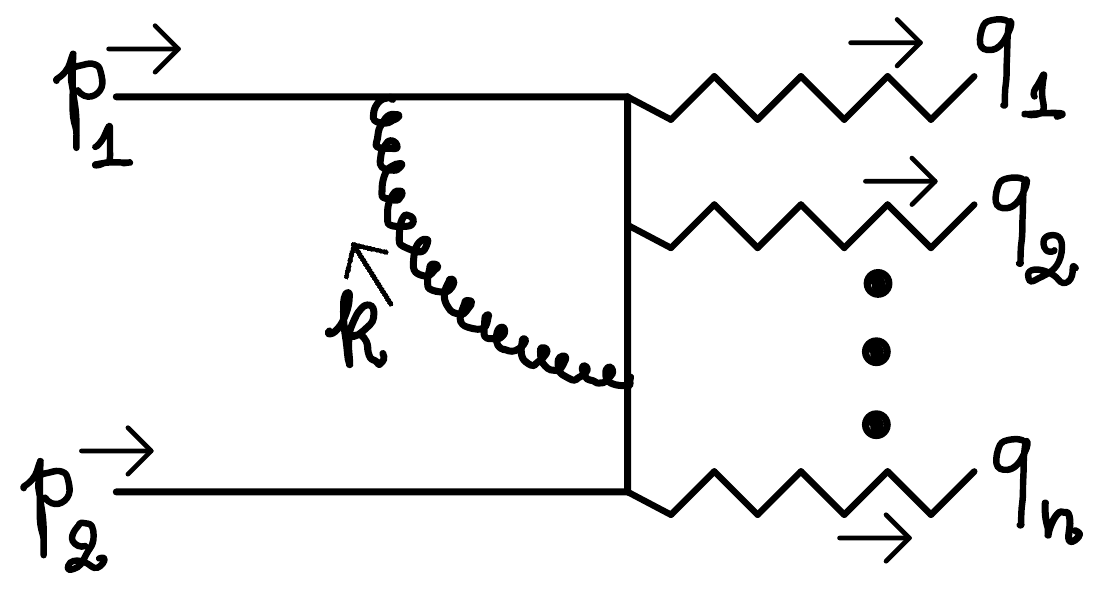}
  \includegraphics[width=6.5cm]{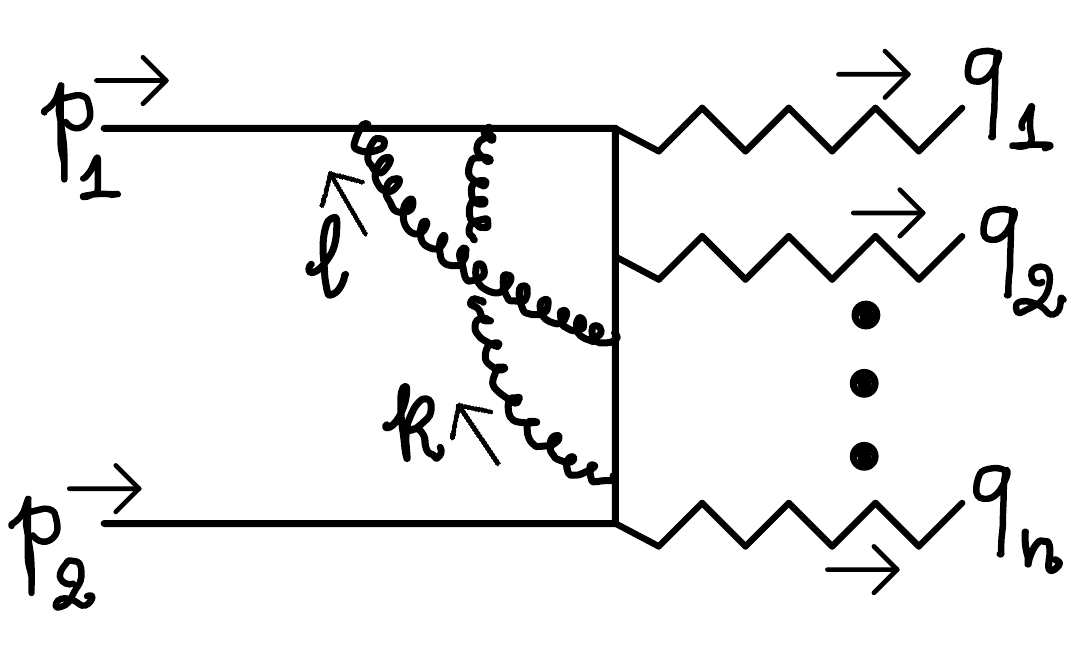}
  \includegraphics[width=8cm]{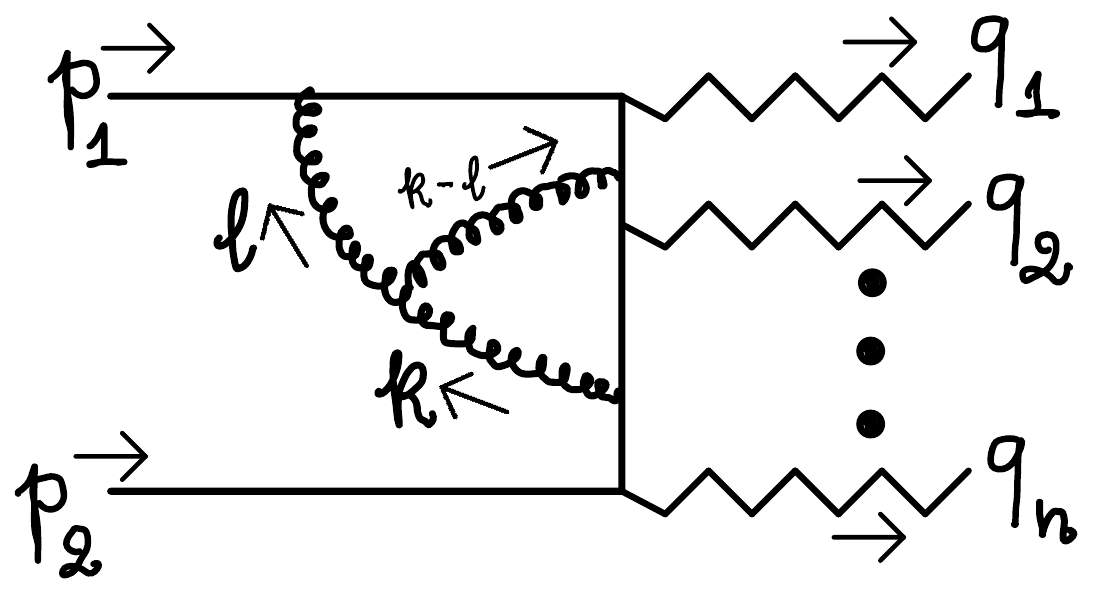}
\end{center}
\caption{\label{fig:loopmomenta} Assignments of loop-momentum variables
in the amplitude through two loops.}
\end{figure}

All Feynman diagrams contain a fermion line connecting the antiquark spinor $\bar v (p_2)$ and the
quark spinor $u(p_1)$.  At one and two loops, we assign the variable $k$ to the momentum of
the gluon flowing out of the quark-antiquark-gluon vertex which is
closest to the incoming antiquark. At two loops, in diagrams with a
triple-gluon vertex we assign a second
loop-momentum variable $l-k$ to the gluon flowing out of the second
quark-antiquark-gluon vertex  closest to the incoming antiquark.
In ``abelian''  diagrams (without a
triple-gluon vertex) we assign a second
loop-momentum variable $l$ to the gluon flowing out of the second
quark-antiquark-gluon vertex  closest to the incoming antiquark.
The assignments of loop-momenta variables are depicted in Fig.~\ref{fig:loopmomenta}.
We will sometimes find it convenient to decompose the loop momenta into components parallel and transverse to
the incoming momenta,
\begin{eqnarray}
  k&=& \frac{k \cdot p_2}{p_1 \cdot p_2} \, p_1
  +\frac{k \cdot p_1}{p_1 \cdot p_2} \, p_2
       + k_\perp\, , \\
   l&=& \frac{l \cdot p_2}{p_1 \cdot p_2} \, p_1
  +\frac{l \cdot p_1}{p_1 \cdot p_2} \, p_2\, .
       + l_\perp \, .
\end{eqnarray}

As in Ref.~\cite{Anastasiou:2020sdt},  we will substitute for the
two-loop integrand an equivalent expression before integration,
\begin{equation}
  \label{eq:loopaverage}
{\cal M}_2(k, l) \to \frac{ 
{\cal M}_2\left(k, l \right)
+{\cal M}_2\left(l, k \right)
}{2} \, .
\end{equation}
In Ref.~\cite{Anastasiou:2020sdt}, a further symmetrization of the
amplitude under $\left( k_\perp, l_\perp\right) \leftrightarrow \left(
  -k_\perp,-l_\perp \right) $ was made. We now find that this symmetrization is not needed for the full amplitude. 
A transverse momentum symmetrization suffices to be applied more finely in certain one-loop vertex subdiagrams. We will cast this refined operation in the form of a
counterterm, whose integral vanishes, in Sec.\ \ref{sec:general-transverse} below.
 
\subsection{Leading Regions and obstacles to local finiteness}

Here, we'll review the list of regions that lead to infrared singularities,  and introduce a convenient notation used to identify them.   After that, we will show how 
Ward identities are invoked, and how to overcome the obstacles to their local implementation.    The regions in question correspond to configurations in 
momentum space where one or both of the loop momenta $k$ and $l$ either become lightlike and collinear to the incoming quark or antiquark, or are soft.   

As in Ref.\ \cite{Anastasiou:2020sdt}, we identify the full list of such regions by the notation $\rm (A,B)$, where by convention we choose $\rm A \in \{1_k,2_k,H_k,S_k\}$,          and
similarly $\rm B \in \{1_l,2_l,H_l,S_l\}$.   As a reminder of the implied ordering, in this paper we provide subscripts to identify the loop momenta, so that, for
example, $\rm (1_k,S_l)$ labels the region where loop $k$ is lightlike and collinear to $p_1$ and $l$ is soft.   We emphasize that these regions refer to
specific diagrams with momentum assignments already made, since in the final expressions we assume the symmetrizations of Eq.\ (\ref{eq:loopaverage}).

\begin{figure}
\begin{equation*}
  \eqs[0.33]{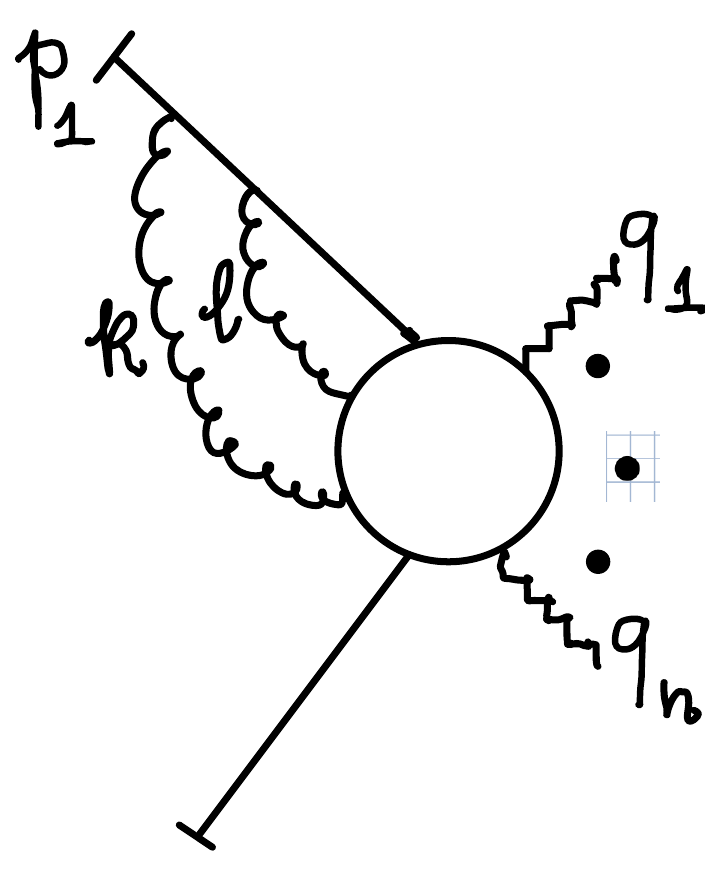}
\quad 
  \mbox{{\Large$ \underrightarrow{\quad k,l \parallel p_1 \quad }$}}
\; 
  \eqs[0.33]{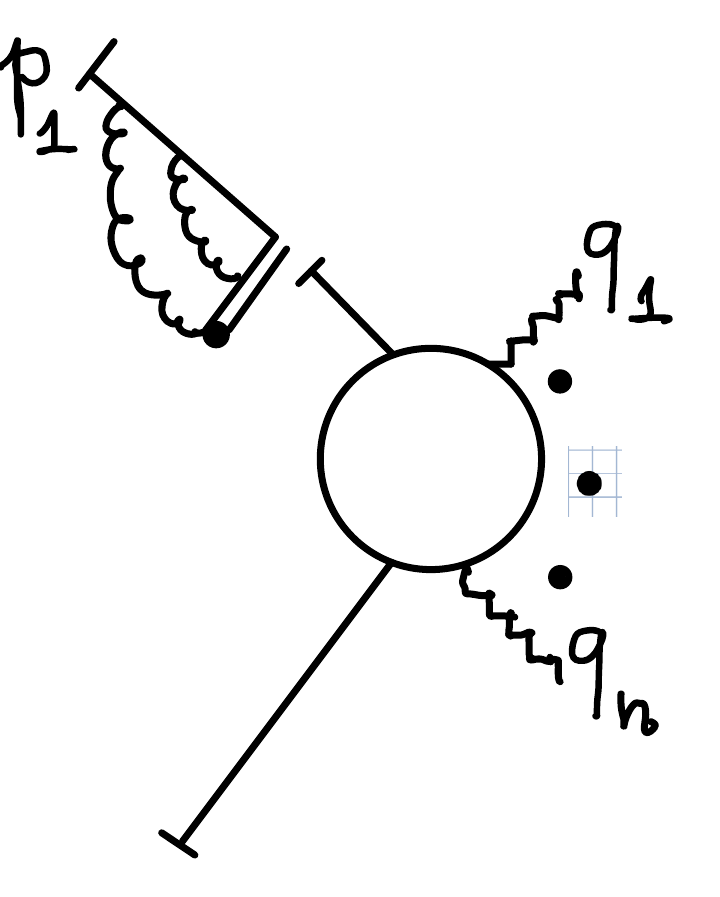}
\end{equation*}
\caption{ Identity for the double collinear region $\rm (1_k,1_l)$.   When the short-distance function is at tree level, this identity is realized for the full singularity locally, with no shifts of loop momentum.}   \label{fig:double-collinear}
\end{figure}

The underlying factorization of the quark pair annihilation amplitudes, $M_{ew}$ in Eq.\ (\ref{eq:M_F-fact}) makes the subtraction scheme of Eq.\ (\ref{eq:cal-M-finite-12}) possible.  In particular for all regions in which each loop momentum, $k$ and $l$, becomes either collinear to $p_1$ or $p_2$, or soft, the matching of singular integrands in all processes is at the local level in the standard forms of the integrand, after the sum over diagrams.   Figure \ref{fig:double-collinear} illustrates this result.   In the limit that $k$ and $l$ are both collinear to $p_1$, their polarizations both become longitudinal, and the sum over all connections of these two gluons to subdiagram $H_{ew}$, with an arbitrary electroweak final state, gives a universal term, independent of the final state.  In this figure, the double line represents a Wilson (eikonal) line in the direction opposite to $p_1$.    At the order to which we work, this identity requires only the tree-level Ward identities of the theory, and holds locally in momenta $k$ and $l$.  Other configurations, where both $k$ and $l$ are lightlike or soft behave in just the same way.   

\begin{figure}
\begin{center}
\begin{equation*}
  \eqs[0.33]{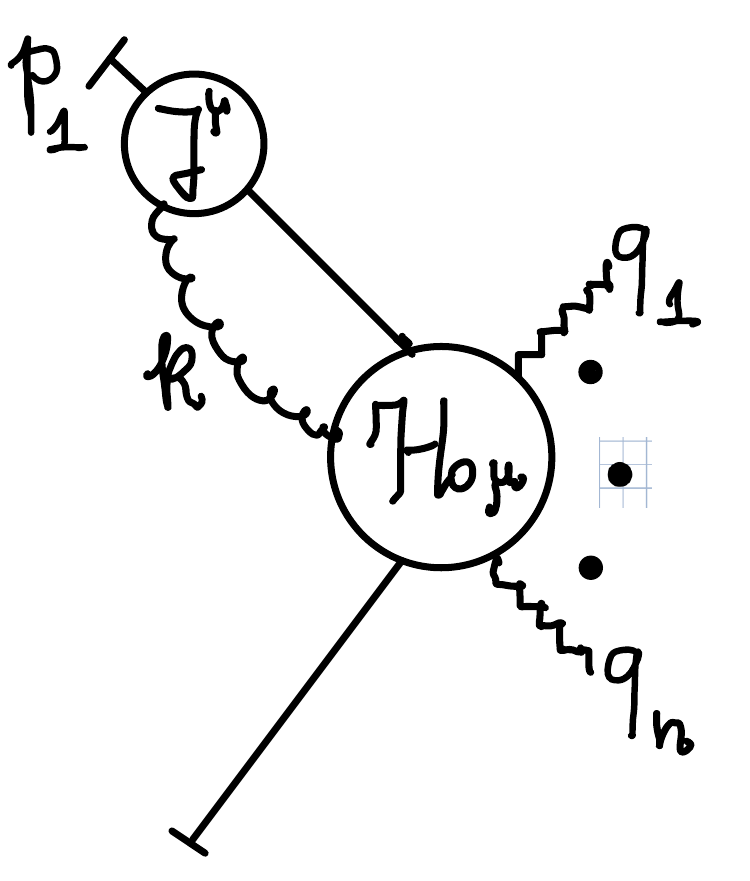}
\quad 
  \mbox{{\Large$ \underrightarrow{\quad k \parallel p_1 \quad }$}}
\; 
  \eqs[0.33]{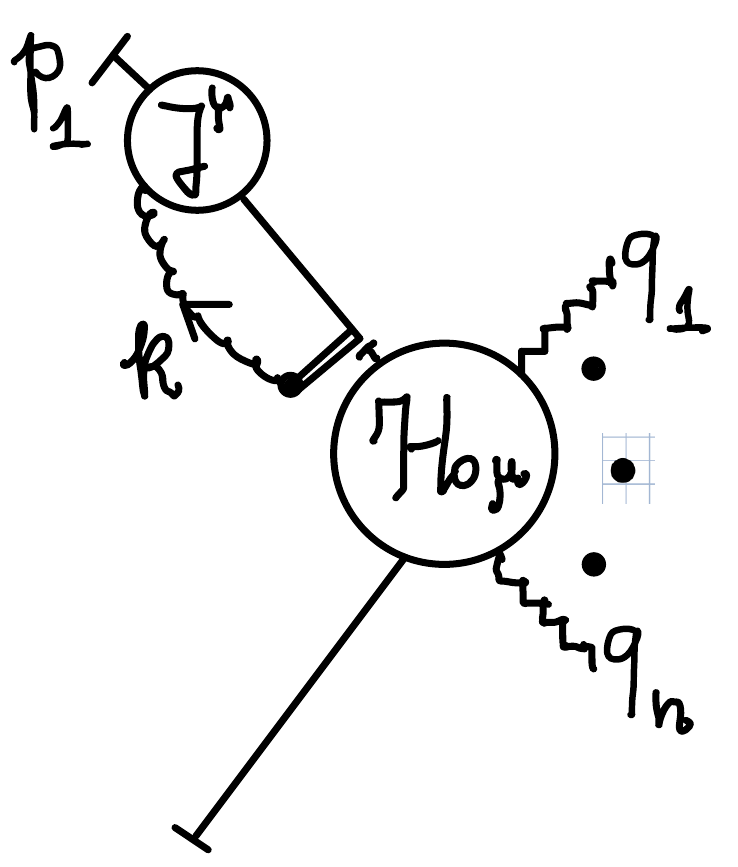}
\end{equation*}
\end{center}
\caption{ Generic diagram and Ward identity for single-collinear regions. At two loops, either the jet subdiagram ${\cal J}^\mu$ or the hard part ${\cal H}_\mu$ is one-loop, while the other is at tree level.  Loop polarizations arise when ${\cal J}^\mu$ is one-loop, and shift mismatches when ${\cal H}_\mu$ is one-loop.  The subdiagram ${\cal J}^\mu$ is single-particle irreducible.}   \label{fig:two-circles}
\end{figure}

The situation becomes more complicated when one of the two gluon lines is collinear to either $p_1$ or $p_2$, while the other line is hard.   At two loops, these difficulties arise in such ``single-collinear" regions in two ways, illustrated for a gluon of momentum $k$ which is parallel to $p_1$, in Fig.\ \ref{fig:two-circles}.   In this figure, we consider the regions $\rm (1_k,H_l)$ when one of subdiagrams ${\cal J}^\mu$ or ${\cal H}_\mu$ is evaluated at one loop, while the other remains at tree level.  Note that we denote these subdiagrams in script, because we are working at the level of integrands.   We define ${\cal J}^\mu$ to be one-particle irreducible, so that it does not include the propagators of the quark with momentum $p_1-k$ or of the gluon of momentum $k$, which are common to all diagrams in the jet.

The first of these obstacles occurs when the jet subdiagram ${\cal J}^\mu$ has one loop, and we encounter the problem of ``loop polarizations" \cite{Anastasiou:2020sdt}.
That is, we encounter terms where  the gluon carries a polarization proportional to the jet function integrand at each point in the loop momentum $l$,
\bea
{\cal J}^\mu(p_1,l)\ \propto \ l^\mu + \dots \, .
\eea
  For components of $l^\mu$ not in the direction of $p_1$, the
  gluon $k$ will then not factor from the hard scattering in the manner illustrated by Fig.\ \ref{fig:two-circles}.
  In integrated form, $\int d^4l {\cal J}^\mu(p_1,l)$, the jet function is obviously no longer a function of $l$. In the full integral over $l^\mu$, such terms integrate to factorized forms.
Nevertheless, this effect is an obstacle to the local realization of the factorization 
of Eq.\ (\ref{eq:M_F-fact}), and hence to the construction of finite integrands in Eq.\ (\ref{eq:cal-M-finite-12}) at the integrand level.   

A second obstacle to local factorization at two loops arises when it is the hard-scattering subdiagram ${\cal H}_\mu$ of Fig.\ \ref{fig:two-circles} that has the additional loop, while ${\cal J}^\mu$ stays at tree level.  In this case, the Ward identity of Fig.\ \ref{fig:double-collinear} requires in general a shift of loop momentum within ${\cal H}^\mu$.   This is the case whenever all lines in the loop carry color charge.   Such a shift is again not consistent with the integrand-level construction of Eq.\ (\ref{eq:cal-M-finite-12}).   We will refer to these below as {\it shift mismatches}.

In addition to overcoming these infrared obstacles, we must also observe that for our finite integrands ${\cal M}^{(1,2)}_{\rm finite}$ to be accessible to numerical integration, they must also be ultraviolet regulated.

In Ref, \cite{Anastasiou:2020sdt}, we provided a construction to solve these problems in a class of processes in massless quantum electrodynamics.
In the following sections, we provide prescriptions for reconciling these features of perturbative QCD with the construction of locally infrared finite and ultraviolet-regulated integrands.   We begin with a discussion of the jet loop diagrams that involve loop polarizations.


\section{V and S subdiagrams and loop polarizations}
\label{sec:VandS}

\begin{figure}
\begin{center}
  \includegraphics[width=4cm]{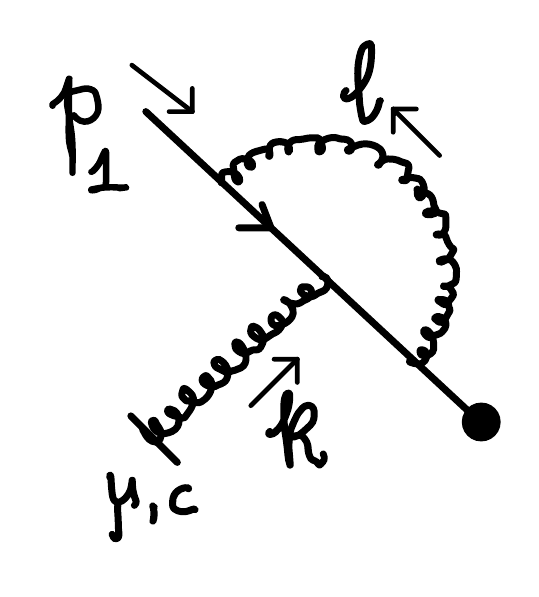}
  \includegraphics[width=4cm]{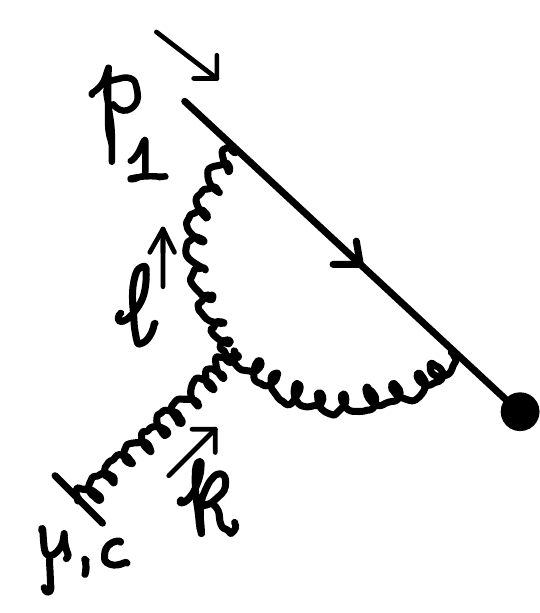}
  \includegraphics[width=4cm]{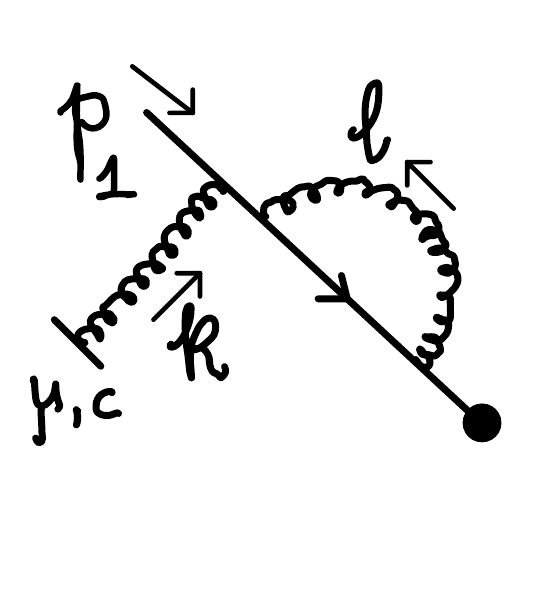}
\end{center}
\caption{ V and S diagrams for the jet $p_1$.
\label{fig:VandS}}
\end{figure}

\begin{figure}
\begin{center}
\includegraphics[width=5cm]{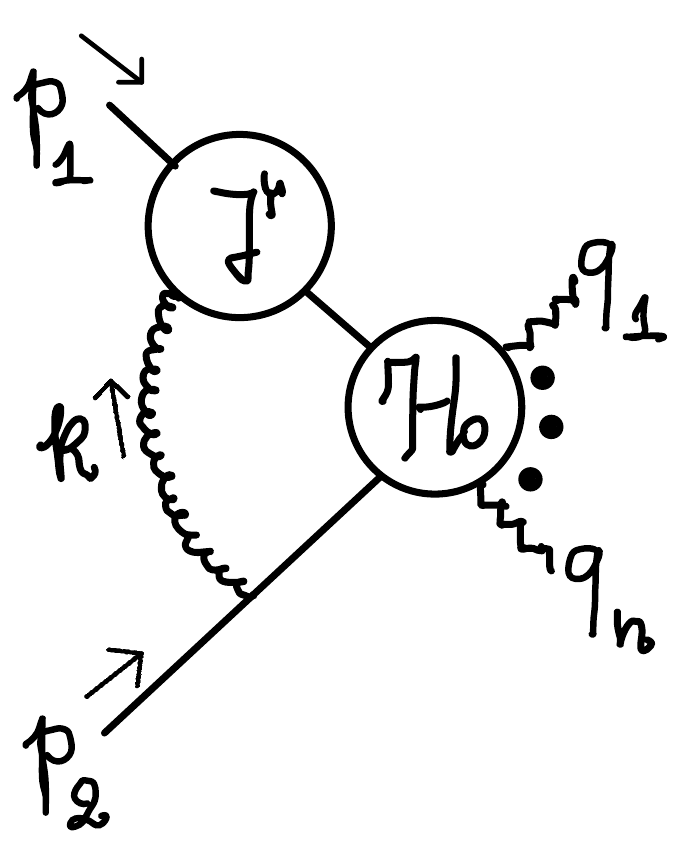}
\end{center}
\caption{ A $V$ diagram on the quark line that is singulart in region $\rm (2_k,H_l)$, where the exchanged gluon is parallel to $p_2$, in addition to $\rm (1_k,H_l)$, where it is parallel to $p_1$.  \label{fig:k1-k2-V}}
\end{figure}

We begin by studying loop polarizations in the one-loop diagrams adjacent to the external line $p_1$, as illustrated by Fig.\ \ref{fig:two-circles}. These are the triangles and the self energy diagrams shown in Fig.\ \ref{fig:VandS}.   Among these diagrams, the special case when gluon $k$ is emitted at a vertex adjacent to the external antiquark is shown in Fig.\ \ref{fig:k1-k2-V}.   In this case, the diagram has a single-collinear divergence in region $\rm (2_k,H_l)$ in addition to $\rm (1_k,H_l)$.

Following Ref.\ \cite{Anastasiou:2020sdt}, we refer to the vertex and self-energy diagrams of Fig.\ \ref{fig:VandS} as V and S type diagrams, respectively.  Note that loop polarizations (${\cal J}^\mu$ one-loop and ${\cal H}_\mu$ tree in Fig.\ \ref{fig:two-circles}) occur only in these diagrams and from regions $\rm (1_k,H_l)$, while $\rm (H_k,1_l)$ corresponds to a one-loop hard scattering subdiagram (${\cal J}^\mu$ tree and ${\cal H}_\mu$ one-loop in Fig.\ \ref{fig:two-circles}) .  As above, ${\cal J}^\mu$ is always a single-particle irreducible diagram.  We discuss both regions because we will need to modify the integrands of certain diagrams to deal with loop polarizations, and we must check that these modifications do not affect other regions.  We begin our discussion with what we will refer to as the ``QED triangle", the vertex diagram in Fig.\ \ref{fig:VandS} with three quark-gluon vertices.   We then go on to the ``QCD triangle", with a three-gluon vertex, and finally the relevant self-energy (Type S) diagrams.

To anticipate, in this and subsequent sections, we will detail the three ingredients of our construction of a locally finite integrand for our amplitudes. First, we will modify Feynman graphs or subgraphs in order to make infrared factorization manifest locally. Then we will introduce IR counterterms, and finally we will introduce ultraviolet counterterms.  

\subsection{Loop polarizations in type V diagrams I:  the QED triangle}
 
A Type V QED-type correction is a vertex adjacent to the incoming quark line, corresponding to integrand factors of the form
\begin{equation}
 \eqs[0.2]{./figures/TypeVqedUp} =  g_s^3 T_c^{(q)} \, \left( \frac
   {C_A}{2} - C_F \right)
 V^{\mu}\left( k,
   l\right) u\left(p_1\right)\, ,
   \label{eq:V-k-l}
\end{equation}
where $T_c^{(q)}$ is the color generator in fundamental representation.   Adopting the convention of Eq.\ (\ref{eq:cal-M-normalization}), the factors of $(2\pi)^{-D}$ associated with loop momenta 
are already accounted for.
A direct application of Feynman rules in the conventions of Refs.~\cite{Ellis:1996mzs,Sterman:1993hfp} yields
\begin{eqnarray}
V^{\mu}\left( k,
  l\right) &=&
\frac{
  \gamma^\nu 
(\slashed k +\slashed l + \slashed p_1)
  \gamma^\mu
(\slashed l + \slashed p_1)
  \gamma_\nu
}{
l^2 \left( l+p_1\right)^2  \left(k+l+p_1\right)^2
}\,.
\label{eq:V-def}
\end{eqnarray}
We will study singularities that arise when terms that result from this subdiagram are inserted into any of the two-loop diagrams in the class under study (quark-antiquark annihilation to color-neutral final states).   Our goal is to identify terms associated with loop polarizations, which factorize after integration, but for which tree-level Ward identities do not immediately result in factorized singular integrands that cancel in the subtracted amplitude, Eq.\ (\ref{eq:cal-M-finite-12}).

To begin this analysis, we write $V^\mu$ as an sum of two terms with differing structure of the collinear singularities,
\begin{eqnarray}
V^{\mu}\left( k,
  l\right) &=&
               V_{k}^{\mu}\left( k,l\right)
               +V_{l}^{\mu}\left( k, l\right)
 \, .
  \label{eq:split-V}
\end{eqnarray}
Here, term $V_l$ produces a single-collinear divergence for $l$
collinear to $p_1$ and $k$ hard (${\rm (H_k,1_l)}$ region), while term
$V_k$  is single-collinear divergent when $l$ is hard, and $k$ is
collinear to either $p_1$ or, in the diagram of Fig.\ \ref{fig:k1-k2-V}, $p_2$ (${\rm (1_k,H_l)}$ and ${\rm
  (2_k,H_l)}$ regions).
Explicitly, $V_{l}$ is given by
\begin{eqnarray}
V_{l}^{\mu}\left( k,l\right) &=&
\frac{
  \gamma^\nu
\slashed k
  \gamma^\mu
(\slashed l + \slashed p_1)
  \gamma_\nu
}{
l^2 \left( l+p_1\right)^2  \left(k+l+p_1\right)^2
},
\label{eq:Vl-def}
\end{eqnarray}
and $V_k$ by
\begin{eqnarray}
V_{k}^{\mu}\left( k,  l\right) &=& 
\frac{
  \gamma^\nu
(\slashed l + \slashed p_1)
  \gamma^\mu
(\slashed l + \slashed p_1)
  \gamma_\nu
}{
l^2 \left( l+p_1\right)^2  \left(k+l+p_1\right)^2\
}\, .
\label{eq:Vk-def}
\end{eqnarray}

Let's see why $V_{l}^\mu$ yields a contribution to two-loop diagrams which is
singular in only the $l \parallel p_1$ single-collinear limit. 
In combination with the Dirac spinor of the incoming quark, as in Eq.\ (\ref{eq:V-k-l}), and also with the numerator
factor of the adjacent quark propagator with momentum $p_1+k$, we have,
\begin{eqnarray}
\left(\slashed k + \slashed{p}_1\right)\, V_{l}^{\mu}\left(
  k,l\right)\, u(p_1) \ &=& \left(\slashed k + \slashed{p}_1\right)\, \,
\frac{
-2 \, (\slashed l + \slashed p_1)
  \gamma^\mu
                                            \slashed k
+2 \, \epsilon \, \slashed k 
  \gamma^\mu
                                            \slashed l
                                            }
{ l^2 \left( l+p_1\right)^2  \left(k+l+p_1\right)^2
}
\, u(p_1)\, .
\label{eq:Vl-2}
\end{eqnarray}
When $k^\mu$ becomes collinear (that is, proportional) to $p_1^\mu$, this expression vanishes by the Dirac equation and $\slashed p_1^2=0$.   This suppression eliminates any singularity from region ${\rm (1_k,H_l)}$.    In the region $\rm (2_k,H_l)$, on the other hand, for the special case of Fig.\ \ref{fig:k1-k2-V}, the matrix $\gamma^\mu$ is contracted with a vector proportional to $p_2^\mu$.  Then, because $k^\mu \propto p_2^\mu$ in this region, $V_l$ vanishes because $\slashed p_2^2=0$.   

The remaining singularity in $V_l(k,l)$ is from region $\rm
(H_k,1_l)$, where $l \parallel p_1$. This singularity factorizes in the sum of Feynman diagrams in the two-loop amplitude as in Fig.\ \ref{fig:two-circles}, and we will cancel it against a two-loop form-factor counterterm.   As we shall see in Sec.\ \ref{sec:construction}, however, this analysis will require treatment for a shift mismatch.    

In contrast to $V_{l}^\mu$,  $V_{k}^\mu$  provides terms
that are all finite as $l \parallel p_1$ but singular as $k \parallel
p_1$ and, in the special case of Fig.\ \ref{fig:k1-k2-V}, $k \parallel  p_2$.
Among these, we will find terms for which the $k \parallel p_1 $ singularity is not factored in the sum of diagrams, due to the presence of loop polarizations. 

For all the diagrams shown in Fig.\ \ref{fig:VandS} and \ref{fig:k1-k2-V}, the singular,
single-collinear region $\rm (2_k,H_l)$, where $k \parallel p_2$,  has
no loop polarization, because the vertex loop is adjacent to the
$p_1$ line. Although free of loop polarizations, these contributions
suffer from a shift mismatch, which in dimensional regularization is
eliminated by the vanishing of a scale-less integral, in this case the
integral of the renormalized one-loop on-shell self-energy diagram on
the incoming quark line.   For any consistent UV regularization, these
contributions cancel upon integration, but they spoil the local
cancellation of singularities and hence for us are problematic.    We will see below, however, that the modification that eliminates loop polarizations in region $\rm (1_k,H_l)$ also eliminates shift mismatches in region $\rm (2_k,H_l)$.
 
To study the role of term $V^\mu_k(k,l)$ in Eq.\ (\ref{eq:Vk-def}), we rewrite Eq.\ (\ref{eq:Vk-def}), acting on the external spinor, as
\bea
 V_{k}^{\mu}\left( k,l\right)\, u(p_1) \ &=&\ -2(1-\epsilon)\, \frac{1}{(k+l+p_1)^2}
\left[ \frac{2(l^\mu+p_1^\mu) \slashed l}{l^2(l+p_1)^2} \ -\ \frac{\gamma^\mu}{l^2} \right]\, u(p_1)\, .
\label{eq:V-k-on-u}
\eea
Here, we see the loop polarization explicitly.   In the following analysis, we will recall that, as for $V_l^\mu$ in Eq.\ (\ref{eq:Vl-2}), in each full diagram, $V^\mu_k(k,l)$ is multiplied from the left by a quark propagator, with numerator $\slashed p_1+\slashed k$.   In single-collinear region $\rm (1_k,H_l)$, this factor gives zero acting on the external Dirac spinor.   After a simple calculation using this result, we find that the only loop polarization in this expression that is singular in $\rm (1_k,H_l)$ is associated with the term in $\slashed l$, which neither commutes nor anticommutes with $\slashed p_1$ in general.   We can thus isolate the singular loop polarization by decomposing $\slashed l$ into light-cone components,
\bea
 V_{k}^{\mu}\left( k,l\right)\, u(p_1) \ &=&\ -2(1-\epsilon)\, \frac{1}{(k+l+p_1)^2}
 \nn\\[2mm]
 &\ & \hspace{-35mm} \times
\left[ \frac{2(l^\mu+p_1^\mu)}{l^2(l+p_1)^2} \left[ \frac{2l\cdot
      {\eta_1}}{2p_1\cdot {\eta_1}}\psla_1 + \frac{2l\cdot
      p_1}{2p_1\cdot {\eta_1}}\slashed{\eta}_1 + \lsla_{\perp(p_1, \eta_1)} \right ] \ -\ \frac{\gamma^\mu}{l^2} \right]\, u(p_1)
\nn\\[2mm]
&=&\ -2(1-\epsilon)\, \frac{1}{(k+l+p_1)^2}
 \nn\\[2mm]
 &\ & \hspace{-35mm} \times
\left[ 2(l^\mu+p_1^\mu)  \left[  \left( \frac{1}{l^2} -
      \frac{1}{(l+p_1)^2} \right)  \frac{\slashed{\eta}_1}{2p_1\cdot
      {\eta_1}} +  \frac{\lsla_{\perp(p_1, \eta_1)}}{l^2(l+p_1)^2} \right ] \ -\ \frac{\gamma^\mu}{l^2} \right]\, u(p_1)\, .
\nn\\
\label{eq:V-k-loop-pol}
\eea
For the light-cone decomposition of the loop momentum, we have used a lightlike auxiliary vector
$\eta_1$, with $\eta_1 \cdot p_1 \neq 0$. For two-loop V-type diagrams,
which are also singular in the region ${\rm (2_k, H_l)}$, we shall
see that we must set $\eta_1 = p_2$. In other V-type diagrams, which are not
singular in this region, we can choose any lightlike $\eta_1$ that is not proportional to $p_1$. 

To arrive to the second expression in Eq.\ (\ref{eq:V-k-loop-pol}), we applied the Dirac equation and the relation, $2p_1\cdot l = (p_1+l)^2-l^2$.   This isolates the singular loop polarization in two terms, each with two, rather than three, propagator denominators.   In our explicit construction below, we will eliminate these terms of the integrand, replacing them with terms that are free of loop polarizations, yet leave the result of the integrals unchanged.   We can see at this stage that this is possible, by noting the integral identities,
\bea
\int \frac{d^Dl}{(2\pi)^D} \frac{  l^\mu}{l^2 (p_1+k+l)^2}\ &=&\ - \frac{1}{2} \int \frac{d^Dl}{(2\pi)^D} \frac{  k^\mu + p_1^\mu }{l^2 (p_1+k+l)^2}\, ,
\nn\\[2mm]
\int \frac{d^Dl}{(2\pi)^D} \frac{  l^\mu}{(p_1 + l)^2 (p_1+k+l)^2}\ &=&\ - \frac{1}{2} \int \frac{d^Dl}{(2\pi)^D} \frac{  2p_1^\mu +  k^\mu }{l^2 (p_1+k+l)^2}
 \, ,
 \label{eq:l-integral-identities}
\eea
which can be proved using changes of variables, $l'=-(l+k+p_1)$ and $l'=-(l+k+2p_1)$, respectively, for $D$ less than four dimensions.   Using these results below, in four dimensions, it will be possible to introduce ultraviolet counterterms to ensure that the resulting integrals remain convergent, while again leaving the results of the integrals unchanged.   Using these identities, we will cancel loop polarizations locally.

\subsection{Loop polarizations in type V diagrams II:  the QCD triangle}

We identify loop polarizations in the QCD triangle in a similar fashion.   The vertex  takes the form
\begin{equation}
  \eqs[0.2]{./figures/TypeVqcdUp} = g_s^3 T_c^{(q)}\, \frac{C_A}{2}\,  W^{\mu}\left( k,l \right) u\left(p_1\right)\, .
\label{eq:W-k-l}
\end{equation}
The momentum dependence of the integrand in the truncated QCD triangle diagram in Fig.\ \ref{fig:VandS} will be written as
\bea
W^\mu(k,l)\ &=&\   
W_{\rm scalar}^\mu(k,l)\, +\, O^\mu(k,l) 
\, ,
\label{eq:W-def}
\eea
where, as for the QED vertex, it is convenient to isolate terms on the basis of their behavior in different regions.
As we will describe, the vectors $W^\mu_{\rm scalar}$ and $O^\mu$ have different behavior in region  $\rm (1_k,H_l)$.   They also give, respectively, self-energy and ghost contributions to the Ward identity for external gluon $k$, which we will also review below. 

The first vector is generated from the ``scalar" term of the three-gluon vertex, and is given by
\bea
  W^{\mu}_{\rm scalar}(k,l)\ =\  
  \frac{
 (2 l - k)^\mu \,  \gamma_\alpha \left (\slashed{l} + \slashed{p}_1 \right ) \gamma^\alpha
  }                      
{l^2 \, (l+p_1)^2 \, (k-l)^2}\, .
\label{eq:W-scalar-def}
\eea
Acting on the Dirac spinor, this simplifies in $D=4-2\epsilon$ dimensions to
\bea
  W^{\mu}_{\rm scalar}(k,l)\, u(p_1)\ =\  
-\,   \frac{
2\left(1 - \epsilon \right) (2 l - k)^\mu \slashed{l} 
  }                      
{l^2 \, (l+p_1)^2 \, (k-l)^2}
\, u(p_1)\, ,
\label{eq:W-scalar-u}
\eea
and we see explicit loop polarizations.   As in the QED vertex in Eq.\ (\ref{eq:V-k-loop-pol}), we isolate loop polarizations that are singular in region $\rm (1_k,H_l)$  by expanding the vector $l^\mu$ in $\lsla$, in terms of its components in the $p_1$, $\eta_1$ and perpendicular directions.   Then, using $p_1^2=0$ and the Dirac equation in Eq.\ (\ref{eq:W-scalar-u}), we  find
\bea
W^\mu_{\rm scalar}(k,l)\, u(p_1)\ &=&\ 
- \, 
 \frac{   2(1-\epsilon)\, (2l-k)^\mu}{(p_1+l)^2l^2(k-l)^2} \, 
\left[ \frac{2l\cdot {\eta_1}}{2p_1\cdot {\eta_1}}\psla_1 +
  \frac{2l\cdot p_1}{2p_1\cdot {\eta_1}}{{\slashed \eta}_1} +
  \lsla_{\perp(p_1, \eta_1)} \right ] u(p_1)
\nn\\[2mm]
&\ & \hspace{-20mm} =\ -\,   \frac{2(1-\epsilon) \, (2l-k)^\mu}{(k-l)^2} \ \Bigg [  \frac{ {{\slashed \eta}_1}}{2p_1\cdot {\eta_1}}\, \left ( \frac{1}{ l^2}\, -\,  \frac{1}{ (p_1+l)^2} \right )
\, +\, \frac{ \lsla_{\perp(p_1, \eta_1)} }{ (p_1+l)^2l^2 } \Bigg ]\, u(p_1)\, .
\nn\\
\label{eq:W-scalar-loop-pol}
\eea
 As in the case of the QED triangle contribution, $V_k(k, l)$ of Eq.~(\ref{eq:V-k-loop-pol}), the $l^\mu$ terms here are not
 singular in region $(H_k,1_l)$.    A singularity in this region must
 arise from the simultaneous vanishing of the denominators $l^2$ and
 $(p_1-l)^2$, but in the final expression for $W_{\rm scalar}^\mu$ the
 only term that has this combination is suppressed by the transverse
 components $l_{\perp(p_1, \eta_1)}$ in the numerator.

The other vector on the right-hand side of Eq.\ (\ref{eq:W-def}), $O^\mu$, is given by the remaining two terms of the three-gluon vertex,
\begin{eqnarray}
  O^{\mu}\left(k, l\right) &=& -\, \frac{
  \left( \eta^{\mu\alpha }(l-2k)^\beta\, +\, \eta^{\beta\mu} ( l+k)^\alpha \right )\, \gamma_\alpha\, (\lsla + \psla_1   ) \gamma_\beta  }                      
                               {l^2 \, (l+p_1)
                               ^2 \, (k-l)^2}
\nn\\[2mm]
  &=& -\,  \frac{
  \gamma^\mu \left(\slashed{l} +\slashed{p}_1 \right) \left( \slashed{l}
                        - 2 \slashed{k} \right)
+  \left(\slashed{l} +\slashed{k} \right) \left( \slashed{l}
                        + \slashed{p}_1 \right)
  \gamma^\mu  }                      
                          {l^2 \, (l+p_1)^2 \, (k-l)^2}\, .
\label{eq:O-def}
\end{eqnarray}
The contribution of vector $O^\mu$ to the vertex is easily seen to be free of loop polarizations for terms that are singular in region $\rm (1_k,H_l)$.   When inserted into the diagram in Fig.\ \ref{fig:VandS}, all of its terms are either explicitly proportional to $k^\mu$ or $p_1^\mu$, or can be shown by simple Dirac algebra to vanish in the limit that $k$ becomes collinear to $p_1$. (The fermion numerator factor $\psla_1+\ksla$ in the figure plays a role as for the QED vertex terms above.)  Explicitly, we find,
\bea
O^\mu(k,l)\ u(p_1) &=&\   \frac{  1  } {(l+p_1)^2l^2(k-l)^2} \,\Bigg [ 2\gamma^\mu \Big [ l^2 + p_1\cdot (l - 2k)  \Big ]
\nonumber\\[2mm]
& & \hspace{10mm} +\ 2p_1^\mu\, (\lsla+\ksla) \, -2\, \gamma^\mu\, \lsla\, \ksla \, + \ksla\, \lsla \gamma^\mu  \Bigg ] u(p_1)\, .
\label{eq:Omu-def}
\eea
  The $O^\mu$ term has a collinear singularity in region $\rm (H_k,1_l)$, and in fact has the entire collinear singularity in this region.   The $W^\mu_{\rm scalar}$ term is suppressed in this region by the factor $\lsla\, u(p_1)$ (see Eq.\ (\ref{eq:W-scalar-u})).

\subsection{Type S diagrams}

The Type S diagram for the $p_1$ line is a quark self energy.   Written in standard form, this diagram also leads to a term with loop polarization, which, of course,
integrates to factorizable form.   This contribution is entirely avoided, however, if we use symmetric integration to reduce the quark self-energy to a scalar integral.
To be specific, we introduce the factors
\begin{equation}
N_{S-q}(k, l) =\frac{  \left( 1-\epsilon \right) }{l^2
\, \left(l+k+p_1 \right)^2}\, ,
\end{equation}
and
\begin{equation}
N_{S-{\bar q}}(k, l) =\frac{  \left( 1-\epsilon \right) }{l^2
\, \left(l+k-p_2 \right)^2}\, ,
\end{equation}
for quark and antiquark, respectively.
We then perform the replacements
\begin{eqnarray}
 \label{eq:Nsquark} 
\eqs[0.2]{./figures/TypeSup}   &=& \eqs [0.2]{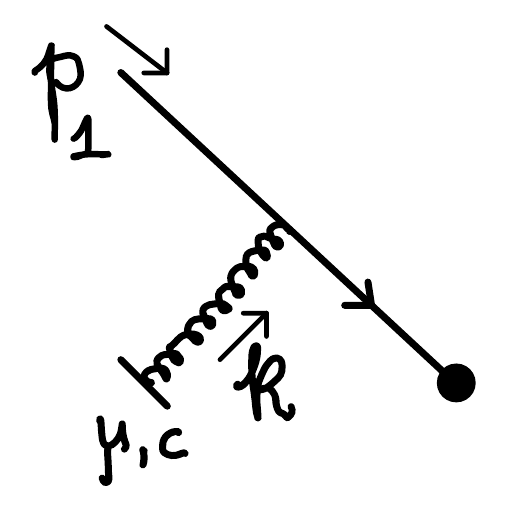} \;
\times \;  i g_s^2 C_F\, N_{S-q}(k, l)
\nn\\[2mm]
                               &=& \  g_s^3 C_F\, N_{S-q}(k, l)\, T_c^{(q)} \, \gamma^\mu \,  u(p_1)
\end{eqnarray}
and, for a self-energy diagram on the incoming antiquark,
\begin{eqnarray}
   \label{eq:Nsantiquark} 
\eqs[0.2]{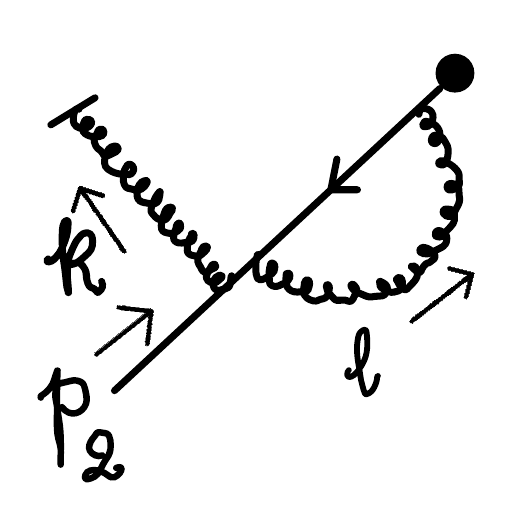}  &=&  \eqs [0.2]{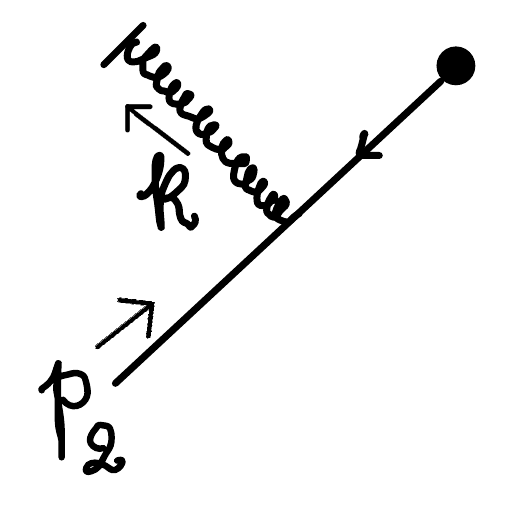} \;
\times \; i g_s^2 C_F\, N_{S-\bar q}(k, l)
\nn\\[2mm]
&=& \  g_s^3 C_F\, N_{S-\bar q}(k, l)\, T_c^{(q)} \, {\bar v}(p_2) \,
    \gamma^\mu \, . 
\end{eqnarray}
The above modifications do not alter the integrated value of the
amplitude.

In summary, after symmetric integration for type S diagrams, the remaining singular loop polarization terms are found in Eqs.\ (\ref{eq:V-k-loop-pol}) and (\ref{eq:W-scalar-loop-pol}).  
Having identified these terms,  we are now ready to show how to rewrite the corresponding contributions to the  integrands in a manner that explicitly
removes all singular loop polarizations at the local level.    As we will see, it is possible to do this without changing the results of integration, by the addition of counterterms, based on the identities in Eq.\ (\ref{eq:l-integral-identities}).  The modified integrands will satisfy the Ward identities of Fig.\ \ref{fig:two-circles} locally, making possible the local cancellation
of these regions in the subtracted amplitudes of Eq.\ (\ref{eq:cal-M-finite-12}).    We emphasize,
that as for the QED amplitudes studied in Ref.\ \cite{Anastasiou:2020sdt}, these counterterms are added to both the electroweak
amplitude in question and to the form factor that defines its subtractions.

\subsection{Counterterms for loop polarizations}
\label{sec:loop-pol-counterterms}

In this section, we will construct the specific subtractions that remove loop polarizations in Eqs.\ (\ref{eq:V-k-loop-pol}) and (\ref{eq:W-scalar-loop-pol}).   
The full set of $V$ and $S$ integrands under consideration appear as a jet integrand, as in Fig.\ \ref{fig:two-circles}, and are combined in the full integrand as
\begin{eqnarray}
{\cal J}_c^\mu(k,l)\ &=&\ g_s^3 T^{(q)}_c\, \left[ \left( \frac{C_A}{2} - C_F \right)\, V^\mu(k,l) + \frac{C_A}{2}\, W_{\rm scalar}^\mu(k,l)  + C_F\, \gamma^\mu\, N_{S-q}(k,l)
\right. \nonumber\\
&\ & \left. \hspace{5mm} +\  \frac{C_A}{2}\, O^\mu(k,l)  \right ]\, ,
\label{eq:Jmu-VWSO}
\end{eqnarray}
where the vectors in square brackets are given, in order from the left, by  Eqs.\ (\ref{eq:V-def}),  (\ref{eq:W-scalar-u}), (\ref{eq:Nsquark}), and (\ref{eq:O-def}), respectively. 
As above, we suppress the external momentum in the arguments of these functions.  These are the integrands in S and V diagrams that contain all loop polarizations, and that we will modify.   We present results for the incoming quark jet; diagrams associated with the antiquark are found in the same way, and can be generated by replacing $p_1$ by $-p_2$, just
as for the quark and antiquark self-energies, Eqs.\ (\ref{eq:Nsquark}) and (\ref{eq:Nsantiquark}).

We begin our analysis of the jet vector integrand by separating terms in the one-loop jet function, Eq.\ (\ref {eq:Jmu-VWSO}) that are divergent in the single-collinear regions $\rm (1_k,H_l)$, and for Fig.\ \ref{fig:k1-k2-V},  $\rm (2_k,H_l)$, and those that are collinear in $\rm (H_k,1_l)$, \footnote{Note that these diagrams have no singularities when loop $l$ is collinear to $p_2$.}  
\bea
{\cal J}_c^\mu(k,l)\ = \ g_s^3 T^{(q)}_c\, \left[ C_F\, {\cal J}_{k,F}^\mu(k,l) + \frac{C_A}{2}\,  {\cal J}_{k,A}^\mu(k,l)\ +  {\cal J}_{l}^\mu(k,l) \right ]\, .
\label{eq:J-mu-c-def}
\eea
As above, ${\cal J}_c^\mu$ is one-particle irreducible, and is associated with a quark and gluon propagator in the jet subdiagram.
Here, all singularities associated with loop $l$ collinear to the quark momentum $p_1$ are contained in the vector
\bea
\nonumber\\[2mm]
{\cal J}_{l}^\mu(k,l)\ &\equiv&\  \left( \frac{C_A}{2} - C_F \right )\, V_l^\mu(k,l) + \frac{C_A}{2}\, O^\mu(k,l)  \, .
\eea
We saw above that the vector integrand factors $V_l^\mu(k,l)$, Eq.\ (\ref{eq:Vl-def}) and $O^\mu$, Eq.\ (\ref{eq:O-def}) are free of singularities in regions $\rm (1_k,H_l)$ and $\rm (2_k,H_l)$, and we shall not modify these terms.\footnote{We also note that in the double collinear region $\rm (2_k,1_l)$, $O^\mu$ is singular, but of course is free of loop polarizations.  The Ward identities necessary to factorize this region are all tree-level, and ensure factorization for the unmodified integrand.   In $\rm (2_k,1_l)$, $V_l^\mu$ is finite.}

As a next step, we examine separately  terms associated with loop polarizations that multiply $C_F$ and $C_A$.   For the $C_F$ terms, we have
\bea
{\cal J}_{k,F}(k,l)\ =\ -\, V_k^\mu(k,l)\ +\ \gamma^\mu\, N_{S-q}(k,l)\, ,
\label{eq:JkF-def}
\eea 
with $V_k^\mu$ given by (\ref{eq:Vk-def}) and $N_{S-q}$ by (\ref{eq:Nsquark}).
For integrands that multiply $C_A/2$, we chose to evaluate the QED vertex integrand at a modified value for loop $l$, replacing $l$ by $-l-p_1$ in Eq.\ (\ref{eq:Vk-def}),
\bea
{\cal J}_{k,A}(k,l)\ =\ V_k^\mu(k,-l-p_1) \ +\  W_{\rm scalar}^\mu(k,l)\, ,
\label{eq:JkA-def}
\eea
leaving $W^\mu_{\rm scalar}(k,l)$ in the form defined by Eq.\ (\ref{eq:W-scalar-def}).   Clearly, this modification does not affect the integral over $l$.  Also, because it acts only on the function $V_k^\mu$, it does not interfere with the factorization in the single-collinear region $\rm (H_k,1_l)$.   As we now describe, this choice of argument also helps control the $p_1$ jet function in the region $\rm (2_k,H_l)$.   

An important feature of the full jet function, ${\cal J}^\mu_c$ is that its entire divergence in the single-collinear region $\rm (2_k,H_l)$ is generated from the unmodified integrand factor
$O^\mu(k,l)$.   This result is ensured by the Abelian-like Ward identities for the remaining coefficients of $C_F$ and $C_A$,
\begin{eqnarray}
k_\mu \int \frac{d^D l}{(2\pi)^D}\,  \left [  {\cal J}_{k,F}^\mu(k,l) - V^\mu _l(k,l) \right ]  u(p_1)\ =\ 0\, ,
\label{eq:J_F-WI}
\end{eqnarray}
and 
\begin{eqnarray}
k_\mu \int \frac{d^D l}{(2\pi)^D}\,  \left [  {\cal J}_{k,A}^\mu(k,l) + V^\mu _l(k,-l-p_1) \right ]  u(p_1)\ =\ 0\, .
\label{eq:J_A-WI}
\end{eqnarray}
These relations can be confirmed in a number of ways. Working them out explicitly, the integrands add locally to integrals that vanish by symmetric integration and the Dirac equation.  Another approach is to observe that the Ward identity for a scalar-polarized gluon acting on a one-loop three-point function adjacent to an external line vanishes, except for ghost contributions, which are given identically by the vector integrand $O^\mu$ identified above. 
We review the role of ghost terms in the full Ward identity in Section~\ref{sec:ghosts}, where we confirm that they factorize locally, and require no modification.
   We note that, although we must include the function $V_l^\mu$ in these relations to make them exact, the contribution of $V_l^\mu$ is integrable in both region $\rm (2_k,H_l)$ and $\rm (1_k,H_l)$, so that even without this term, contraction with vector $k_\mu$ gives a result that is collinear finite in either region.

Our aim, then, is to identify a vector integrand for the one-loop jet function, whose integral is exactly the same as the integral specified by ${\cal J}^\mu_{k,F}$ and ${\cal J}^\mu_{k,A}$, and which is both locally free of loop polarizations in region $\rm (1_k,H_l)$  and is locally integrable in the region $\rm (2_k,H_l)$.   We start with ${\cal J}^\mu_{k,F}$.
Combining the $V_k^\mu$ and $N_{S-q}$ terms from Eqs.\ (\ref{eq:V-k-on-u}) and (\ref{eq:Nsquark}) respectively, ${\cal J}^\mu_{k,F}$. is given by
\bea
{\cal J}_{k,F}^\mu(k,l) \ &=&\ - \, V_k^\mu(k,l) \  + \ \gamma^\mu\, N_{S-q}(k,l)
\nn\\[2mm]
&=&   \frac{2(1-\epsilon)}{(k+p_1+l)^2}\, \left [ \frac{2(l^\mu+p_1^\mu)\, \lsla}{ l^2(l+p_1)^2} - \frac{ \gamma^\mu}{2l^2} \right ]\, .
\label{eq:JkF-explicit}
\eea
We note that the self energy, $\gamma^\mu N_{S-q}$ simply gives a factor of $1/2$ for the coefficient of the $\gamma^\mu$ term, relative to the result for $V_k^\mu$ alone.
We seek a subtraction that integrates to zero, yet cancels all singular loop polarizations in $\rm (1_k,H_l)$ and any singularities in $\rm (2_k,H_l)$ in (\ref{eq:JkF-explicit}).   To find such a subtraction, we identify singular loop polarizations in region $\rm (1_k,H_l)$ and singular behavior in $\rm (2_k,H_l)$.
Using the same expansion of $\lsla$ as in Eq.\ (\ref{eq:V-k-loop-pol}), and applying it as well to the matrix $\gamma^\mu$, we have
\bea
\left [ -V_k^\mu(k,l) \, +\, \gamma^\mu\, N_{S-q}(k,l) \right ] u(p_1)\ &=&\ 2(1-\epsilon)\, \frac{1}{(k+l+p_1)^2}
 \nn\\[2mm]
 &\ & \hspace{-40mm} \times
\left[ 2(l^\mu+p_1^\mu)   \left ( \left \{  \frac{1}{l^2} -
      \frac{1}{(l+p_1)^2} \right \}   \frac{{{\slashed
          \eta}_1}}{2p_1\cdot \eta_1} + \frac{\lsla_{\perp(p_1,
        \eta_1)}}{l^2 \, (l+p_1)^2}
  \right ) \right. \nonumber \\ && \hspace{-3cm} 
-\left. \frac{p_1^\mu}{l^2} \frac{{{\slashed \eta}_1}}{2p_1\cdot \eta_1} \,
- \frac{\gamma_{\perp(p_1, \eta_1)}^\mu}{2l^2} \right]\, u(p_1)\, .
\label{eq:JkF-expand}
\eea
Terms proportional to $\psla_1$ have vanished when acting on Dirac
spinor $u(p_1)$, leaving only terms proportional to ${\slashed
  \eta}_1$ and transverse $\gamma_{\perp}$ functions.   In the overall
integral, the factors $\gamma_\perp$ anticommute with the factor
$\psla_1+\ksla$ from the quark propagator adjacent to ${\cal J}^\mu$,
which eliminates the $\rm (1_k,H_l)$ single-collinear divergence.   In
diagrams  with a divergence in 
the region $\rm (2_k,H_l)$  (Fig.~\ref{fig:k1-k2-V}), we can set the reference vector $\eta_1 =
p_2$. Then, for $k \parallel p_2$,  the explicit $\lsla_\perp$  term is odd in
$l_\perp$.   With this assignment of $\eta_1$, this term is finite in both regions.

Examining the $\eta_1$ dependence of Eq.\ (\ref{eq:JkF-expand}), we realize that we can cancel it by a counterterm with the same $l^\mu$ and $p_1^\mu$ dependence, adding only $k^\mu$ terms,
 \bea
\delta {\cal J}^\mu(k,l)\ &=&\  \frac{2(1-\epsilon)}{(p_1+k+l)^2}\,  \left[  \frac{2 l^\mu+p_1^\mu + k^\mu}{l^2} - \frac{2(l+p_1)^\mu + k^\mu}{(l+p_1)^2} \right ]\, \frac{\slashed{\eta}_1}{2 p_1 \cdot \eta_1}\, ,
\label{eq:delta-J-CF}
\eea 
where the integrals of both terms indeed vanish by Eq.\ (\ref{eq:l-integral-identities}), so that
\bea
\int d^D l\ \delta {\cal J}^\mu(k,l)\ &=&\ 0\, .
\eea
We next show that a subtraction of the same functional form will remove loop polarizations for the $C_A$ terms in the jet function, ${\cal J}^\mu_{kA}$, defined in Eq.\ (\ref{eq:JkA-def}), to which we now turn.

Combining  the $C_A$ parts of the jet function as defined in (\ref{eq:JkA-def}) we find
\bea
 {\cal J}^\mu_{k,A}(k,l)\ &=&\ \frac{-2(1-\epsilon)}{(k-l)^2l^2(p_1+l)^2}\, \left( (4l-k)^\mu\, \lsla - l^2 \gamma^\mu 
 \right )\, .
\label{eq:hatJ-def}
\eea 
The same procedure as for the $C_F$ terms results in a counterterm that is identical to (\ref{eq:delta-J-CF}), but evaluated with a change of variables, $l$ to $-l-p_1$,
\bea
2\delta {\cal J}^\mu\left(k,-l-p_1\right)\ &=&\ \frac{-4(1-\epsilon)}{(k-l)^2}\,  \frac{\slashed{\eta}_1}{2 p_1 \cdot \eta_1}\, \left( \frac{2l^\mu - k^\mu}{l^2} \ -\ \frac{2l^\mu - k^\mu + p_1^\mu}{(p_1+l)^2}  \right )\, .
\nonumber\\
\label{eq:delta-J-CA}
\eea 
Combining the jet integrand with these two counterterms, we achieve an
integrand that is locally free of loop polarizations and also of
singular behavior in region $\rm (2_k,H_l)$ by choosing $\eta_1 = p_2$
for the diagrams of Fig.~\ref{fig:k1-k2-V}.
We represent the subtracted jet by
\begin{eqnarray}
g_s^3T^{(q)}_c\, {\cal J}^{\mu}\left( k,
  l\right)   &\to& \overline{{\cal J}}^\mu_c(k,l) \, .
                 \label{eq:bar-J}
\eea
The modified jet vertex integrand,  $\overline{\cal J}_c^\mu\left(k,l \right)$ is given by
                   \bea
                   \label{eq:Jbar}
\overline{\cal J}_c^\mu(k,l) &=& \ g_s^3 T^{(q)}_c\, \Bigg[ C_F\, \left( {\cal J}_{k,F}^\mu(k,l) - \delta {\cal J}^\mu(k,l) \right ) 
\nn\\[2mm]
&\ &  \hspace{5mm} +\ \frac{C_A}{2}\, 
\left(  {\cal J}_{k,A}^\mu(k,l) - 2\, \delta {\cal J}^\mu(k,-l-p_1) \right ) \ +  {\cal J}_{l}^\mu(k,l) \Bigg]
\nn\\[2mm]
&=& \ g_s^3 T^{(q)}_c\, \Bigg[ C_F\, \left( -V_k^\mu(k,l) + \gamma^\mu N_{S-q}(k,l) - \delta {\cal J}^\mu(k,l) \right ) 
\nn\\[2mm]
&\ &  \hspace{5mm} +\ \frac{C_A}{2}\, 
\left(  V_k^\mu(k,-l-p_1) + W^\mu_{\rm scalar}(k,l) - 2\, \delta {\cal J}^\mu(k,-l-p_1) \right ) 
\nn\\[2mm] 
&\ &  \hspace{5mm} +\   \left( \frac{C_A}{2} - C_F \right )\, V_l^\mu(k,l) + \frac{C_A}{2}\,  O^\mu(k,l) \Bigg]\, ,
\eea
     where in the second equality we have written the integrand out in terms of the functions introduced in the previous subsection.  This is our final form for the jet function integrands.

     
Let us summarize our considerations so far.  To eliminate local loop
     polarizations in the $p_1$ jet function, given the momentum flows
     identified in Sec.~\ref{subsec:momenta}, we have modified the
     standard integrand for the jet subdiagram in three ways.   First,
     we have chosen to evaluate the function $V^\mu_k$ with a shift in
     momentum $l \rightarrow - l - p_1$, but only in the terms that
     multiply color factor $C_A/2$, as in Eqs.\ (\ref{eq:JkA-def}) and
     (\ref{eq:Jbar}).    Second, we have identified explicit
     counterterms $\delta J^\mu$, in Eqs.\ (\ref{eq:delta-J-CF}) and
     (\ref{eq:delta-J-CA}).   Third, we evaluate the  self-energy subdiagram according to (\ref{eq:Nsquark}).

For practical purposes, it is natural to combine the shift of
     arguments in the $(C_A/2) V^\mu_k$ terms with $\delta J^\mu$ into
     a single counterterm, which can be added directly to the standard
     integrand.  This provides a straightforward procedure that is
     readily implemented algorithmically.   To be specific, let us
     define a ``canonical" jet function, $J^\mu_{c,{\rm canonical}}$,
     as the integrand found directly from standard Feynman rules
     applied with the momentum flows assigned as in
     Sec. \ref{subsec:momenta}, and shown in Eqs.\  (\ref{eq:V-k-l}) and (\ref{eq:W-k-l}).   The modified jet function of Eq.\ (\ref{eq:Jbar}) that avoids loop polarizations can
     then be written equivalently in terms of ${\cal J}_{c,{\rm canonical}}^\mu(k,l)$, as  
\begin{equation}
\overline{\cal J}_c^\mu(k,l) ={\cal J}_{c,{\rm canonical}}^\mu(k,l) + \Delta_1 {\cal J}_c^\mu(k,l)\, .
\label{eq:add-Delta1}
\end{equation}
The term ${\cal J}_{c,{\rm canonical}}^\mu(k,l)$  is naturally produced in a
     conventional generation of the Feynman diagrams for the
     electroweak process, in which S-type diagrams are treated as in Eq.\ (\ref{eq:Nsquark}),
     \begin{eqnarray}
       {\cal J}_{c,{\rm canonical}}^\mu(k,l) \equiv\ g_s^3 T^{(q)}_c\,
       \Bigg[
       C_F\, \gamma^\mu N_{S-q}(k,l) + \left(\frac{C_A}{2} -C_F
       \right) \, V^\mu(k, l) +\frac{C_A}{2} \, W^\mu(k, l) 
       \Bigg]. \nonumber \\ &&
       \label{eq:J-canonical-def})
       \end{eqnarray}
     The counterterm in Eq.\ (\ref{eq:add-Delta1}) is then
     \begin{eqnarray}
       \label{eq:FRuleLPquark}
       \Delta_1 {\cal J}_c^\mu(k,l) &\equiv&\eqs[0.25]{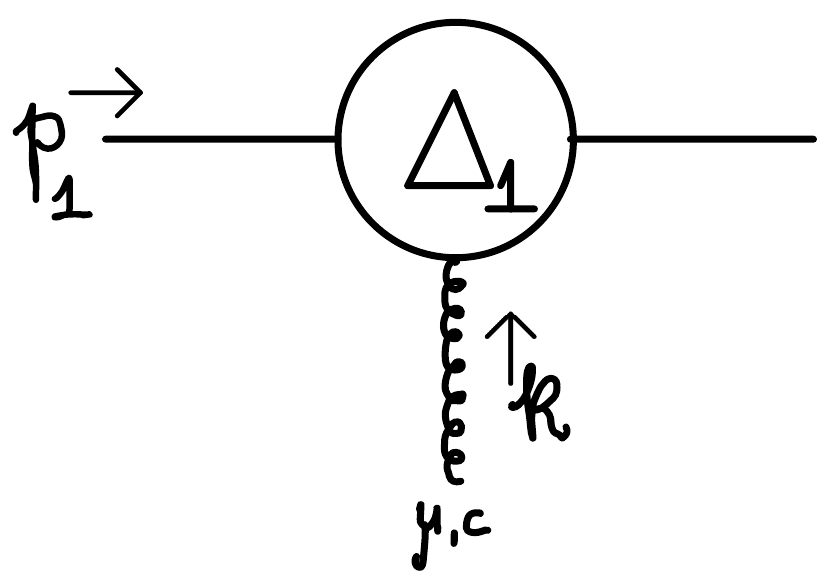}
                                             =
                                             \ g_s^3 T^{(q)}_c\,
                                             \Bigg[
                                             \frac{C_A}{2} \,  \left\{
         V_k^\mu(k,-l-p_1) -V_k^\mu(k,l)
                                      \right\}
                                              \nonumber \\ && \hspace{-1cm}
       - C_A \, \delta {\cal J}^\mu(k,-l-p_1)
       - C_F \, \delta {\cal J}^\mu(k,l)
                                     \Bigg] \, .
     \end{eqnarray}
                                    This term can be thought of as an additional
                Feynman rule.
                
We eliminate loop polarizations from the jet function of
                the incoming antiquark by introducing an analogous additive term, obtained directly from
                Eq.~(\ref{eq:FRuleLPquark}) by
                                                              exchanging
                                                              the
                                                              momenta
                                                              labels, $
                                                              k
                                                              \leftrightarrow
                                                              l$, by substituting $p_1  \to -
                p_2$ (as noted after Eq.\ (\ref{eq:Jmu-VWSO})), and defining an appropriate auxiliary vector $\eta_1 \to \eta_2$ 
                                                              with
                                                              $p_2
                                                              \cdot
                                                              \eta_2
                                                              \neq 0$.   The result is,
                                                              \begin{equation}
                  \Delta_2 {\cal J}_c^\mu(l,k)  \equiv \eqs[0.25]{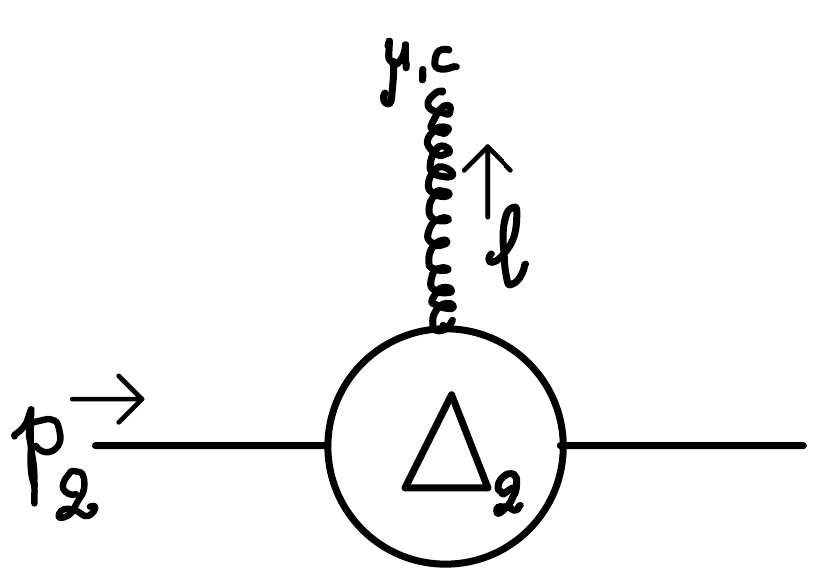} = \Delta_1
                    {\cal J}_c^\mu(l,k) \Bigg|_{
                      \stackrel{\eta_1 \to \eta_2}{   p_1 \to -p_2}}\, .
                    \label{eq:FRuleLPqbar}
                \end{equation}
                                     
\subsection{Single-collinear region $\rm (2_k,H_l)$ and its Ward identity}
\label{sec:single-co-wi}

To confirm the arguments after Eq.\ (\ref{eq:JkF-expand}) concerning the finiteness in the $\rm (2_k,H_l)$ of
 the modified jet diagrams, we examine the subtracted integrands explicitly, acting on $(p_1)$.  We start by combining terms that appear with 
 explicit coefficient $C_A/2$ in the first form of
 Eq.\ (\ref{eq:Jbar}), 
\begin{eqnarray}
  \label{eq:Vtype_Ca}
&& \left [ {\cal J}_{kA}^\mu(k,l) - 2\, \delta {\cal J}^{\mu}(k,- l - p_1) \right ] \, u(p_1)\ =\
\frac{ 2 \, (1-\epsilon) }{l^2 \left(l+p_1\right)^2 \left( k-l \right)^2 }
                           \nonumber \\
                       &&
                          \hspace{1cm}
                          \times
                          \Bigg\{
                           l^2 \gamma^\mu_{\perp(p_1, \eta_1)}
                          +
                          (k - 4 l)^\mu \slashed{l}_{\perp(p_1, \eta_1)}
                          - \frac{l \cdot p_1}{ p_1 \cdot \eta_1} k^\mu \slashed{\eta}_1
                          \Bigg\}\, u(p_1)\, .
\end{eqnarray}
This expression does not include $V^\mu_l$ and $O^\mu$ terms associated with ${\cal J}_l^\mu$ of Eq.\ (\ref{eq:J-mu-c-def}),
which requires no modifications in region $\rm (2_k,H_l)$.
Similarly, the terms that contribute to the $C_F$ color factor in the first form of
 Eq.\ (\ref{eq:Jbar}) are
\begin{eqnarray}
\label{eq:Vtype_Cf}
&& \left [ {\cal J}^\mu_{kF} -  \delta {\cal J}^\mu \left(k, l\right) \right ]\, u(p_1)\
   =\
   \frac{ 2 \left( 1- \epsilon \right) } {
l^2 \left(l+p_1\right)^2 \left( k+l+p_1\right)^2  
   } 
\nonumber \\
  &&
     \hspace{1cm}     
     \times
     \left\{
     -\frac 1 2 \left( l+p_1\right)^2
     \gamma^\mu_{\perp(p_1, \eta_1)}  
+2 \left( l+p_1\right)^\mu \slashed{l}_{\perp(p_1, \eta_1)} 
     - \frac{l \cdot p_1}{ p_1 \cdot \eta_1} k^\mu \slashed{\eta}_1
     \right\}\, u(p_1)\, .
     \nonumber\\
\end{eqnarray}
In Eqs~(\ref{eq:Vtype_Ca}) and (\ref{eq:Vtype_Cf}),  the subscript $\perp$ denotes the projection of vectors on the
transverse space of $p_1$ and $\eta_1$.
Our aim is to study these contributions to the jet function in the region ${\rm (2_k, H_l)}$ for diagrams
     where the gluon of momentum $k$ attaches the jet to the incoming antiquark that
     carries momentum $p_2$ (Fig.~\ref{fig:k1-k2-V}).  For these diagrams, we exploit the freedom to
     choose the auxiliary vector $\eta_1$ in $\delta J^\mu$, and set it to
     \begin{equation}
\eta_1 = p_2\, .
     \end{equation}
With this assignment, the first and third terms in the curly brackets of
Eqs~(\ref{eq:Vtype_Ca})-(\ref{eq:Vtype_Cf}) vanish  when contracted with a vector which is parallel to
$p_2$, as in the $k \parallel p_2$ limit.
In contrast, the second term, proportional to $\slashed{l}_\perp$
     {\it still} leads to a $k \parallel p_2$ singularity for fixed $\lsla_\perp$.
In this limit, however the integrand is odd under a reflection of the
     loop momenta on the transverse plane. 
    As has been proposed in Ref.~\cite{Anastasiou:2020sdt}, we could remove this remaining singularity 
by performing a global symmetrization of the amplitude under 
\begin{equation}
\label{eq:globalsymmetrization}
\left(k_\perp, l_\perp \right) \leftrightarrow \left(-k_\perp, -l_\perp \right).
\end{equation}
In what follows, we propose a refined approach, which does not require
     the above symmetrization to be applied to the entire amplitude
     integrand.
     
\subsection{General transverse subtraction}
\label{sec:general-transverse}

 The amplitudes in the class of electroweak production processes that we consider here, with color neutral final states, exhibit singularities in just two possible collinear directions, which define a unique transverse space. 
However, in processes with more than two ``jets'' a different symmetrization would be needed for each pair of collinear directions at two loops.  As such, the global amplitude symmetrization of Eq.~(\ref{eq:globalsymmetrization}) needs to be refined.  
In anticipation of a future generalization of our method to processes with final-state collinear singularities, we present here an improved solution, which implements this symmetrization selectively, on  V-type diagrams that generate collinear singularities in jet pairs.

We decompose the $l$ loop momentum into components parallel and transverse to
the incoming momenta
\begin{eqnarray}
   l&=& \frac{l \cdot p_2}{p_1 \cdot p_2} \, p_1
  +\frac{l \cdot p_1}{p_1 \cdot p_2} \, p_2
       + l_\perp , 
\end{eqnarray}
and from $l$ we define a dual loop-momentum $\tilde l$ with transverse components reflected,
\begin{eqnarray}
   \tilde l &=& \frac{l \cdot p_2}{p_1 \cdot p_2} \, p_1
  +\frac{l \cdot p_1}{p_1 \cdot p_2} \, p_2
       - l_\perp .
\end{eqnarray}
                In the subset of $V-$type diagrams that  are singular in the $k \parallel p_2$ limit, we symmetrize the term proportional to $\slashed{l}_\perp$ in Eqs.~(\ref{eq:Vtype_Ca}) and (\ref{eq:Vtype_Cf}).  
                For these diagrams, we
                execute operationally the transverse momentum symmetrization by adding the following
                counterterm
 \begin{eqnarray}
                  \label{eq:FRuleLPquarkTrans}
       \Delta_{1\perp} {\cal J}_c^\mu(k,l) &\equiv&
                                                 \ g_s^3 T^{(q)}_c\,
                                                    \left(1-\epsilon \right)
  \frac{   -2\,  \slashed{l}_\perp}{l^2 (l+p_1)^2} \, 
                                             \Bigg\{
C_F \, \left[ 
  \frac{
 \left( l+p_1\right)^\mu
  }{
\left( k+l+p_1\right)^2 
  }
  +
  \frac{
 \left( \tilde l+p_1\right)^\mu
  }{
\left( k+ \tilde l+p_1\right)^2 
  }
     \right]
   \nonumber \\
  &&
- C_A \, \left[ 
  \frac{
 l^\mu
  }{
 \left( k-l\right)^2 
  }
  +
  \frac{
 \tilde l^\mu 
  }{
 \left( k- \tilde l \right)^2 
  }
  \right]
     \Bigg\}\, .
\end{eqnarray}
This counterterm is odd under $l_\perp \to l_\perp$ and integrates to
zero. Due to the overall $\slashed{l}_\perp$ factor, it does not
produce a singularity in either the $l \parallel p_1$ or the $k
\parallel p_1$ limits. Finally, in the $k \parallel p_2$ limit, the
counterterm is contracted with a vector parallel to $k^\mu$ and the
reflected vector $\tilde l$ is equivalent to $l$. Explicitly, In this
     limit, we have for the singular behavior,
\begin{eqnarray}     
 \lim_{k \parallel p_2}\Delta_{1\perp} {\cal J}_c^\mu(k,l) &=&
                                                 \ g_s^3 T^{(q)}_c\,
                                                    \frac{   -4\,  \left(1-\epsilon \right) \,  \slashed{l}_\perp}{l^2 (l+p_1)^2} \, 
                                                                  \Bigg\{
C_F \, \frac{
 \left( l+p_1\right)^\mu
  }{
\left( k+ l+p_1\right)^2 
                                                                  }
                                                                  -
                                                                  C_A \frac{
 l^\mu
  }{
 \left( k-l\right)^2 
  }                                                                                                                               
                                                               \Bigg\}\, ,
                                                               \nonumber
  \\ &&
\end{eqnarray}
matching and cancelling exactly the remaining singularities of
        Eqs.~(\ref{eq:Vtype_Ca}) and (\ref{eq:Vtype_Cf}).

       In practice, for diagrams of the type in Fig.~\ref{fig:k1-k2-V} we combine
        $\Delta_{1} {\cal J}_c^\mu(k,l)$ and $\Delta_{1\perp} {\cal
        J}_c^\mu(k,l)$  into a single counterterm which we implement
        in the form of a Feynman rule
\begin{equation}
  \label{eq:FRuleLPquarkComb}
  {\tilde \Delta}_{1} {\cal J}_c^\mu(k,l)
  \equiv
\eqs[0.25]{Delta1T.pdf}
  \equiv 
   \Delta_{1} {\cal J}_c^\mu(k,l) \Bigg|_{\eta_1 = p_2}  + \Delta_{1\perp} {\cal J}_c^\mu(k,l).
  \end{equation}
The Feynman rule of Eq.~(\ref{eq:FRuleLPquarkComb}) replaces the rule of
        Eq.~(\ref{eq:FRuleLPquark}) in the subset of diagrams where the gluon with momentum $k$
        connects to the $p_2$ external leg.
        For the analogous case of cancelling the $k \parallel p_1$
        singularity in diagrams with loop polarisations in the $p_2$ jet
        function, we introduce a counterterm,
\begin{equation}
  \label{eq:FRuleLPqbarComb}
  {\tilde \Delta}_{2} {\cal J}_c^\mu(l,k)
  \equiv
\eqs[0.25]{Delta2T.pdf}
  \equiv 
   \Delta_{2} {\cal J}_c^\mu(l,k) \Bigg|_{\eta_2 = p_1}  +
   \Delta_{1\perp} {\cal J}_c^\mu(l,k) \Bigg|_{p_1 \to -p_2}.
  \end{equation}      
        

\section{Eliminating shift mismatches: local Ward identities for QCD hard parts}
\label{sec:construction}

 In addition to the V and S diagrams discussed in the previous section, single collinear regions also occur in two-gluon ladders, and in diagrams where one or both of the gluons attach to the hard scattering subdiagram through off-shell quark lines.  Such configurations  factorize directly in QED, as shown in \cite{Anastasiou:2020sdt}, by the application of the basic Ward identities.    The implementation of Ward identities for QCD is somewhat more complex than in QED at the two loop level, because of the shift mismatches identified above.   In both theories, Ward identities require shifts in loop momenta whenever all lines in the loop are charged in the theory.  In QED, this problem does not arise at two loops for diagrams like those in Fig.\ \ref{fig:two-circles}.   For S and V diagrams, an appropriate average over loop momenta compensated for these shifts, as shown above, to give a local realization of the Ward identities.   

For diagrams in which the hard loop momentum is connected to the electroweak subprocess, we also seek to implement the factorization of collinear configurations locally in momentum space (without shifting loop momenta) so that form factor subtractions precisely match the singular behavior of the sums of these diagrams.   In this case, there can be many diagrams, depending on the number of electroweak bosons produced.   We provide a prescription to modify the integrands of a subset of the diagrams, which provides this desired locality.

\subsection{Factorization of single-collinear limits: the form factor}

The motivation for our prescription can be made clear by examining the behavior of diagrams in the simplest case, the vector electroweak form factor itself.  What we will see is that even in the form factor the factorization of the one-loop collinear region from the one-loop hard subdiagram requires a momentum shift in the hard loop.   This will require us to redefine integrals for the nonabelian theory for the two-loop ``regular" diagrams of the form factor \cite{Anastasiou:2020sdt}, in addition to the S and V diagrams.   Such redefinitions are not necessary in the case of QED for this class of diagrams, because the Ward identities require a shift in QED only for diagrams with a fermion loop,    Once we have identified the necessary modification of the form factor integrals, the extension to two-loop diagrams in general electroweak production will be straightforward.

Employing the same notation as above to identify singular regions,  we consider here the regions $\rm (1_k,H_l)$ and $\rm (2_k,H_l$), when the off-shell loop momentum $l$ flows through the electroweak vertex.  
The diagrams we treat are shown in Figs.\ \ref{fig:regular-collinear-p1} and \ref{fig:regular-collinear-p2}.     In the single-collinear regions in question, the loop momentum $k$ is either collinear to the quark momentum $p_1$, in region $\rm (1_k,H_l)$, or to the antiquark momentum $p_2$, in $\rm (2_k,H_l)$.    Both of these regions are present in diagram (a), shown in both figures.   In contrast, diagrams 5b, 5c and 5d have a single-collinear divergence only in $\rm (1_k,H_l)$, and 6b, 6c and 6d only in $\rm (2_k,H_l)$.   
\begin{figure}[ht]
  \begin{center}
  \begin{tabular}{cccc}
$
    \begin{array}{c}
\eqs[0.20]{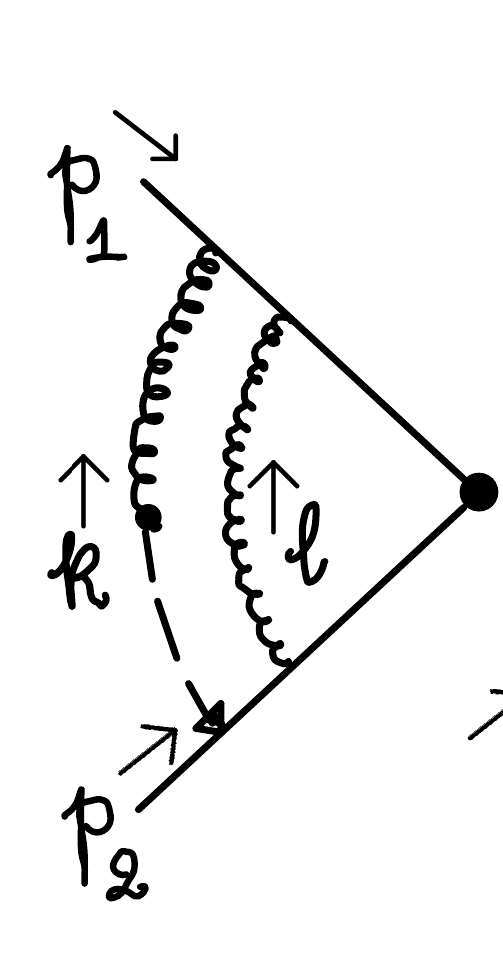}
      \\
      C_F^2
      \\
      \\
      \mbox{(a)}
    \end{array}
    $
    &
$
    \begin{array}{c}
\eqs[0.20]{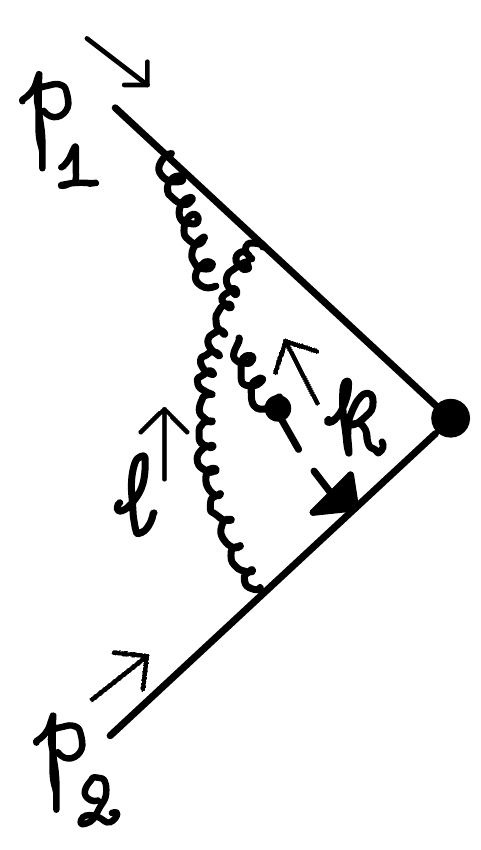}
      \\
      C_F\left( C_F - \frac{C_A}{2} \right)
      \\
      \\
      \mbox{(b)}
    \end{array}
    $
    &
$
    \begin{array}{c}
\eqs[0.20]{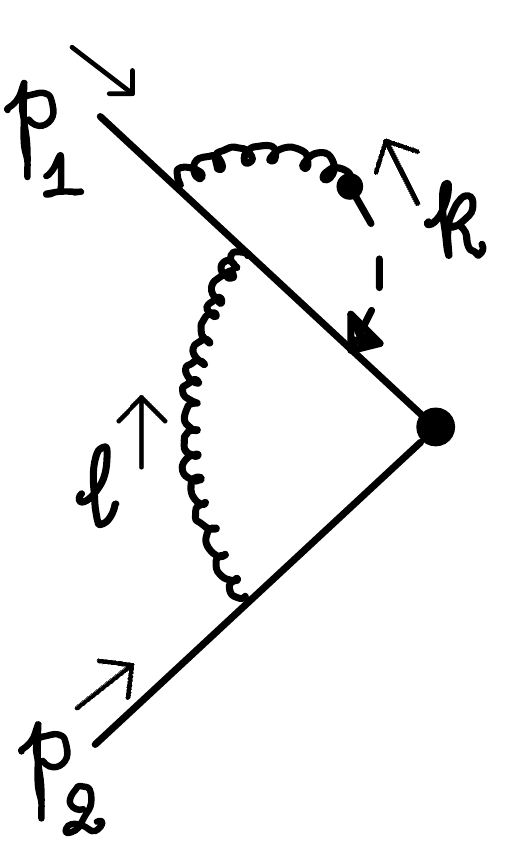}
      \\
      C_F\left( C_F - \frac{C_A}{2} \right)
      \\
      \\
      \mbox{(c)}
    \end{array}
    $
    &
      $
    \begin{array}{c}
\eqs[0.20]{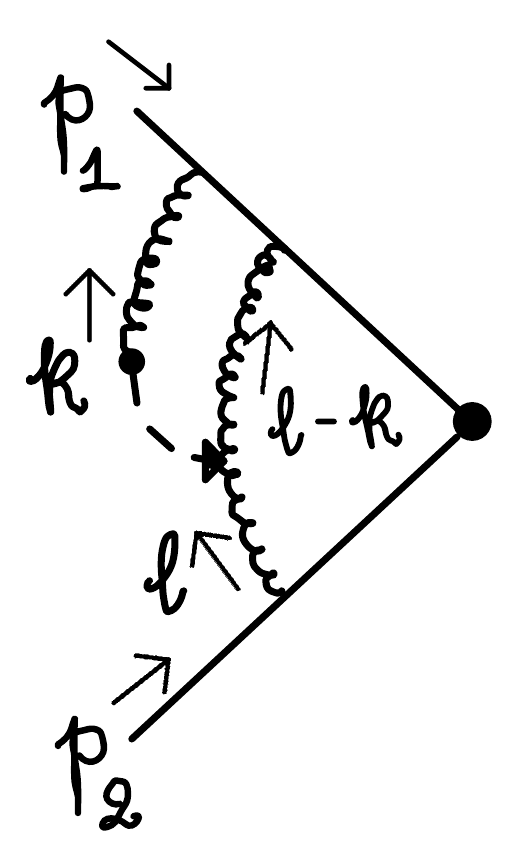}
      \\
      \frac{C_F C_A}{2}
      \\
      \\
      \mbox{(d)}
    \end{array}
    $
  \end{tabular}
  \end{center}
\caption{ Two-loop form factor diagrams singular in region $\rm (1_k,H_l)$, along with their color factors (we suppress the factor $-g_s^4$ that multiplies the overall color factors exhibited here and in subsequent figures.). The special graphical notation applies to this region, and is explained in the text. \label{fig:regular-collinear-p1}}
\end{figure}

 \begin{figure}[ht]
  \begin{center}
  \begin{tabular}{cccc}
$
    \begin{array}{c}
\eqs[0.20]{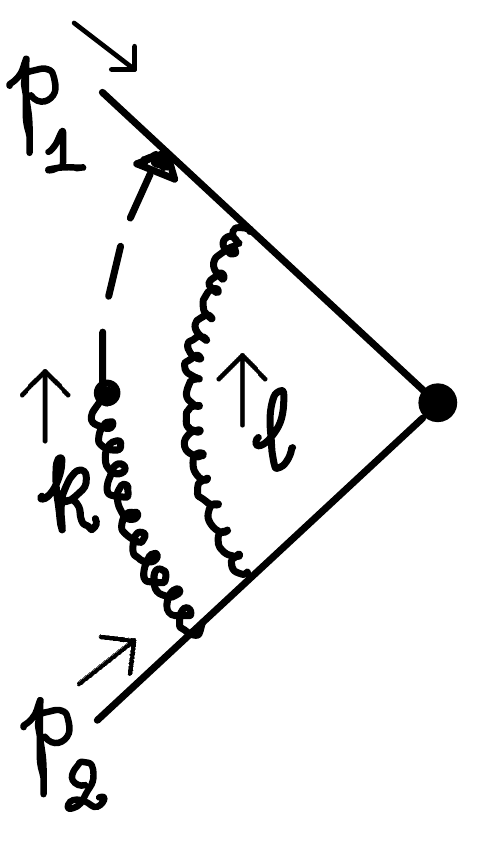}
      \\
      C_F^2
      \\
      \\
      \mbox{(a)}
    \end{array}
    $
    &
$
    \begin{array}{c}
\eqs[0.20]{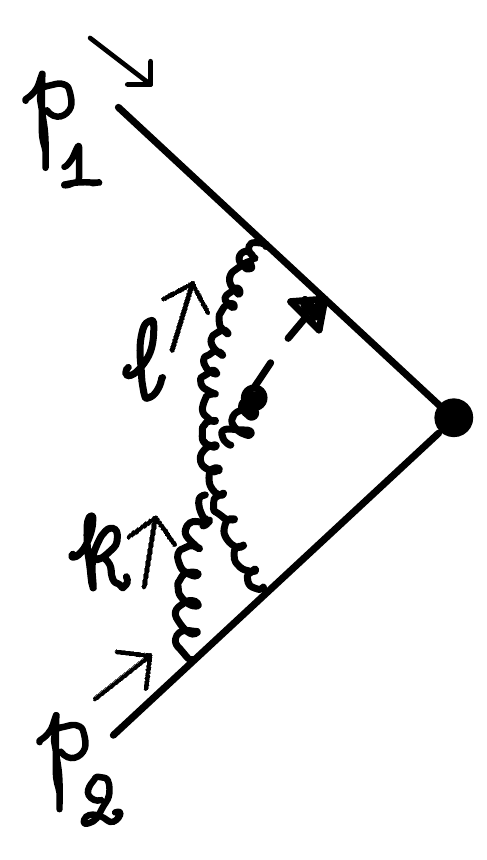}
      \\
      C_F\left( C_F - \frac{C_A}{2} \right)
      \\
      \\
      \mbox{(b)}
    \end{array}
    $
    &
$
    \begin{array}{c}
\eqs[0.20]{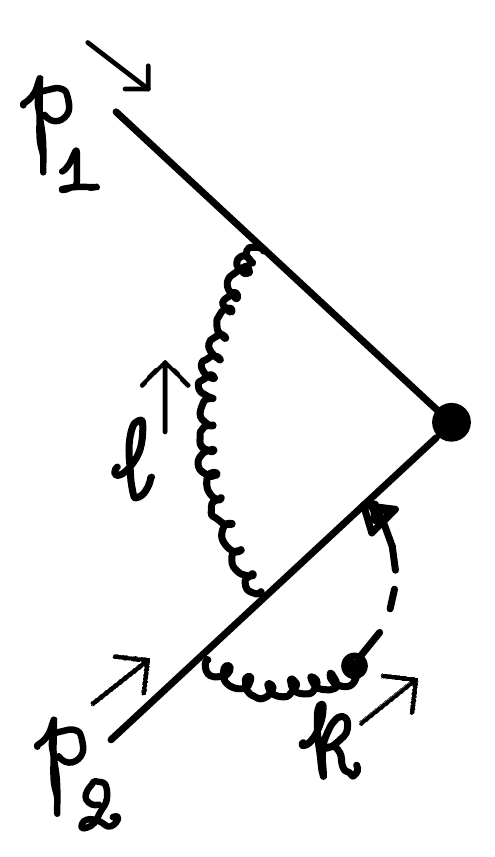}
      \\
      C_F\left( C_F - \frac{C_A}{2} \right)
      \\
      \\
      \mbox{(c)}
    \end{array}
    $
    &
      $
    \begin{array}{c}
\eqs[0.20]{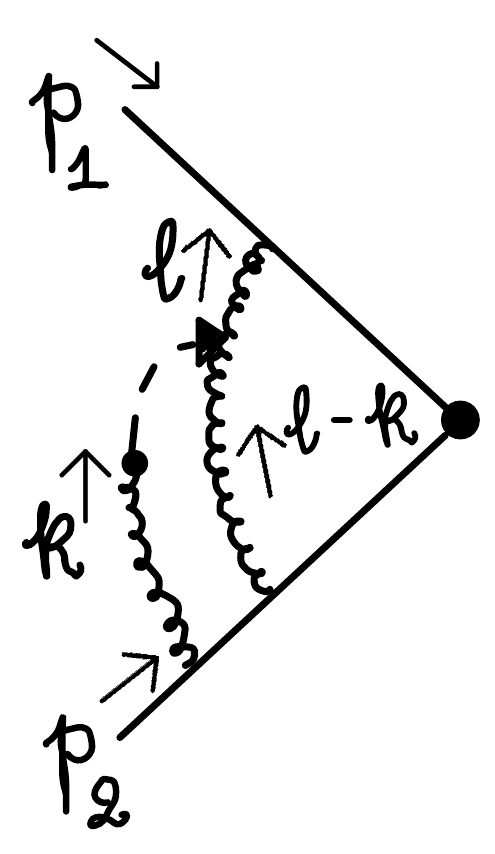}
      \\
      \frac{C_F C_A}{2}
      \\
      \\
      \mbox{(d)}
    \end{array}
    $
  \end{tabular}
  \end{center}
\caption{ Two-loop form factor diagrams singular in region $\rm (2_k,H_l)$, along with their color factors.  \label{fig:regular-collinear-p2}}
\end{figure}

In the diagrams of Figs.\ \ref{fig:regular-collinear-p1}, 
we have introduced a notation for the gluon propagator that reflects the behavior of the integrand in the corresponding region.    The dot, dashed line and arrow reflect  the following ``collinear approximations" on the polarization tensor of the collinear gluon $k$:
\bea
\frac{\eta^{\mu\nu}}{k^2}\  &\rightarrow&  \frac{1}{k^2}\  \frac{p_2^\mu (-k)^\nu}{p_2\cdot (- k)}\quad {\mathrm{in\ Region}}\ {\rm (1_k,H_l)}\, ,
\nonumber\\[2mm]
\frac{\eta^{\mu\nu}}{k^2}\  &\rightarrow&   \frac{1 }{k^2}\ \frac{p_1^\mu k^\nu}{p_1\cdot k} \quad {\mathrm{in\ Region}}\ {\rm (2_k,H_l)}\, ,
\label{eq:co-approx}
\eea
where in both of these expressions index $\mu$ is summed against the vertex adjacent to the external line to which $k$ becomes collinear, while $\nu$ is summed against the hard subdiagram.   Again notice that diagram (a) contains both singular regions.   In Figs.\ \ref{fig:regular-collinear-p1} and \ref{fig:regular-collinear-p2}, we also exhibit the color factor associated with each diagram.

We can combine the integrands of Figs.\  \ref{fig:regular-collinear-p1} and \ref{fig:regular-collinear-p2} at fixed values of the loop momenta $k$ and $l$ by applying the identities,
\bea
\frac{1}{ \rlap{/}{p} } \, \left(  \rlap{/}{k} \right )\, \frac{1}{ \rlap{/}{p} + \rlap{/}{k} } \ =\ \frac{1}{ \rlap{/}{p} }\ -\  \frac{1}{ \rlap{/}{p} + \rlap{/}{k} }\, ,
\label{eq:Feyn-iden}
\eea
and 
\bea
k^\nu \left( 2l_\nu - k_\nu \right)\, \frac{1}{l^2}\, \frac{1}{(l-k)^2} \ =\  \   \frac{1}{(l-k)^2}  \ -\ \frac{1}{l^2}   \, .
\label{eq:tHooft-iden}
\eea
The first of these is the lowest-order QED Ward identity, and the second is the lowest-order QCD Ward identity in axial gauge.   
The remaining terms in the lowest-order Feynman gauge QCD Ward identity, associated with ghost contributions, factorize independently, as it is discussed in Section~\ref{sec:ghosts}.

\begin{figure}[ht]
   \begin{center}
  \begin{tabular}{ccccc}
$
    \begin{array}{c}
\eqs[0.20]{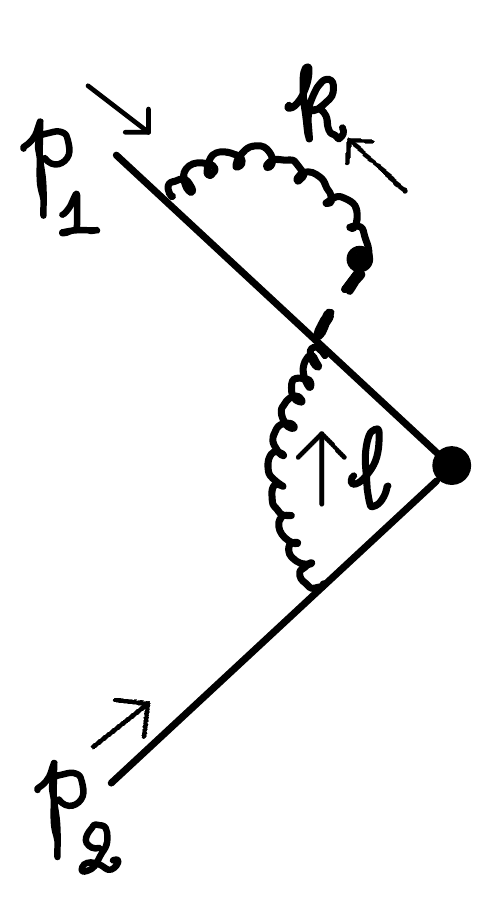}
      \\
      C_F^2
    \end{array}
    $
    &
      $
      \begin{array}{c}
        \mbox{{\Large $+$}}
        \\
        {}
      \end{array}
    $
      &
$
    \begin{array}{c}
\eqs[0.20]{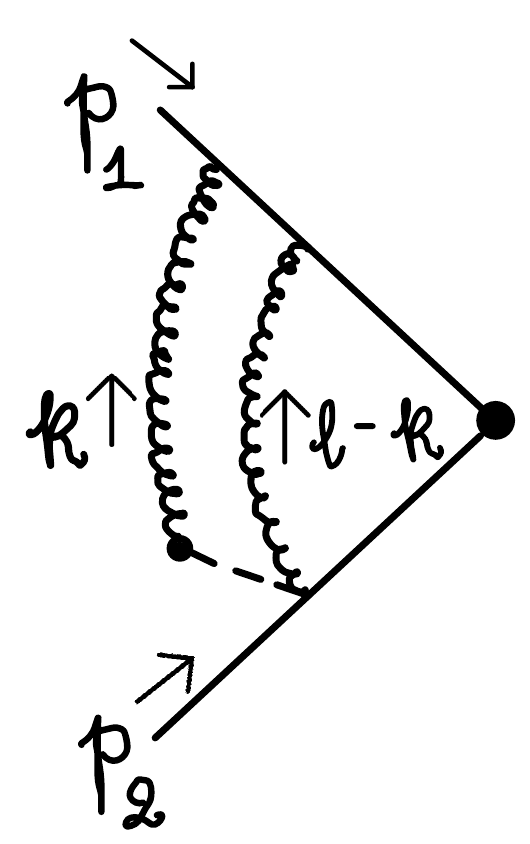}
      \\
      \frac{C_F C_A}{2}
    \end{array}
    $
    &
      $
      \begin{array}{c}
        \mbox{{\Large $-$}}
        \\
        {}
      \end{array}
    $
      &
$
    \begin{array}{c}
\eqs[0.20]{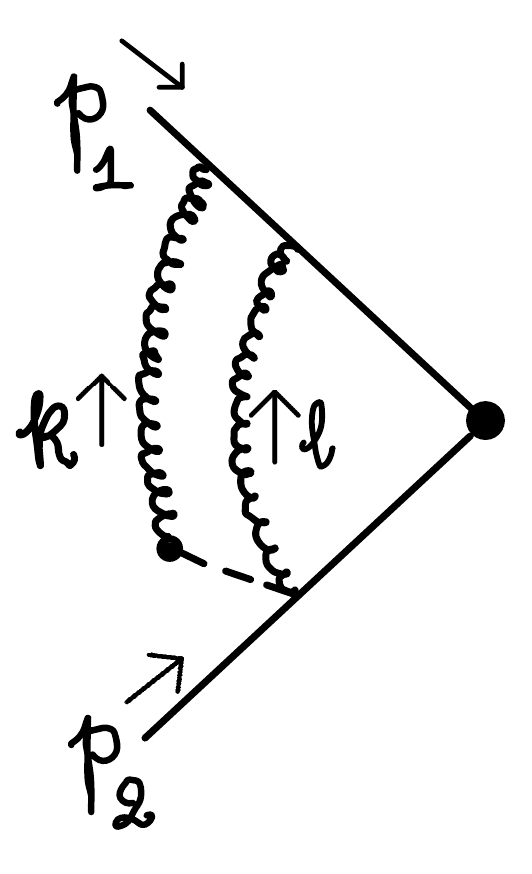}
      \\
      \frac{C_F C_A}{2}
      \\
    \end{array}
    $
  \end{tabular}
  \end{center}
\caption{ Sum of the integrands of Fig.\ \ref{fig:regular-collinear-p1}, neglecting ghost contributions, which factorize independently.  After integration, the non-factoring $C_FC_A$ terms cancel. 
\label{fig:regular-wi-p1}}
\end{figure}

\begin{figure}
     \begin{center}
  \begin{tabular}{ccccc}
$
    \begin{array}{c}
\eqs[0.20]{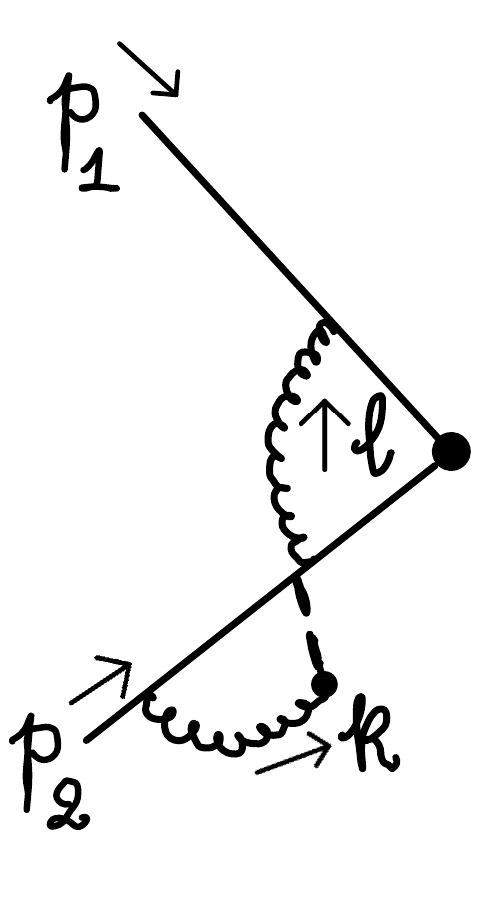}
      \\
      C_F^2
    \end{array}
    $
    &
      $
      \begin{array}{c}
        \mbox{{\Large $+$}}
        \\
        {}
      \end{array}
    $
      &
$
    \begin{array}{c}
\eqs[0.20]{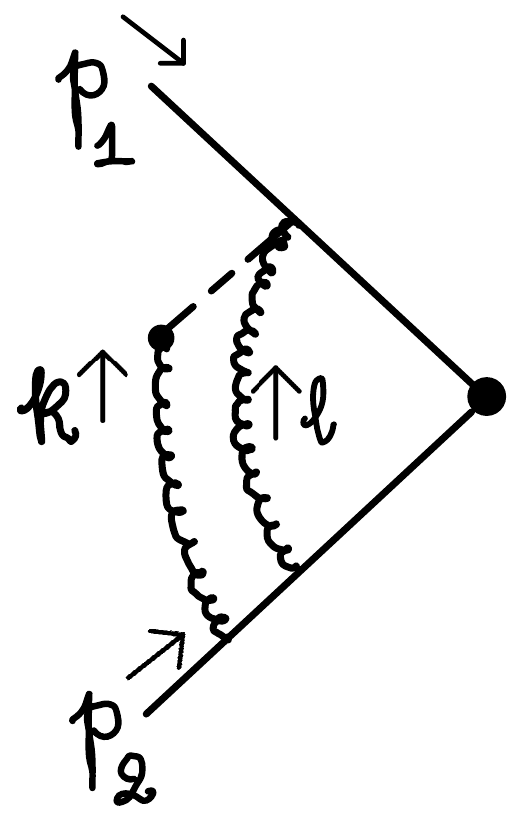}
      \\
      \frac{C_F C_A}{2}
    \end{array}
    $
    &
      $
      \begin{array}{c}
        \mbox{{\Large $-$}}
        \\
        {}
      \end{array}
    $
      &
$
    \begin{array}{c}
\eqs[0.20]{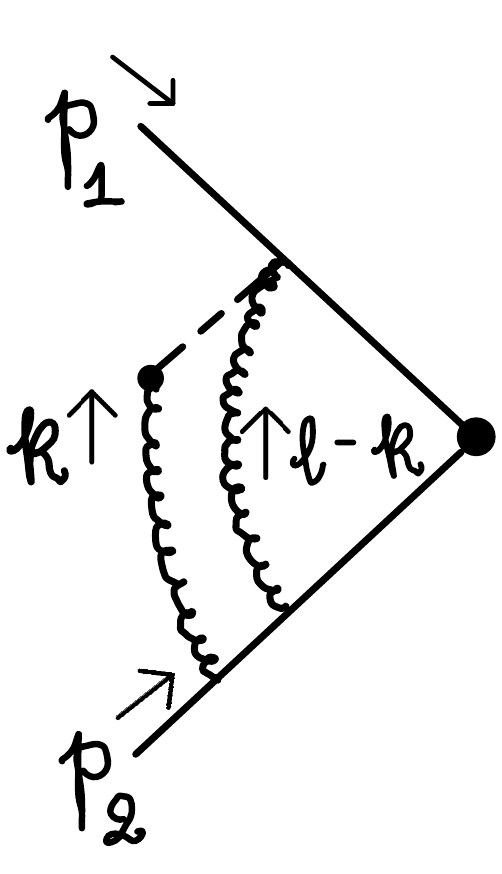}
      \\
      \frac{C_F C_A}{2}
      \\
    \end{array}
    $
  \end{tabular}
  \end{center}
\caption{ Sum of the integrands of Fig.\ \ref{fig:regular-collinear-p2}, neglecting ghost contributions. After integration, the non-factoring $C_FC_A$ terms cancel. \label{fig:regular-wi-p2}}
\end{figure}

Using Eqs.\ (\ref{eq:Feyn-iden}) and (\ref{eq:tHooft-iden}) and the explicit color factors for each diagram, the sums of the integrands of Figs.\ \ref{fig:regular-collinear-p1} and \ref{fig:regular-collinear-p2} are shown in Figs.\ \ref{fig:regular-wi-p1} and \ref{fig:regular-wi-p2}.  The dot followed by a dashed line represents the factors $p_i^\mu / [p_i\cdot (\pm k)]$, which remain after the application of the Ward identity.\footnote{The graphical notation of these figures is inspired by (but not identical to) the notation introduced by `t Hooft long ago in an early analysis of perturbative gauge theory~\cite{tHooft:1971akt}.}    (We again remind the reader that we suppress the ghost terms in Feynman gauge, which we discuss  below in Sec.\ \ref{sec:ghosts}.)     In each case, the algebra results in two terms.   In one of these, with color factor $C_F^2$, the $k$ and $l$ loop momenta have manifestly factorized.   The second term, consisting of two contributions, both with color factor $C_FC_A/2$, is not factorized at fixed $k$ and $l$.  The integral of the two terms, however, does vanish after a shift in loop momentum $l$, to $l'=l-k$.  

To make this cancellation local, we add a single infrared counterterm, consisting of the two-loop ladder times the (nonstandard) color factor $C_FC_A/2$.  The additive counterterm  is proportional to the difference between the two integrands corresponding to two routings of momentum $k$ through the inner triangle diagram.   Explicitly, we define
\bea
\Delta_\Gamma\ \equiv\  g_s^4 \frac{1}{2}C_AC_F\ \frac{1}{k^2}\,  \bar v(p_2) \gamma^\beta  \frac{1}{-\rlap{p}{/}_2 + \rlap{k}{/} } \gamma^\alpha\
\Bigg [ \left( \frac{1}{l^2}\right )\, \frac{1}{- \rlap{p}{/}_2 + \rlap{k}{/} + \rlap{l}{/} }\, \Gamma\,  \frac{1}{\rlap{p}{/}_1 + \rlap{k}{/}  + \rlap{l}{/} } 
\nonumber\\[2mm]
\ -\  \left ( \frac{1}{(l-k)^2} \right )\,  \frac{1}{- \rlap{p}{/}_2 + \rlap{l}{/}  }\, \Gamma\,  \frac{1}{\rlap{p}{/}_1 + \rlap{l}{/}   }    \Bigg ]\
\gamma_\alpha  \frac{1}{ \rlap{p}{/}_1 + \rlap{k}{/} } \gamma_\beta\, u(p_1)\, ,
\nonumber\\
\label{eq:form-factor-difference}
\eea
where $\Gamma$ is the electroweak Dirac matrix of the form factor.    We will refer to these as shift counterterms.
 After integration over $l$, these two terms cancel.   They also cancel at large loop momentum $l$, and hence do not require separate UV counterterms.

 \begin{figure}
   \begin{equation*}
     \mbox{
       {\Large
         $
\frac{C_A}{2\, C_F} \; \cdot \; 
         $
     }}
     \eqs[0.20]{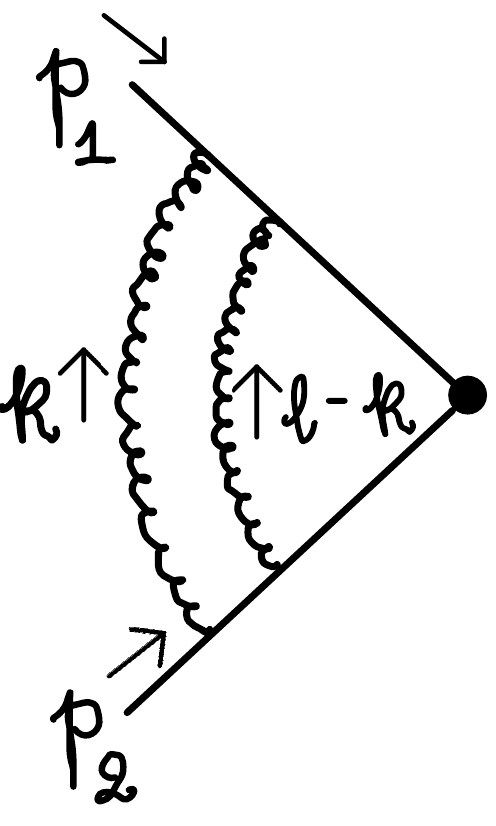}
     \quad \;
     \mbox{{\Large $-$ }}
     \quad \;
          \mbox{
       {\Large
         $
\frac{C_A}{2\, C_F} \; \cdot \; 
         $
     }}
     \eqs[0.20]{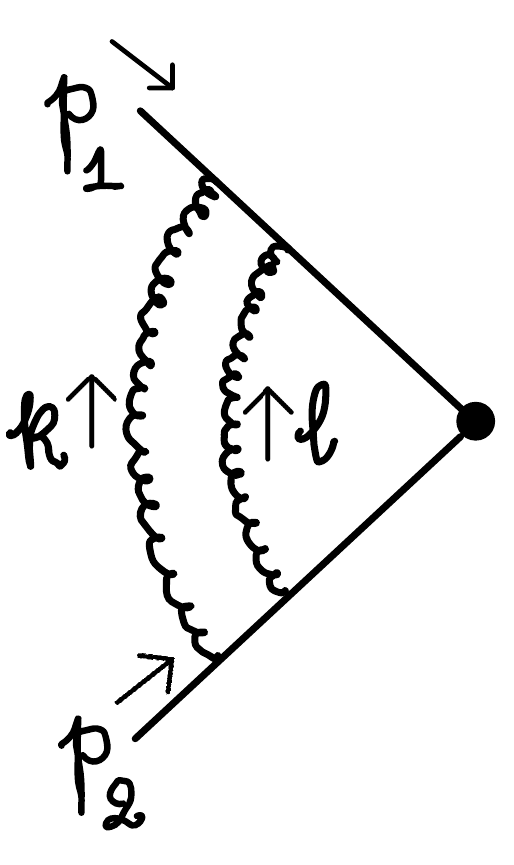}
   \end{equation*}
\caption{ Infrared counterterm for both Fig.\ \ref{fig:regular-collinear-p1} and \ref{fig:regular-collinear-p2}, given explicitly in Eq.\ (\ref{eq:form-factor-difference}).   When integrated over loop momentum $l$, the combination vanishes, and is
UV convergent.  \label{fig:regular-ir-uv}}
\end{figure}

Once the ladder diagram has been modified by the addition of the non-standard color factors in the form of Fig.\ \ref{fig:regular-ir-uv}, with the counterterms of Eq.\ (\ref{eq:form-factor-difference}), the form factor integral itself factorizes at the level of integrands in the single-collinear regions
$\rm (1_k,H_l)$ and $\rm (2_k,H_l)$.   In either case, the approximations of Eq.\ (\ref{eq:co-approx})  apply in these regions.   The application of the identity, Eq.\ (\ref{eq:Feyn-iden}) to line $k$ in the integrand of the counterterm, Eq.\ (\ref{eq:form-factor-difference}) after the application of (\ref{eq:co-approx}) leads to an expression that is the negative of the  unwanted terms in Fig.\ \ref{fig:regular-wi-p1} and \ref{fig:regular-wi-p2}, which differ by the same shift of loop $l$.  For example, In region $\rm (1_k,H_l)$, we have
\bea
\Delta_\Gamma\, \big |_{\rm (1_k,H_l)}\ &=&  g_s^4 \frac{1}{2}C_AC_F\ \frac{1}{k^2}\,  \bar v(p_2) \frac{\left ( - \ksla \right)}{p_2\cdot (-k)}  \frac{1}{-\rlap{p}{/}_2 + \rlap{k}{/} } \gamma^\alpha
\nn\\
&\ & \hspace{5mm} \times\
\Bigg [ \left( \frac{1}{l^2}\right )\, \frac{1}{- \rlap{p}{/}_2 + \rlap{k}{/} + \rlap{l}{/} }\, \Gamma\,  \frac{1}{\rlap{p}{/}_1 + \rlap{k}{/}  + \rlap{l}{/} } 
\nonumber\\[2mm]
&\ & \ -\  \left ( \frac{1}{(l-k)^2} \right )\,  \frac{1}{- \rlap{p}{/}_2 + \rlap{l}{/}  }\, \Gamma\,  \frac{1}{\rlap{p}{/}_1 + \rlap{l}{/}   }    \Bigg ]\
\gamma_\alpha  \frac{1}{ \rlap{p}{/}_1 + \rlap{k}{/} } \left ( \psla_2 \right )\, u(p_1)
\nn\\
&=&   g_s^4 \frac{1}{2}C_AC_F\ \frac{1}{k^2}\,  \bar v(p_2) \frac{1}{p_2\cdot k}  \gamma^\alpha
\nn\\[2mm]
&\ & \hspace{5mm} \times\
\Bigg [ \left( \frac{1}{l^2}\right )\, \frac{1}{- \rlap{p}{/}_2 + \rlap{k}{/} + \rlap{l}{/} }\, \Gamma\,  \frac{1}{\rlap{p}{/}_1 + \rlap{k}{/}  + \rlap{l}{/} } 
\nonumber\\[2mm]
&\ & \ -\  \left ( \frac{1}{(l-k)^2} \right )\,  \frac{1}{- \rlap{p}{/}_2 + \rlap{l}{/}  }\, \Gamma\,  \frac{1}{\rlap{p}{/}_1 + \rlap{l}{/}   }    \Bigg ]\
\gamma_\alpha  \frac{1}{ \rlap{p}{/}_1 + \rlap{k}{/} } \left ( \psla_2 \right )\, u(p_1)\, ,
\nonumber\\
\label{eq:form-factor-difference-H1}
\eea
which cancels the two unfactorized terms of Fig.\ \ref{fig:regular-wi-p1}.

We note that the shift counterterms themselves are also singular in the double collinear limits $\rm (1_k,1_l)$ and $\rm (2_k,2_l)$, but it is easy to check that these contributions factor independently.   To show this, we observe that in  $\rm (1_k,1_l)$ and $\rm (2_k,2_l)$ the approximation of Eq.\ (\ref{eq:co-approx}) holds for the ``inner" gluon, carrying momentum $l$, and for the ``outer" gluon, of momentum $k$ in Fig.\ \ref{fig:regular-collinear-p1}.   The result then follows by applying the identity of Eq.\ (\ref{eq:Feyn-iden}), first to the vertex at which outer gluon ($k$) attaches, which cancels the propagator between the two vertices.   The Ward identity (\ref{eq:Feyn-iden}) can then be applied to the resulting vertex as well, because in the region of interest, $l$ and $k$ are parallel.  This gives the  factorized form.   The same procedure demonstrates factorization in $\rm (2_k,2_l)$.   Finally, we note that the shift terms are not singular in the mixed double-collinear regions, $\rm (2_k,1_l)$ and $\rm (1_k,2_l)$.

\subsection{Single-collinear limits: general electroweak amplitudes}
\label{sec:single-collinear-ew}

We are now ready to study the single collinear limits of two-gluon loops in multi-boson electroweak production.   Our goal is to develop a modification of the integrand that results in the same local factorization of the collinear gluon at the integrand level that we found for the form factor above.  
  \begin{figure}[ht]
    \begin{center}
      \begin{tabular}{cc}
        $
        \begin{array}{c}
\eqs[0.38]{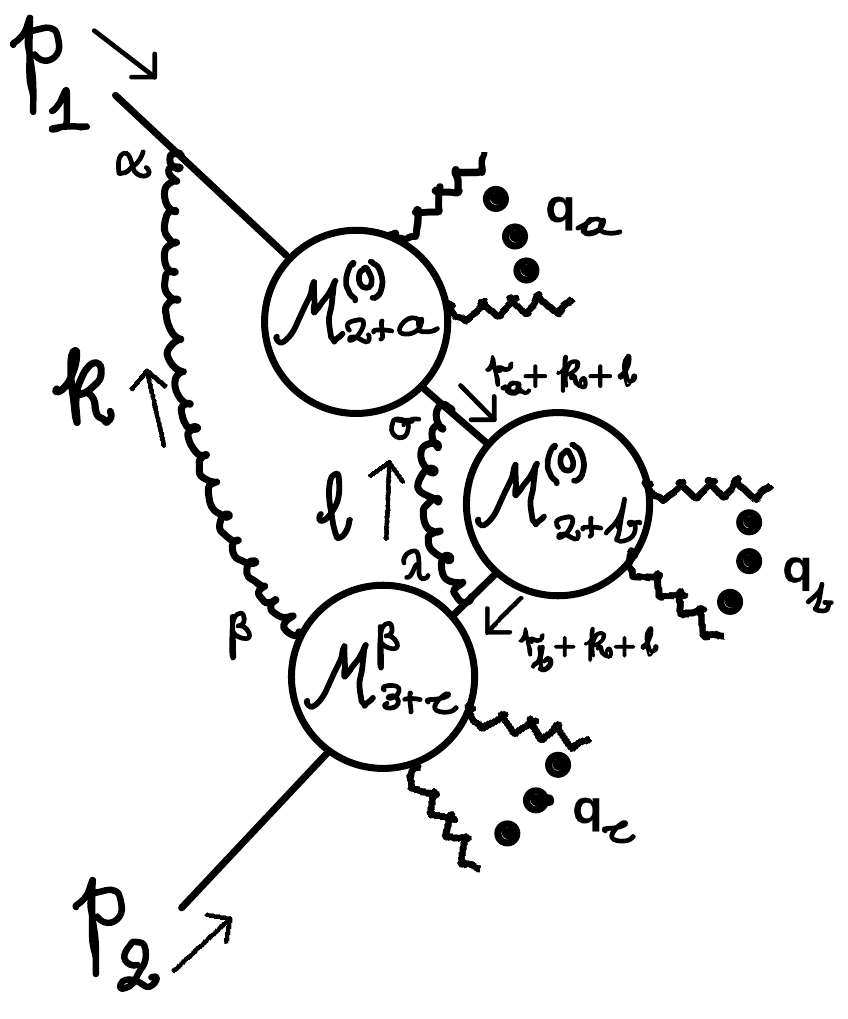}
          \\
          \mbox{\large (a) }
        \end{array}
        $
        &
 $
        \begin{array}{c}
\eqs[0.38]{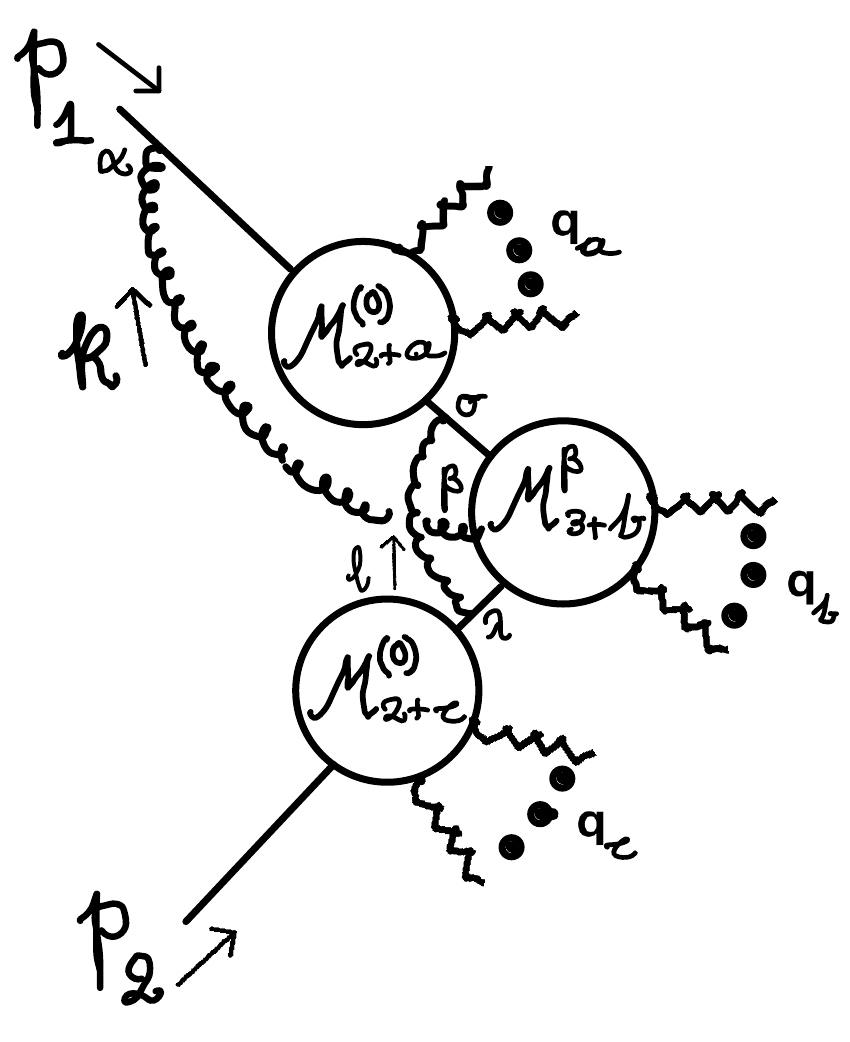}
          \\
          \mbox{\large (b) }
        \end{array}
        $
        \\
 $
        \begin{array}{c}
\eqs[0.38]{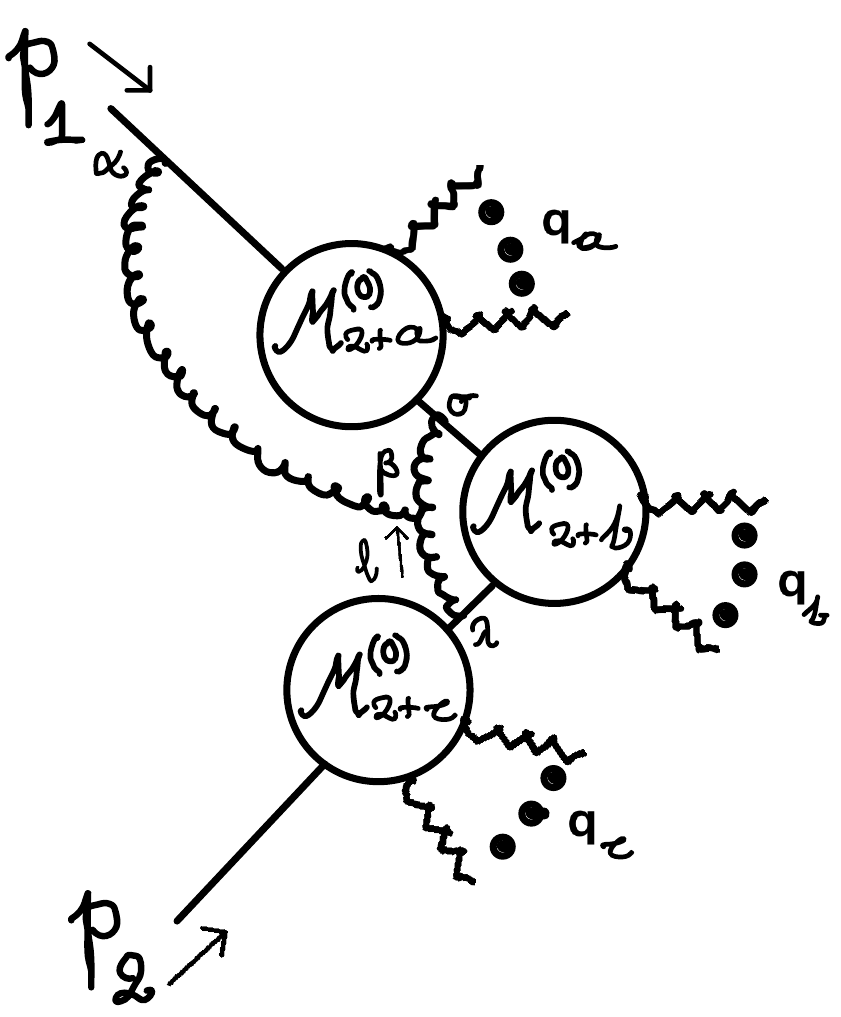}
          \\
          \mbox{\large (c) }
        \end{array}
        $
        &
           $
        \begin{array}{c}
\eqs[0.38]{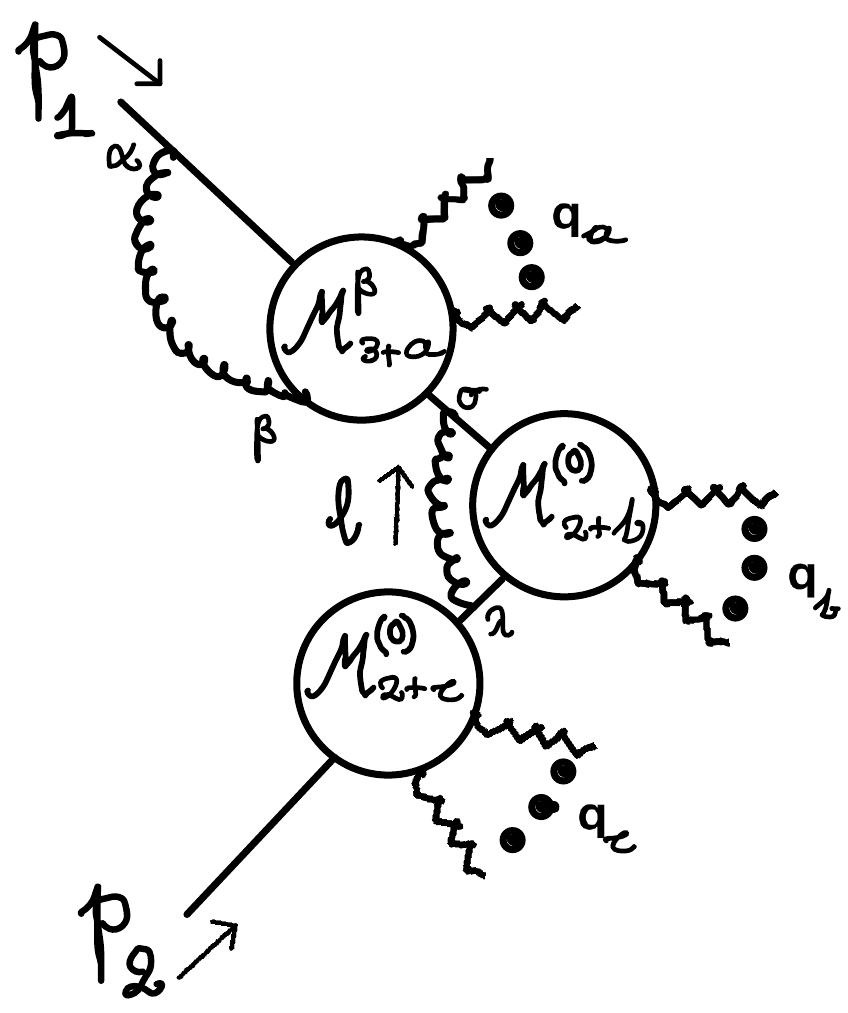}
          \\
          \mbox{\large (d) }
        \end{array}
        $
        \end{tabular}
  \end{center}
\caption{ General two-loop corrections with no fermion loops or self energies that are singular in region $(1_k,H_l)$.   Diagrams  (a) are uncrossed ladders, (b) crossed ladders, (c) have three-gluon vertices, and (d) are diagrams without QCD ladder structure.   \label{fig:two-gluons-ew}}
\end{figure}

For this discussion, we restrict ourselves to diagrams without fermion loops, so that both loop momenta flow through gluon lines.   The set includes ``ladder-like" diagrams with only two gluon lines, which may be crossed or planar, depending on the placement of the quark-gluon vertices, and also diagrams with a three-gluon vertex, which have three gluon lines.  Other diagrams, with gluon self-energies or a separate, 4-point or higher fermion loop, can be treated in just the same way as in QED \cite{Anastasiou:2020sdt}.

 In a general electroweak amplitude, such as for WWZ production,  diagrams with single-collinear limits must have at least one of the gluons attached to a quark-gluon vertex adjacent to the external quark or antiquark line.  Denoting the momentum of this gluon as $k$, we study here the regions where $k$ becomes collinear to the momentum of the relevant external line, while the other gluon carries an off-shell loop momentum.   In the notation introduced above, these regions are denoted by $\rm(1_k,H_l)$ and $\rm (2_k,H_l)$. 
 
Included in these diagrams are vertex and self-energy corrections to one-loop diagrams, in which a gluon connects one of the incoming fermions to a fermion in the hard scattering.  Also included are box and higher-point diagrams for which an additional gluon is attached by at least one vertex to an off-shell quark line.

The full set of diagrams is illustrated in Fig.\ \ref{fig:two-gluons-ew}, where Fig.\ \ref{fig:two-gluons-ew}a shows diagrams with color factors $C_F^2=\sum_{a,b}\, T_b T_a T_a T_b$, the ``uncrossed ladders".   Those in Fig.\ \ref{fig:two-gluons-ew}b are crossed ladders, all with color factors $C_F^2-C_FC_A/2=\sum_{a,b}\, T_b T_a T_b T_a$ and those in Fig.\ \ref{fig:two-gluons-ew}c have a three-gluon vertex, and color factor $C_FC_A/2$..  In the figures, the functions ${\cal M}^{(0)}_{2+d}(r,q_d,r_d)$ represent the integrands of subdiagrams with an incoming quark line of momentum $r$, $d$ external electroweak bosons, whose momenta are denoted collectively by $q_d$, and an outgoing quark line of momentu $r_d$. 
Here, $d=a$ or $b$, and we define
\bea
r_a\ &=&\ p_1-\sum_a q_a\, ,
\nn\\[2mm]
r_b\ &=&\ p_1-\sum_a q_a - \sum_b q_b\, .
\label{eq:r-defs}
\eea
These diagrams are trees with only electroweak vertices, zeroth order in $\alpha_s$. In the same spirit, ${\cal M}^\beta_{3+c}(r_b,k,q_d,p_2)$ is the integrand of an order $g_s$ tree subdiagram with incoming quark line of momentum $r_b$, an outgoing gluon of momentum $k$, external electroweak bosons with momenta $q_d$ and an incoming antiquark of momentum $p_2$.    All the diagrams  ${\cal M}$ are defined to include their adjacent fermion propagators. 

As in the case of the form factor above, we will seek an expression for the integrand in which the sum of all connections of a collinear gluon to the hard subdiagram factorizes at the integrand level, without shifts in the loop momenta.     
We can easily give a corresponding prescription for adding terms that integrate to zero, but which provide the desired local factorization.   The general case is illustrated in Fig.\ \ref{fig:planars}, the direct generalization of Fig.\ \ref{fig:regular-ir-uv} for the form factors.
 The diagrams that we modify are those with color factors $C_F^2$, Fig.\ \ref{fig:two-gluons-ew}a, corresponding to the generator products $\sum_{a,b}\, T_b T_a T_a T_b$, the ``uncrossed ladder" diagrams referred to above.   To all such diagrams, we add a shift counterterm, which integrates to zero.   The integrands of other diagrams are unchanged.

The shift term is the difference between the two momentum-space integrands where the outer gluon momentum ($k$ here) flows through the inner loop either through the inner gluon, or through the quark lines that complete the inner loop.  

All of the diagrams shown in Fig.\ \ref{fig:planars}, which are in one-to-one correspondence to those in Fig.\ \ref{fig:two-gluons-ew}a, are singular in region $\rm (1_k,H_l$).   In this region, the Ward identity insures that the collinear gluon factorizes.   All of the diagrams are finite in region $(2_k,H_l)$ except for the single diagram where the vertex with index $\beta$ attaches adjacent to the external antiquark line, just as in the form factor discussion.

The prescription that provides local validity for the Ward identities when the outer gluon line $k$ becomes collinear to $p_1$ for diagrams in Fig.\ \ref{fig:planars} is a simple generalization of the form factor prescription given in Eq.\ (\ref{eq:form-factor-difference}).   The verification of the Ward identities proceeds in exactly the same manner.    

The shift terms just described for the diagrams illustrated in Fig.\ \ref{fig:planars} are given in the notation described above, and in the figure, by
\bea
 \Delta_{\rm planar}\ &=&\ \frac{1}{2}C_FC_A \ \left ( \frac{-i\eta_{\alpha\beta} }{k^2 + i\epsilon} \right )\ \bar v(p_2) {\cal M}^\beta_{3+c} \left( r_b+k,k,q_c,p_2\right) 
 \nonumber\\
&\ & \times  \; (-ig_s\gamma^\lambda) \, \Bigg [ \frac{-i\eta_{\lambda\sigma}} {l^2 + i\epsilon}\  {\cal M}^{(0)}_{2+b} \left (r_a+k+l,q_b,r_b+k\right)
\nonumber\\[2mm]
&\ & \hspace{10mm} -\ \frac{-i\eta_{\lambda\sigma}}{(l-k) ^2 + i\epsilon}\  {\cal M}^{(0)}_{2+b} \left (r_a+l,q_b,r_b+k\right) \Bigg ] (-ig_s\gamma^\sigma)\, 
\nonumber\\[2mm]
&\ & \times  \ {\cal M}_{2+a}^{(0)} \left ( p_1+k,q_a,r_a+k\right ) (-ig_s\gamma^\alpha )\,  u(p_1)\, .
 \label{eq:non-standard-color}
\eea 
where momentum $r_a+k$, flows out of the subdiagram with integrand ${\cal M}^{(0)}_{2+a}$ in each of the diagrams of 
Fig.\ \ref{fig:two-gluons-ew},
above the loop carrying momentum $l$ in the figure, and $r_b+k$ flows into the subdiagram ${\cal M}^\beta_{3+c}$ in the ladder diagrams of Fig.\ \ref{fig:two-gluons-ew}a.   The cancellation that ensures local
factorization when the diagrams of Fig.\ \ref{fig:planars} are combined with the remainder of the diagrams that are singular in $\rm (1_k,H_l)$ is identical to the cancellations described above for the form factor,
and the Ward identities are realized locally in momentum space in the same fashion.   

To be explicit, in region $(1_k,H_l)$, we can again use the collinear approximation given in Eq.\ (\ref{eq:co-approx}), allowing us to use the identity of Eq.\ (\ref{eq:Feyn-iden}) repeatedly in the subdiagrams ${\cal M}^\beta_{3+c} \left( r_b,k,q_c,p_2\right)$, giving the result,
\bea
 \Delta_{\rm planar} \Big |_{(1_k,H_l)}\ &=&\ \frac{1}{2}C_FC_A \ \left ( \frac{-i }{k^2 + i\epsilon} \right )\ \bar v(p_2) {\cal M}^{(0)}_{2+c} \left( r_b,q_c,p_2\right) \, \frac{1}{p_2\cdot k}
 \nonumber\\
&\ & \times  \; (-ig_s\gamma^\lambda) \, \Bigg [ \frac{-i\eta_{\lambda\sigma}} {l^2 + i\epsilon}\  {\cal M}^{(0)}_{2+b} \left (r_a+k+l,q_b,r_b+k\right)
\nonumber\\[2mm]
&\ & \hspace{10mm} -\ \frac{-i\eta_{\lambda\sigma}}{(l-k) ^2 + i\epsilon}\  {\cal M}^{(0)}_{2+b} \left (r_a+l,q_b,r_b+k\right) \Bigg ] (-ig_s\gamma^\sigma)\, 
\nonumber\\[2mm]
&\ & \times  \ {\cal M}_{2+a}^{(0)} \left ( p_1+k,q_a,r_a\right ) (-ig_s\psla_2 )\,  u(p_1)\, .
 \label{eq:non-standard-color-in-1kHl}
\eea 
As desired, these terms cancel the nonfactoring terms, analogous to the second and third diagrams of Fig.\ \ref{fig:regular-wi-p1} for the form factor, which result from making the approximation in Eq.\ (\ref{eq:co-approx}) for region $(1_k,H_l)$ in the crossed ladders and three-gluon vertex diagrams of Figs.\ \ref{fig:two-gluons-ew}b and c.  

 In closing this subsection, we observe that the same procedure that demonstrates factorization of the shift counterterms in double collinear limits $\rm (1_k,1_l)$ and $\rm (2_k,2_l)$  for the form factor applies to the general electroweak amplitudes discussed here.   The only difference is to apply the identity of Eq.\ (\ref{eq:Feyn-iden}) repeatedly, first to the outer gluon, and then to the resulting vertex into which both (collinear) gluon momenta flow.

 \begin{figure}[ht]
   \begin{equation*}
     \mbox{
       {\Large
         $
\frac{C_A}{2\, C_F} \; \cdot \; 
         $
     }}
     \eqs[0.35]{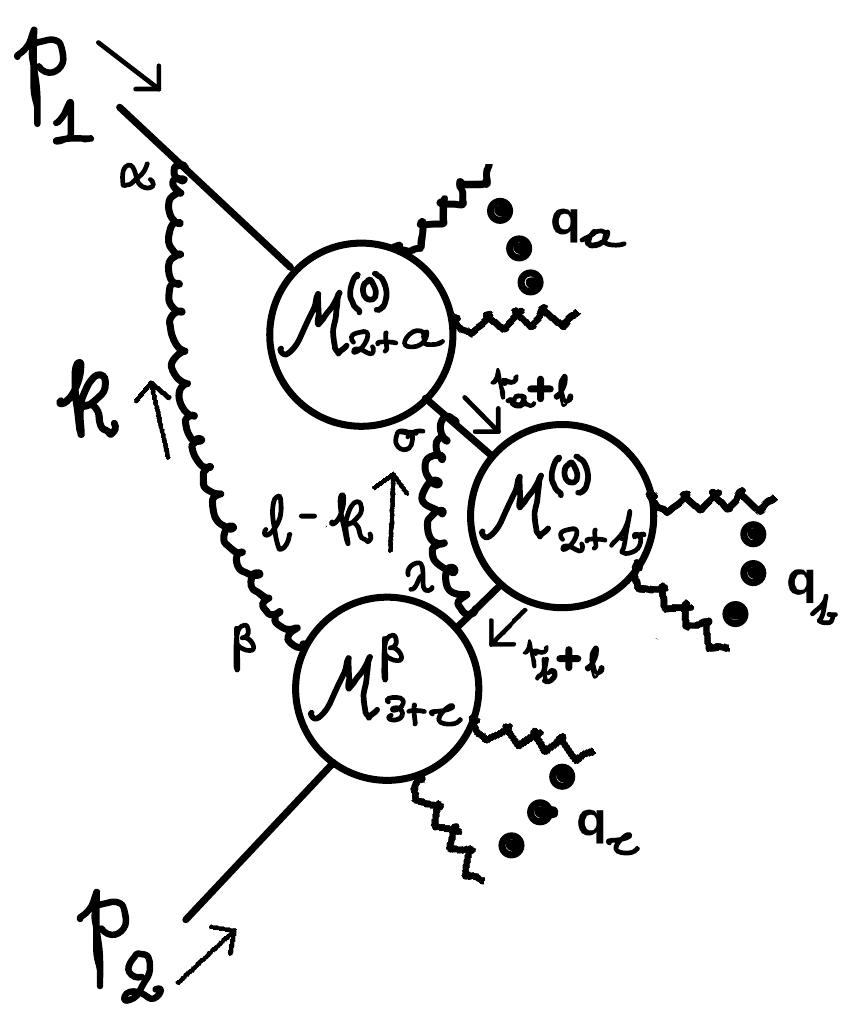}
     \quad \;
     \mbox{{\Large $-$ }}
     \quad \;
          \mbox{
       {\Large
         $
\frac{C_A}{2\, C_F} \; \cdot \; 
         $
     }}
     \eqs[0.35]{dtermA.pdf}
   \end{equation*}
\caption{ Shift term $\Delta_{\rm planar}$, Eq.\ (\ref{eq:non-standard-color-in-1kHl}), for uncrossed gluonic loops. All diagrams are assigned color factor $C_FC_A/2$.  Each pair of diagrams integrates to zero in loop momentum $l$, but enables local factorization in region $(1_k,H_l)$. \label{fig:planars}}
\end{figure}

\subsection{Local factorization for ghost terms}
  \label{sec:ghosts}
  
In the foregoing, we have split the treatment of diagrams with three-gluon vertices into ``scalar" and ``ghost" components.  For V type diagrams discussed in Sec.\ \ref{sec:VandS}, these were the scalar  term  $W^\mu_{\rm scalar}$ and the ghost term $O^\mu(k,l)$, given in Eqs.\ (\ref{eq:W-def}) and (\ref{eq:O-def}), respectively.   The $O^\mu$ term for the three-gluon QCD vertex on the quark line, in particular, is just one of the diagrams that contributes singularities in the single-collinear region $\rm (2_k,H_l)$, where we expect a factorization of the type shown in Fig.\ \ref{fig:two-circles}.  We have set aside contributions of this type until now, and we must still show that their factorization requires no shifts of loop momentum, and hence no additional counterterm.  That is, we will verify that the factorization of the ghost contributions is already local at the order to which we work.  The contributions we have set aside are all in the diagrams of Fig.\ \ref{fig:two-gluons-ew}c for the region $\rm (2_k,H_l)$, with a three-gluon vertex connecting a collinear gluon to the hard scattering.   Precisely analogous arguments apply to $\rm (1_k,H_l)$.   

 This decomposition into scalar and ghost terms for the diagrams of Fig.\ \ref{fig:two-gluons-ew}c originates with the contraction of a tree triple-gluon vertex with a longitudinal polarization from one of the gluons,  
\begin{eqnarray}
  \eqs[0.18]{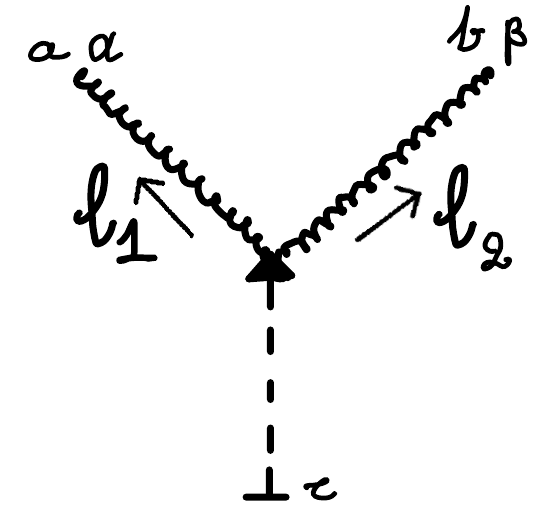} &\equiv& \frac{ \, g_s  f^{abc}}{l_1^2 \, l_2^2} \left[
    \eta^{\alpha \beta} \left( l_2 - l_1 \right)^\mu
    -\eta^{\beta \mu } \left( 2 l_2 +l_1 \right)^\alpha
     +\eta^{\mu \alpha } \left( 2 l_1 + l_2\right)^\beta
  \right] 
  \left(l_1 + l_2 \right)_\mu
  \nonumber \\
                  &=& \frac{ g_s  f^{abc}}{l_1^2 \, l_2^2}  \left[ l_1^\alpha l_1^\beta  -l_2^\alpha l_2^\beta 
                      + \eta^{\alpha \beta} \left( l_2^2 - l_1^2 \right)      
                      \right]\, ,
\end{eqnarray}  
where in the cases we will consider, $l_1+l_2=\pm k$.
The first two terms in the second equality can be interpreted as ghost-gluon vertices multiplied by the momentum of the outgoing ghost.  These are our ghost terms, referring to their role in the Ward identity, which we exhibit them graphically as, for example,
\begin{equation}
  \eqs[0.18]{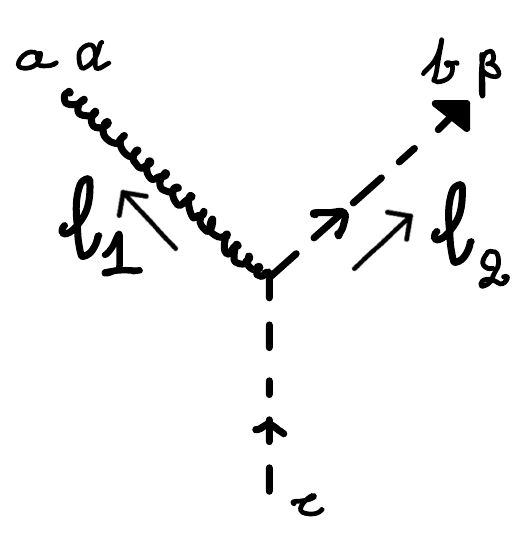} \equiv
   \frac{(-i)}{l_1^2} \frac{(i)}{l_2^2} l_2^\beta
  \left( g_s f^{abc} (-l_2^\alpha) \right)\, ,
\end{equation}
where the term in parentheses is the standard QCD ghost-gluon vertex, and where the arrow at the end of the ghost line is 
\begin{equation}
\eqs[0.12]{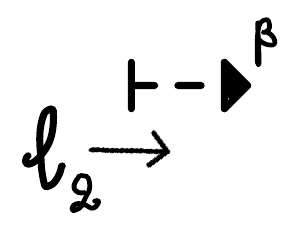} \equiv  l_2^\beta \, . 
\end{equation}
The contraction of the triple gluon vertex is then
\begin{eqnarray}
\label{eq:GGG=ghost+scalar}  
  \eqs[0.17]{Lgg} &=&   \eqs[0.17]{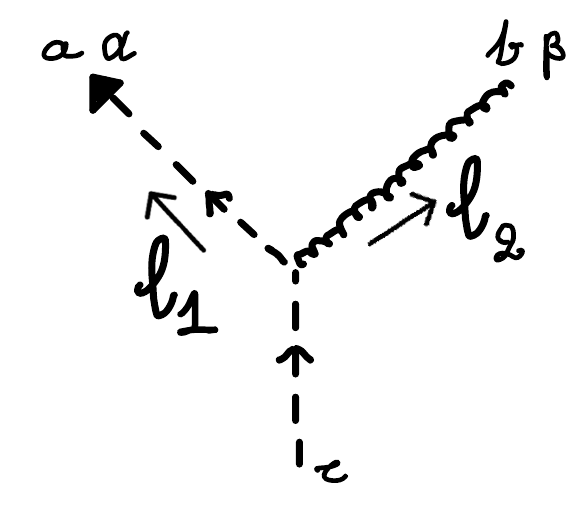} + \eqs[0.17]{Lgg_gh2} 
                      \nonumber \\ &&
\hspace{1cm}                                      
                                      + g_s  f^{abc}  \eta^{\alpha \beta}  \left[ 
                                      \frac{1 }{l_1^2 }
                                      -
                                      \frac{1}{l_2^2 }
                                      \right]
                                      \, .
\end{eqnarray}
The final terms in the right-hand side are the contribution of what we have called the scalar part of the three-gluon vertex.

To understand factorization for ghost terms in  $\rm (2_k,H_l)$, we must consider all diagrams that have three-gluon vertices and are singular in this region.  To anticipate, we check numerically,  in Sec.\ \ref{sec:check},  the absence of singular-collinear divergences in $\rm (2_k,H_l)$ for the sum of all diagrams in the diphoton amplitude, including the ghost contributions.
 We would like to demonstrate here the mechanism of this cancellation analytically in this case.   This will make the pattern clear, and the result applies immediately to amplitudes with arbitrary numbers of massive electroweak bosons at this order in QCD.
Specifically, we are going to confirm that the collinear singularity from ghost terms factorizes, and is thus cancelled by the corresponding IR contribution to the form factor in the finite amplitude we construct in  Eq.\ (\ref{eq:cal-M-finite-12}).

Specializing to diphoton production, there  are five  diagrams  with three-gluon vertices that become singular in the region $\rm (2_k,H_l)$ (plus five more with the photons exchanged.)
To see how factorization works for the ghost  terms, we must combine all five diagrams.
Suppressing the external antiquark line, in the $k \parallel p_2$ collinear limit,  these diagrams contribute
\begin{eqnarray}
 \lim_{k \parallel p_2}\left. {\cal M}_2 \right|_{\rm ghost} 
  &\sim&
         \eqs[0.22]{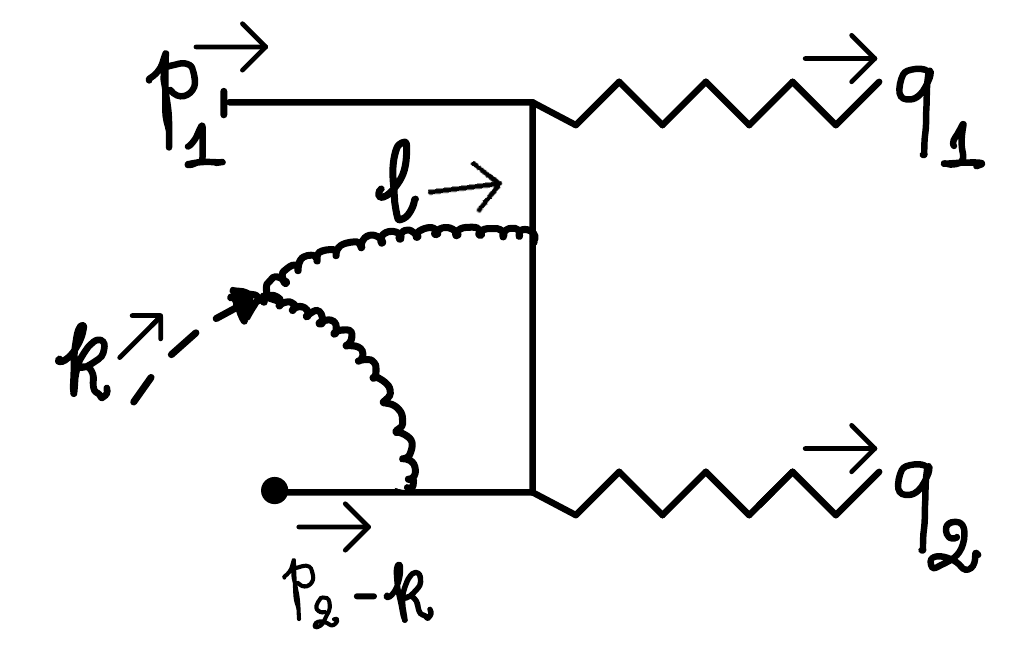}
         +         \eqs[0.22]{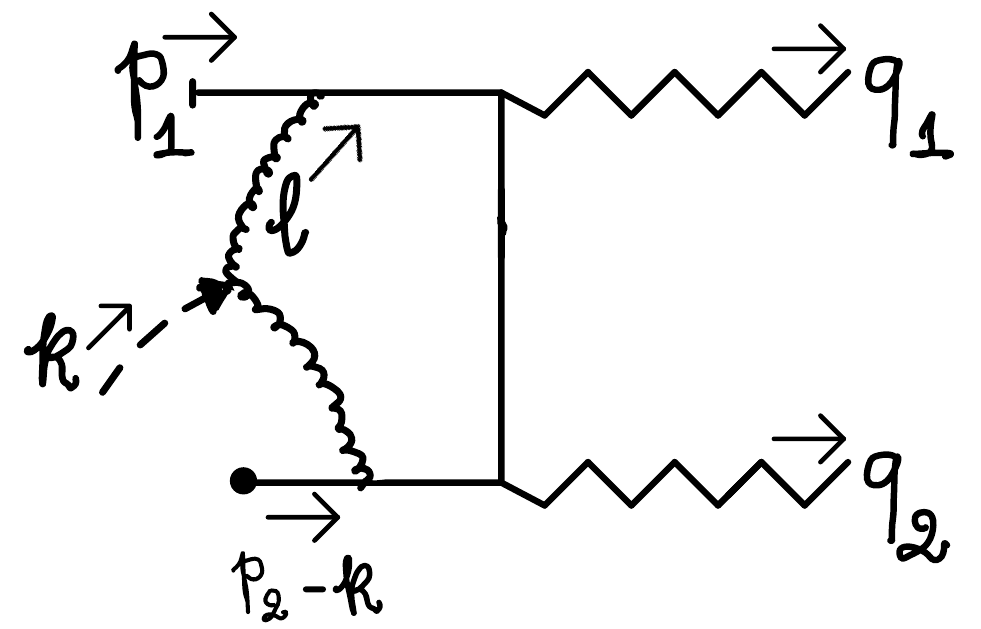}
         \nonumber \\
  && \hspace{1cm}
     +         \eqs[0.22]{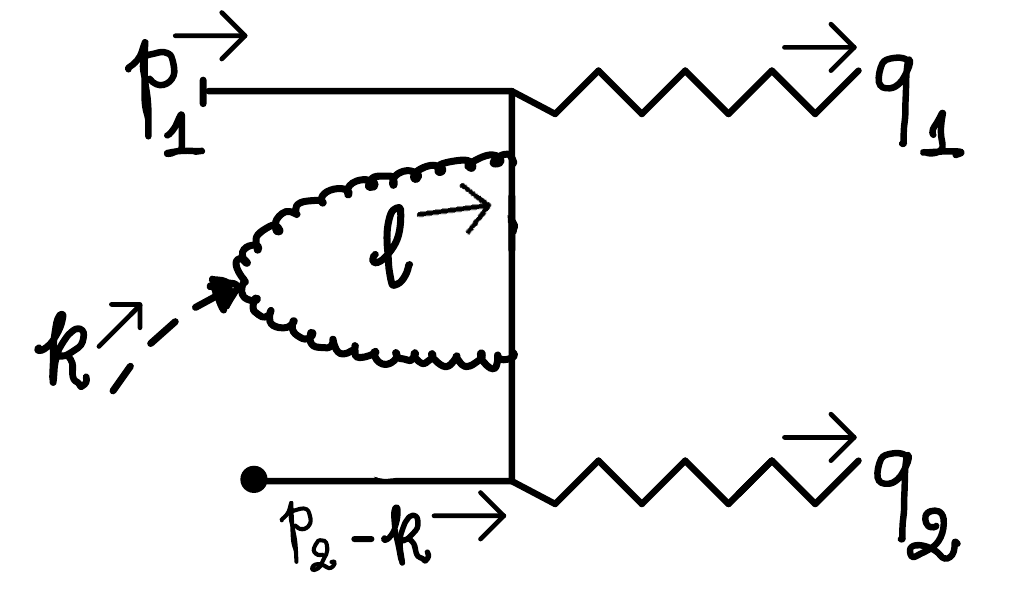}
     +         \eqs[0.22]{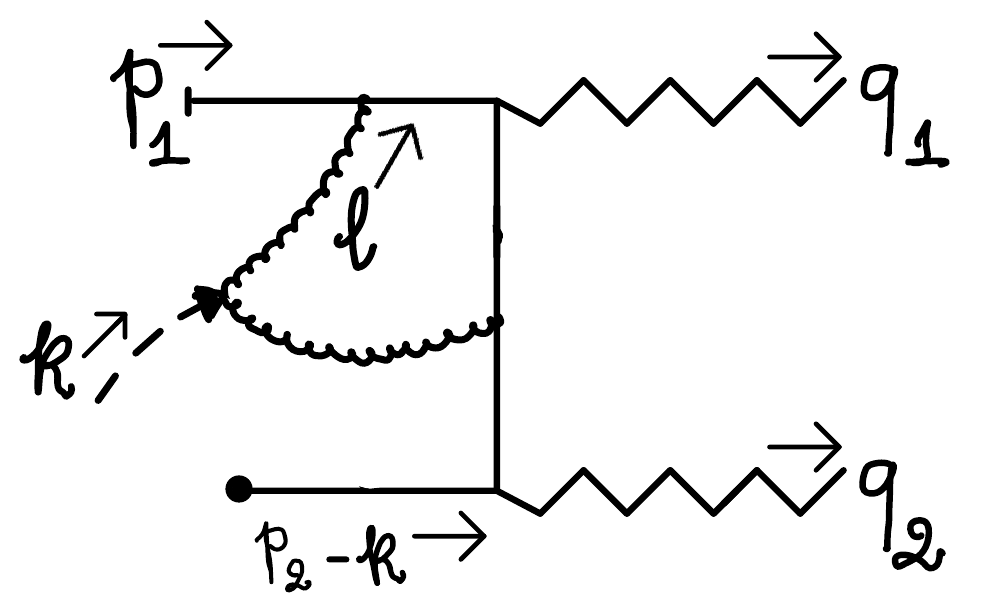}
     +         \eqs[0.22]{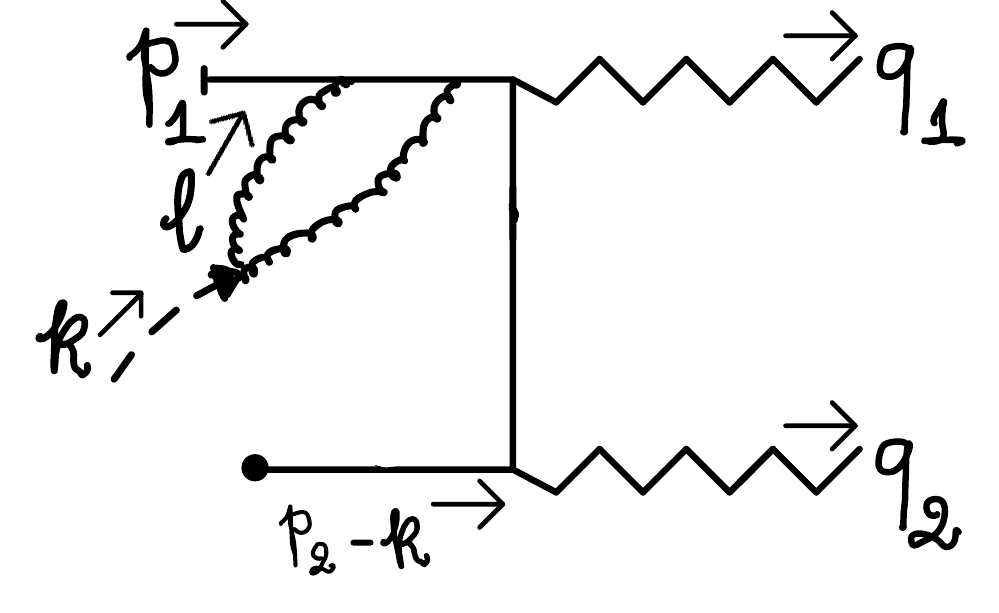}\, .
     \nonumber \\
  &&
\end{eqnarray} 
After applying the identity of Eq.~\eqref{eq:GGG=ghost+scalar} and keeping the ghost contributions,
we have
\begin{eqnarray}
  \label{eq:ghost_diphoton_B}
&&\lim_{k \parallel p_2}\left. {\cal M}_2 \right|_{\rm ghost} \sim 
   \eqs[0.20]{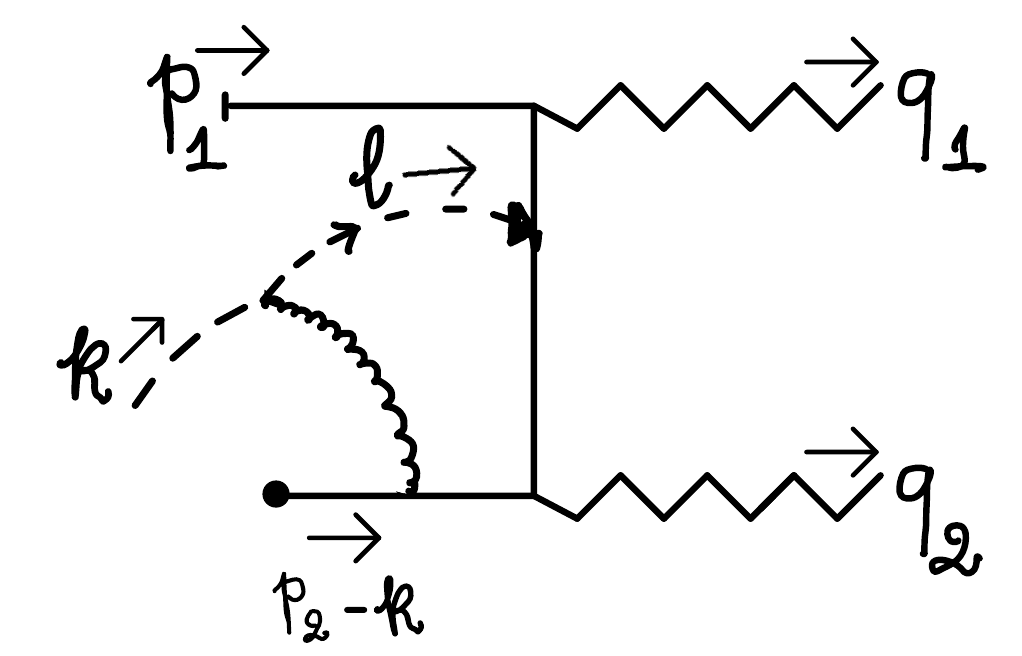}
    +  \eqs[0.20]{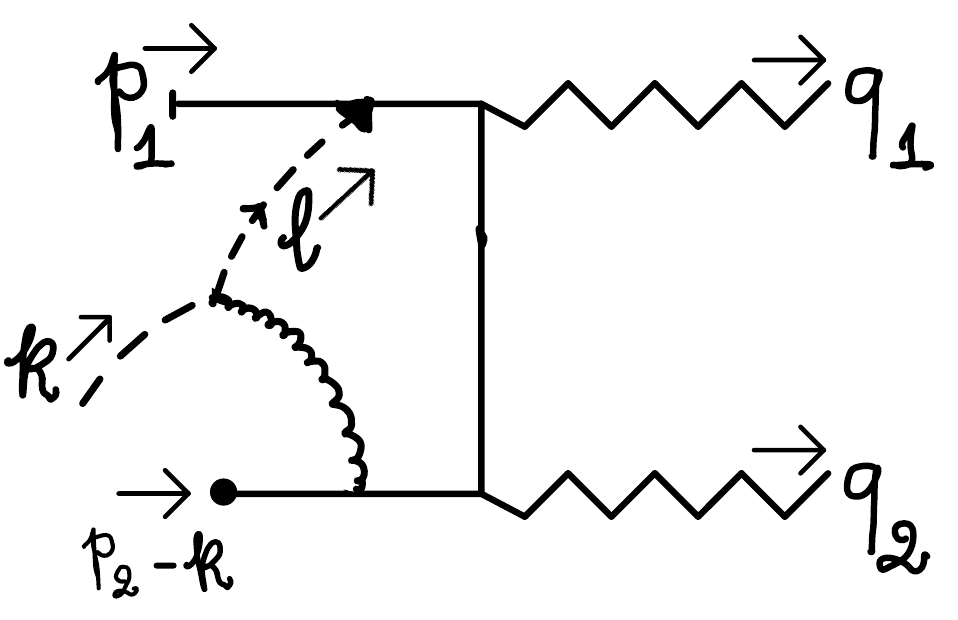}
   \nonumber \\ &&
   +\eqs[0.20]{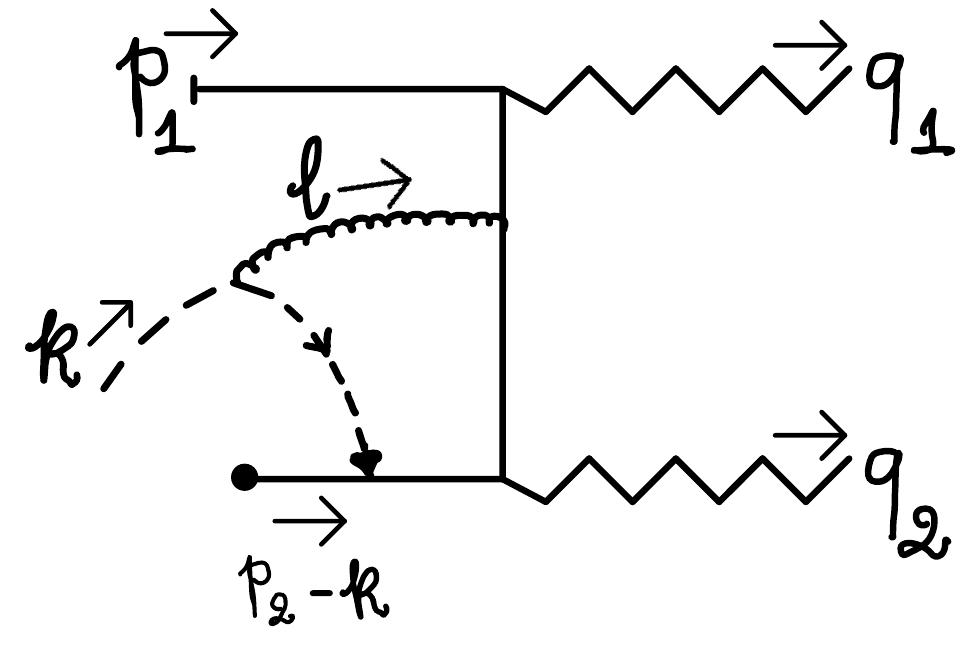}
                   +  \eqs[0.20]{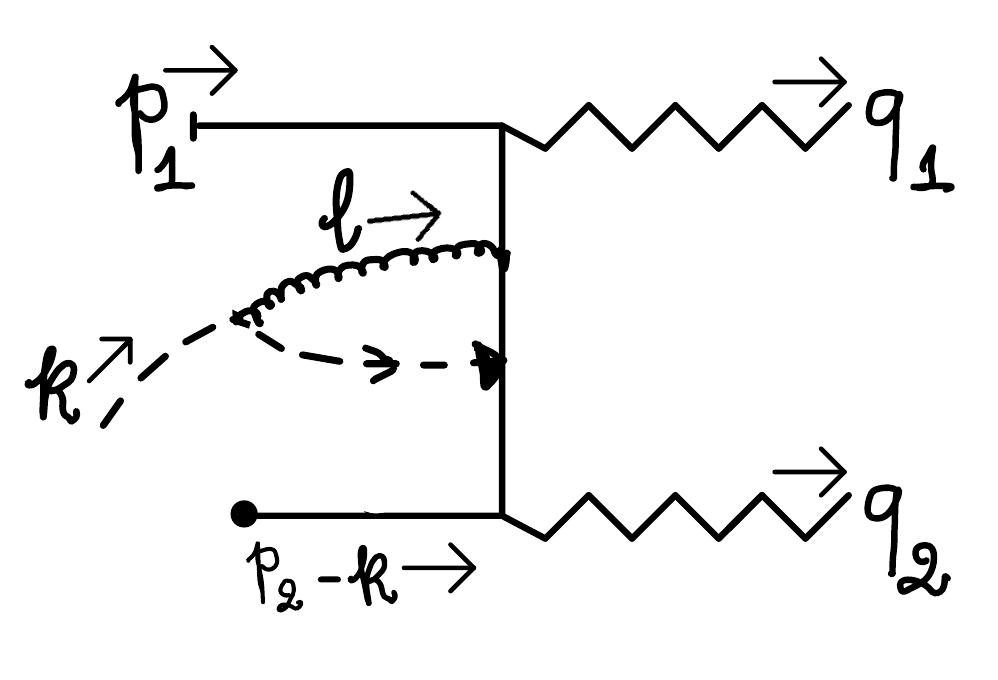}
   + \eqs[0.20]{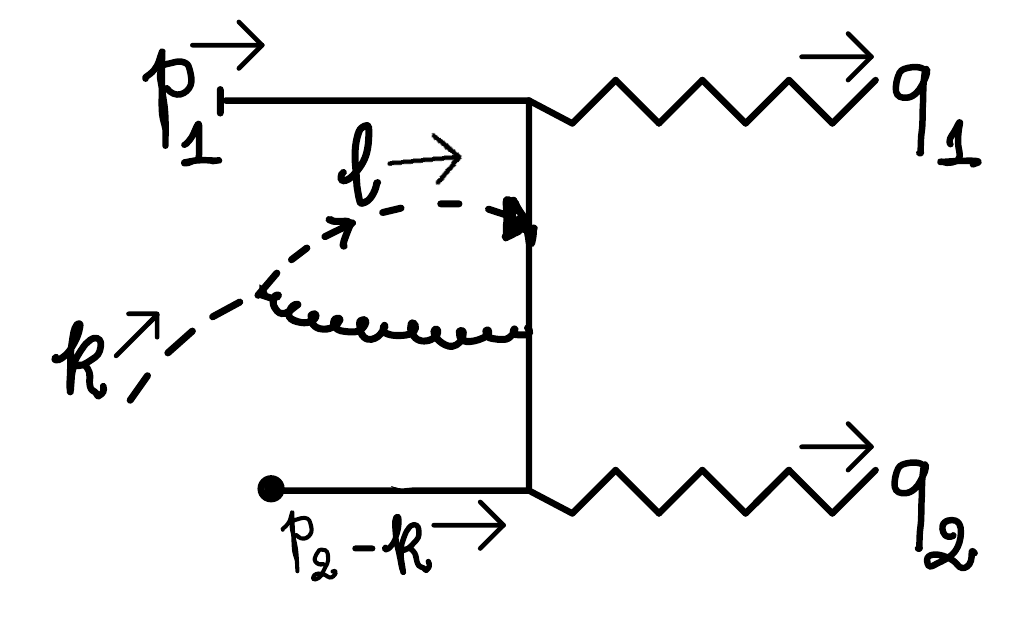}
                   +  \eqs[0.20]{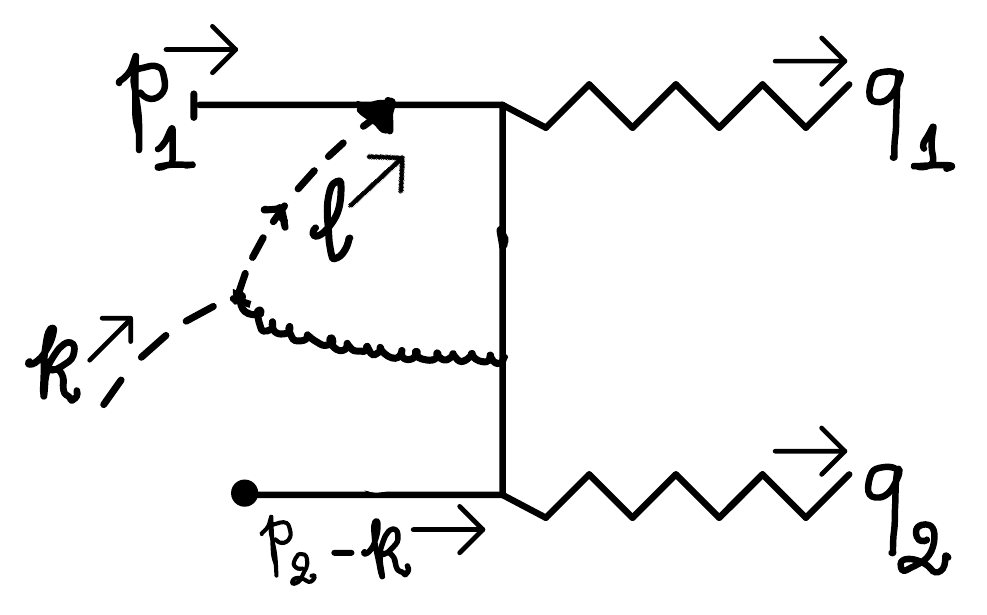}
                   \nonumber \\ &&
      + \eqs[0.20]{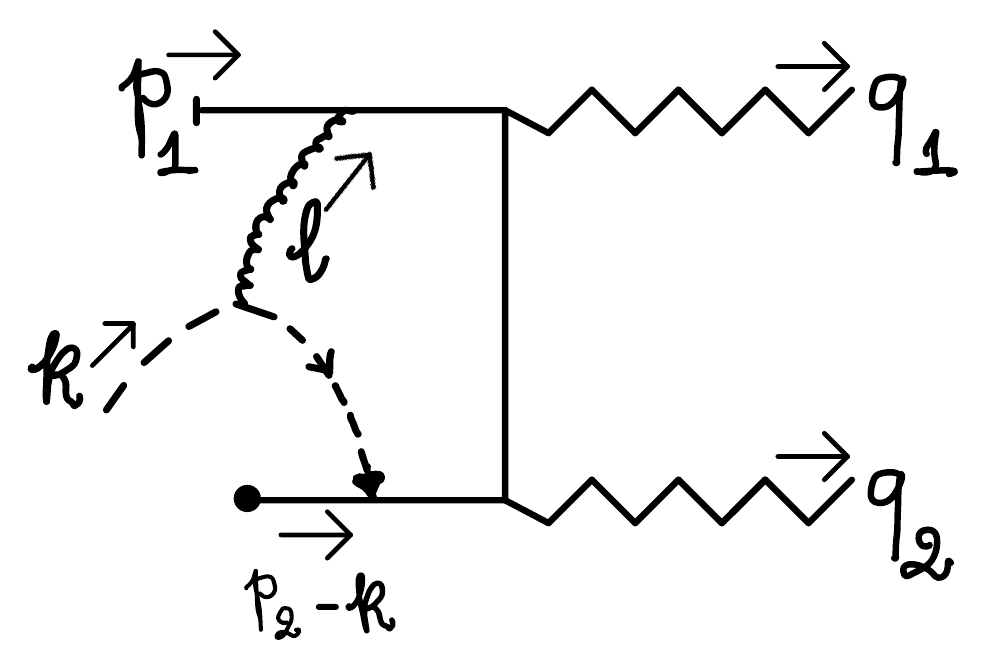}
                                   +  \eqs[0.20]{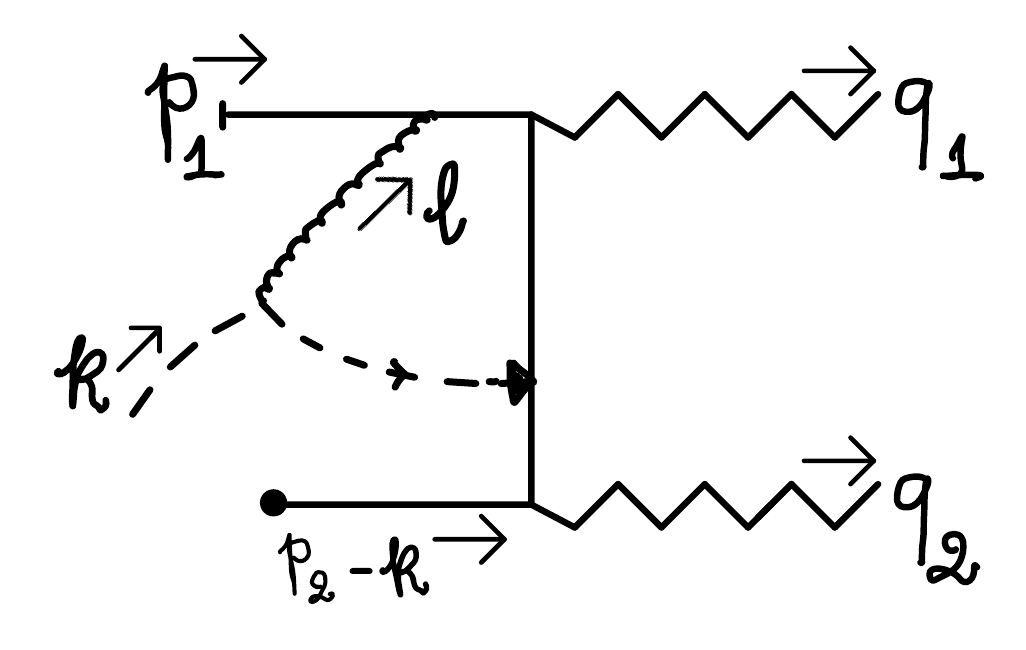}
                                       + \eqs[0.20]{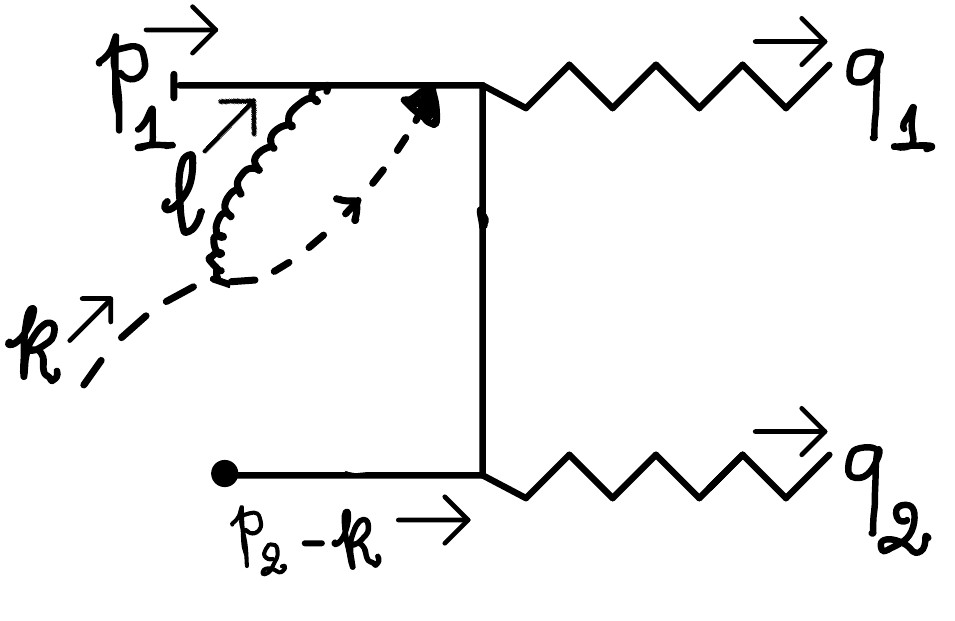}
                                   +  \eqs[0.20]{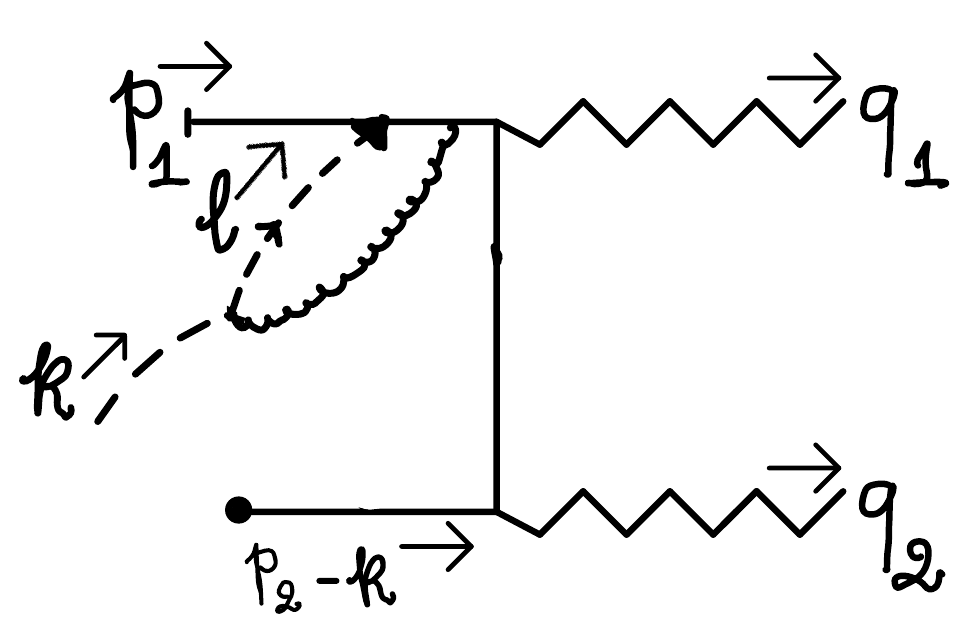}\, .
                                   \nonumber \\
  &&
\end{eqnarray}
In the above, the vertex of a quark an antiquark and a ghost with an arrow is substituted by
\begin{eqnarray}
\eqs[0.15]{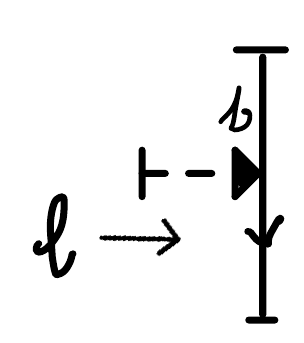} = -i g_s T_b^{(q)} \, \slashed l \,.  
\end{eqnarray}
We also introduce a quark-antiquark-ghost vertex (without an arrow) that reads
\begin{eqnarray}
 \eqs[0.15]{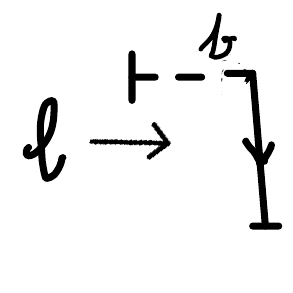} = g_s T_b^{(q)} \, .  
\end{eqnarray}
  
As usual, the key to factorization in the collinear region is the scalar polarization of the collinear gluon where it attaches to the off-shell ``hard" subdiagram.
For the terms considered here, the results we are after follow primarily from the tree-level gluon quark identity of Eq.\ (\ref{eq:Feyn-iden}), which we  represent pictorially as
\begin{eqnarray}
  \label{eq:Feyn_iden_picture}
 \eqs[0.17]{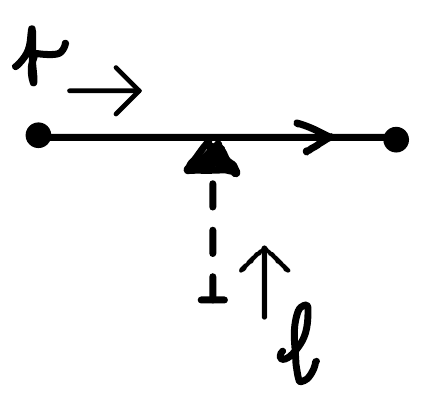} &=& \eqs[0.17]{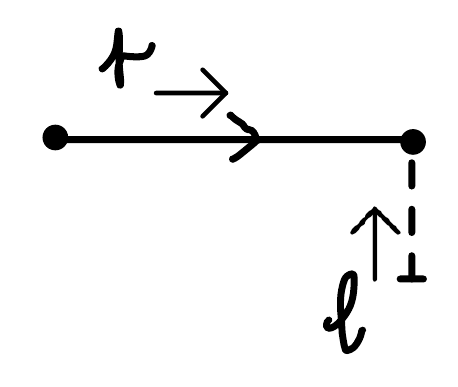} - \eqs[0.17]{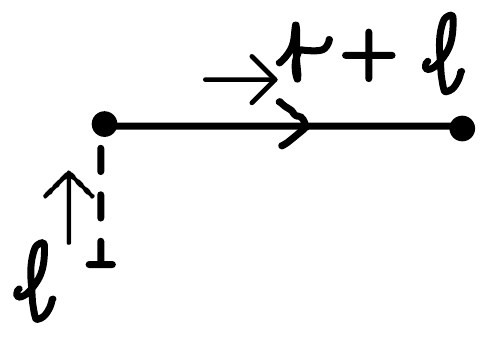}\, .
\end{eqnarray}
With the application of Eq.~\eqref{eq:Feyn-iden}, the expression of Eq.~\eqref{eq:ghost_diphoton_B} simplifies to
\begin{eqnarray}
  \label{eq:ghost_diphoton_C}
&&\lim_{k \parallel p_2}\left. {\cal M}_2 \right|_{\rm ghost} \sim 
   \eqs[0.20]{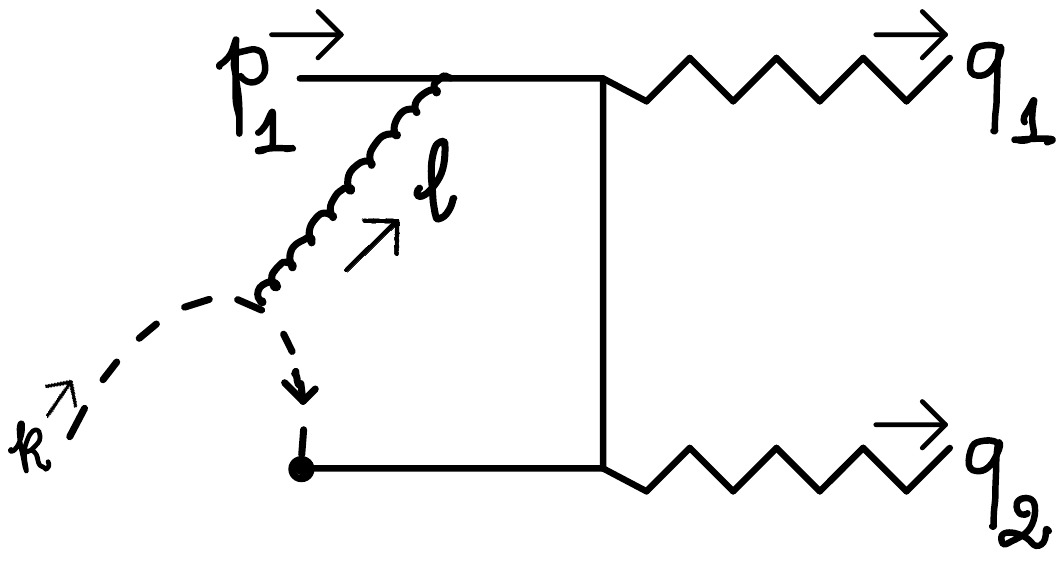} +\eqs[0.20]{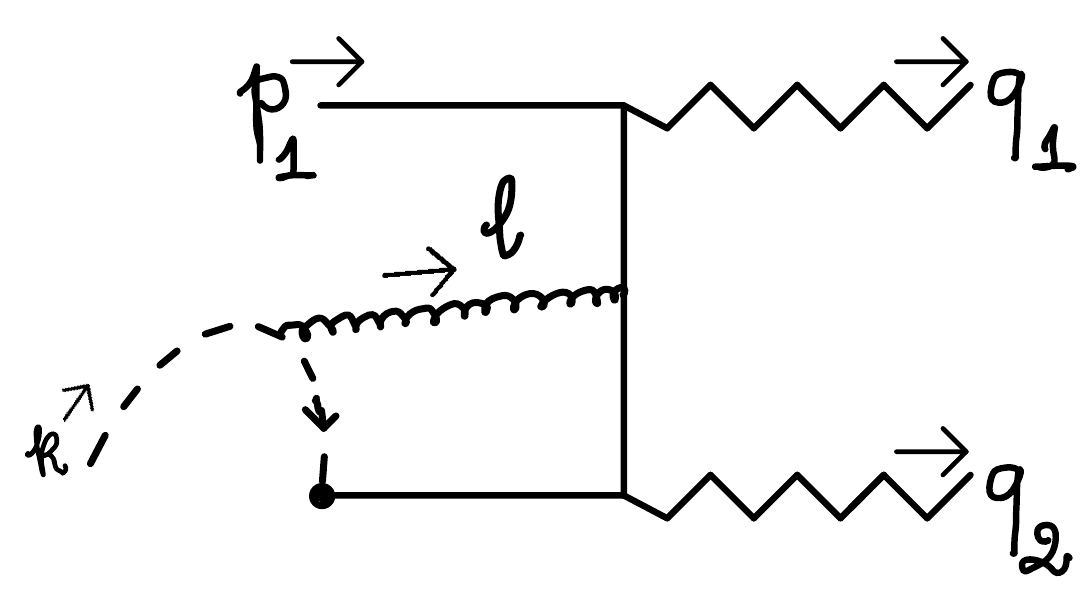} + \eqs[0.20]{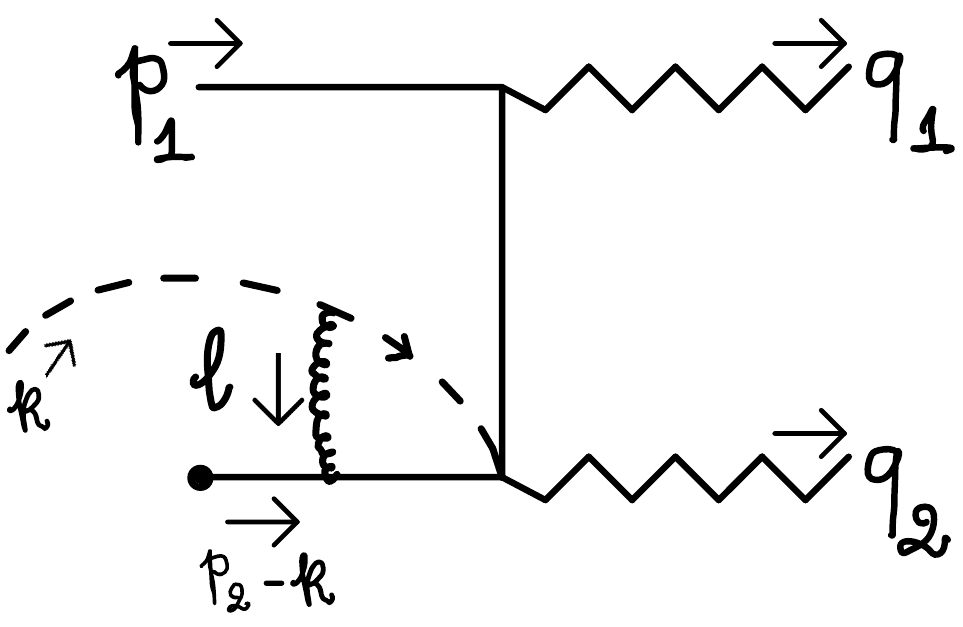}
                   \nonumber \\ &&
                                   + \eqs[0.20]{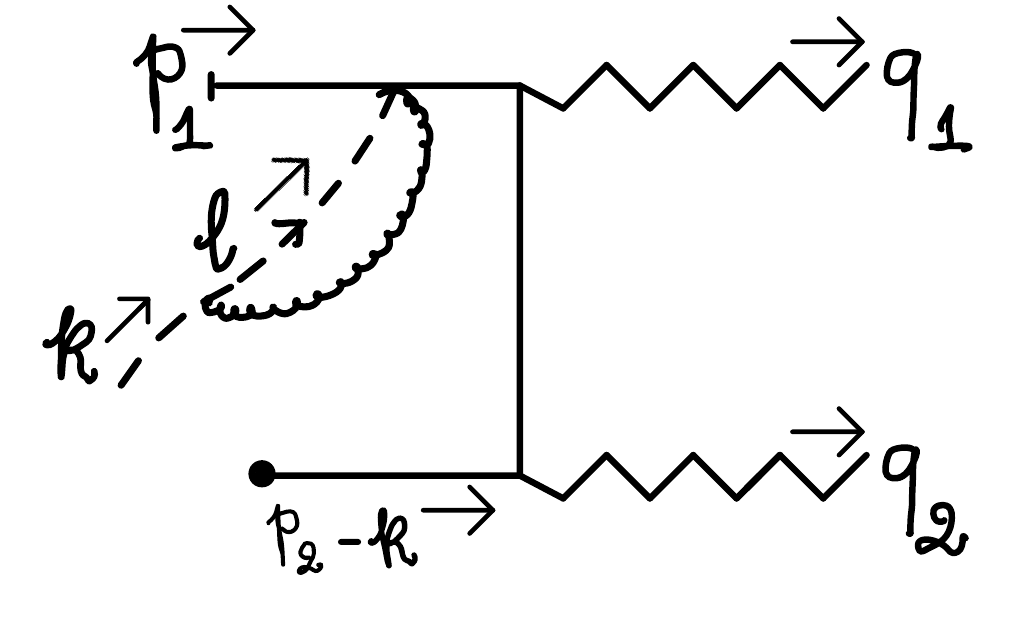}
                                   - \eqs[0.20]{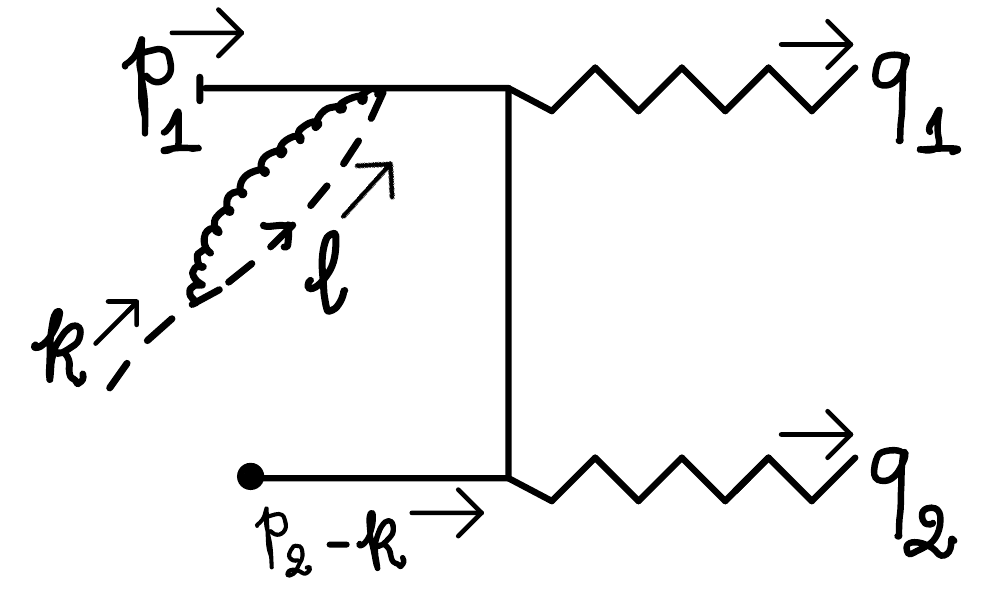}
                                   + \eqs[0.20]{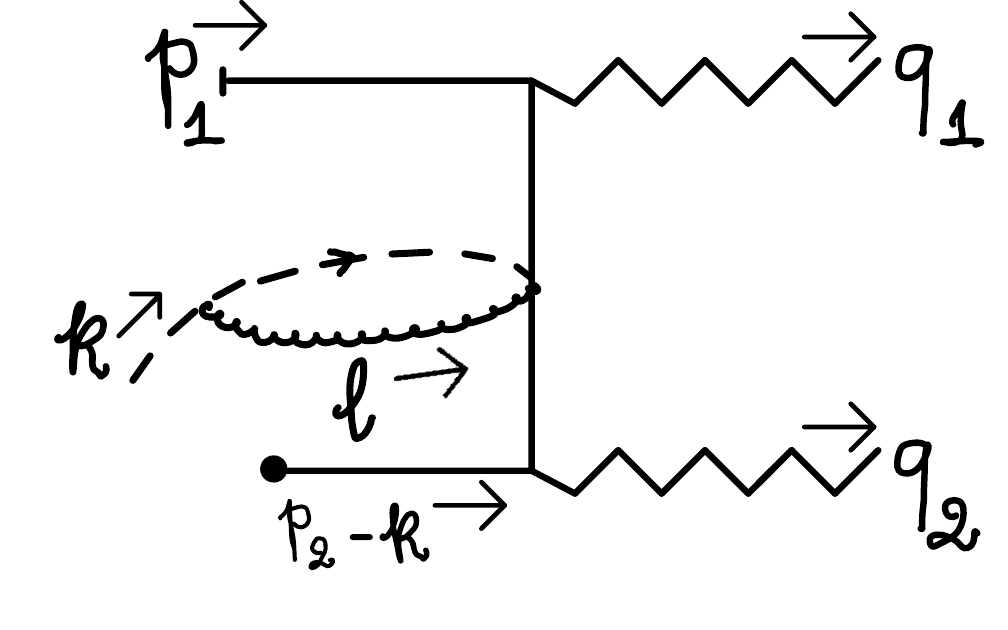}
                                   - \eqs[0.20]{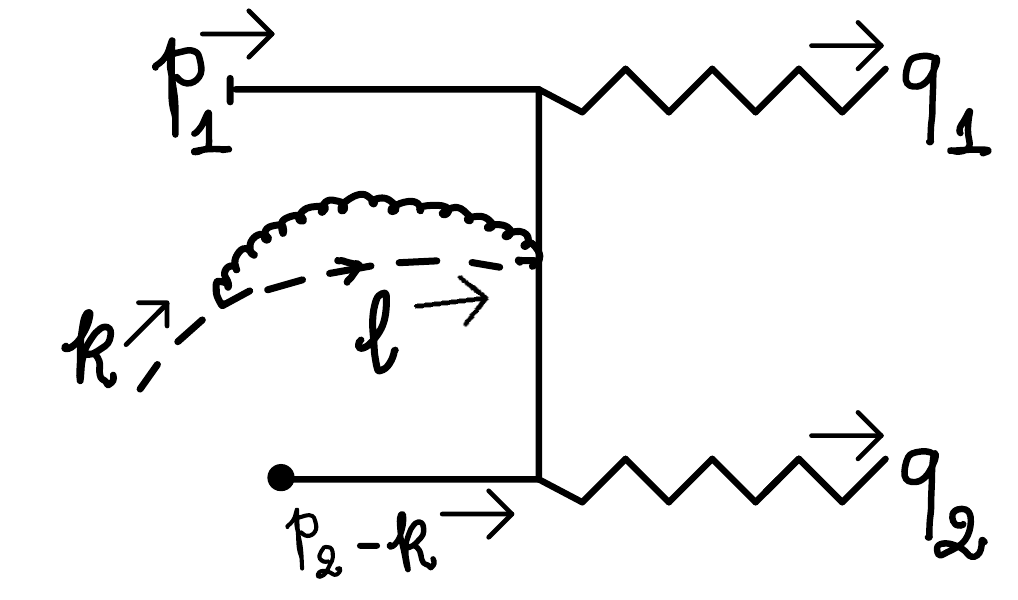}\, .
                                   \nonumber \\
  &&
\end{eqnarray}
     The first and second diagrams in the right-hand side are not singular in the $k  \parallel p_2$ limit, as the $1/(\slashed p_2 - \slashed k )$ denominator is cancelled.  The third diagram contributes to the singularity, but it is factorized.  The remaining diagrams include differences of self-energy subgraphs,
\begin{eqnarray}
  \label{eq:formghostSE_step1}
  \eqs[0.13]{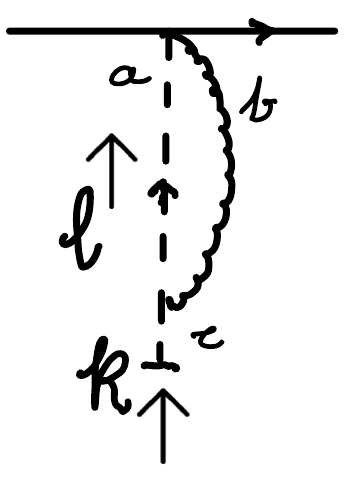} - \eqs[0.13]{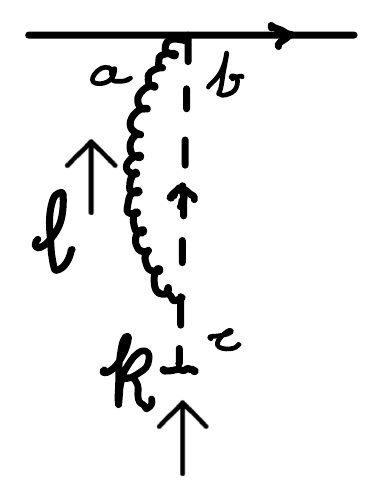} &=&
g_s^3 \left( i f^{bac} T_a^{(q)} T_b^{(q)}\right) \frac{ \slashed l }{l^2 (l-k)^2}
                                              -g_s^3 \left( i f^{abc} T_a^{(q)} T_b^{(q)}\right) \frac{ \left(\slashed k - \slashed l \right)}{l^2 (l-k)^2}
                                              \nonumber \\
  &=&        g_s^3 \frac{C_A}{2} T_c^{(q)}       \frac{ \slashed k }{l^2 (l-k)^2}                                
 \end{eqnarray}  
      which, at the integrand level, each equals a ghost self-energy correction multiplied by a momentum vector. Indeed, a direct computation gives
      \begin{eqnarray}
         \label{eq:formghostSE_step2}
\eqs[0.13]{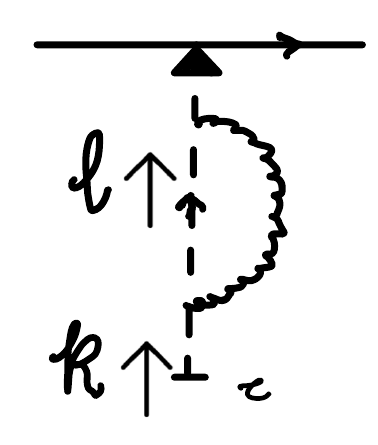} &=& g_s^3 C_A T_c^{(q)} \frac{k \cdot l }{ l^2 \left( l-k \right)^2 k^2} \, \slashed k \, .  
      \end{eqnarray}
Symmetrizing Eq.~\eqref{eq:formghostSE_step2}  over loop momentum flows, $l \leftrightarrow k-l$, we observe that it matches the right-hand side  of 
Eq.~\eqref{eq:formghostSE_step1}. We obtain,                           
\begin{eqnarray}
  \label{eq:formghostSE}
  \eqs[0.13]{selffg} - \eqs[0.13]{selfgf} = \frac 1 2 \eqs[0.13]{selfghostL}
  + \frac 1 2 \eqs[0.13]{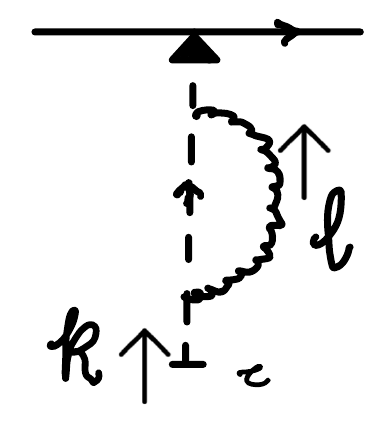}
  \equiv \eqs[0.13]{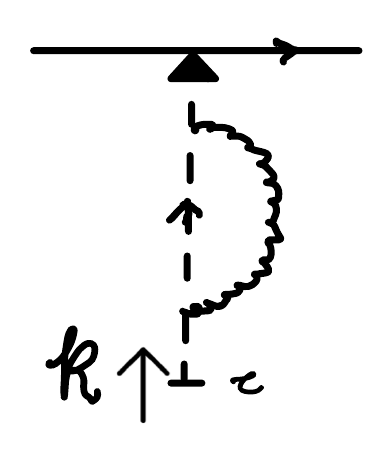}
\end{eqnarray}
The above identity casts the collinear limit in the form
\begin{eqnarray}
  \label{eq:ghost_diphoton_D}
\lim_{k \parallel p_2}\left. {\cal M}_2 \right|_{\rm ghost} &\sim& 
   \eqs[0.20]{diphgh_2} +\eqs[0.20]{diphgh_3} + \eqs[0.20]{diphgh_4}
                   \nonumber \\ &&
                   + \eqs[0.20]{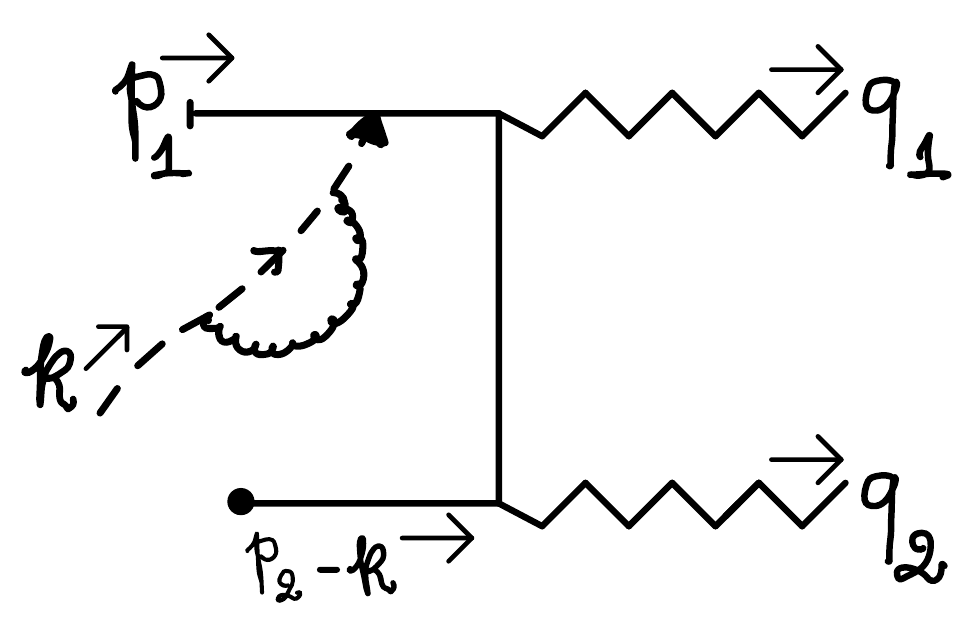}
                   + \eqs[0.20]{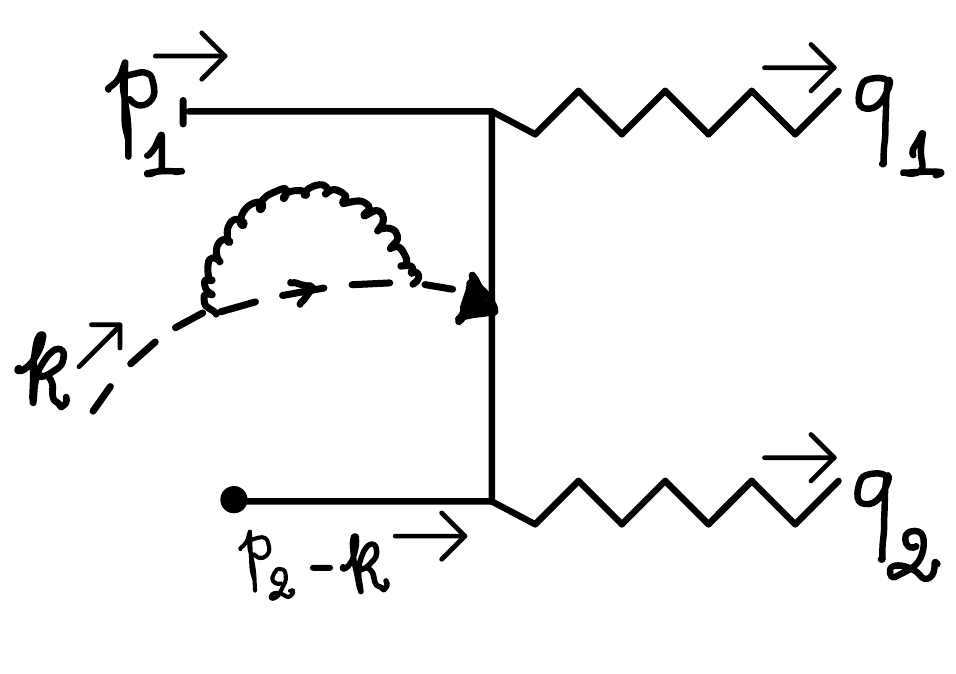} \, .
                                   \nonumber \\
  &&
\end{eqnarray}
The last two diagrams with ghost self-energy insertions simplify with the application of Eq.~\eqref{eq:Feyn_iden_picture} and the use of the Dirac equation on the quark $u(p_1)$.  We find,
\begin{eqnarray}
  \label{eq:ghost_diphoton_E}
\lim_{k \parallel p_2}\left. {\cal M}_2 \right|_{\rm ghost} &\sim& 
   \eqs[0.20]{diphgh_2} +\eqs[0.20]{diphgh_3} 
                   \nonumber \\ &&
+ \eqs[0.20]{diphgh_4}
                   + \eqs[0.20]{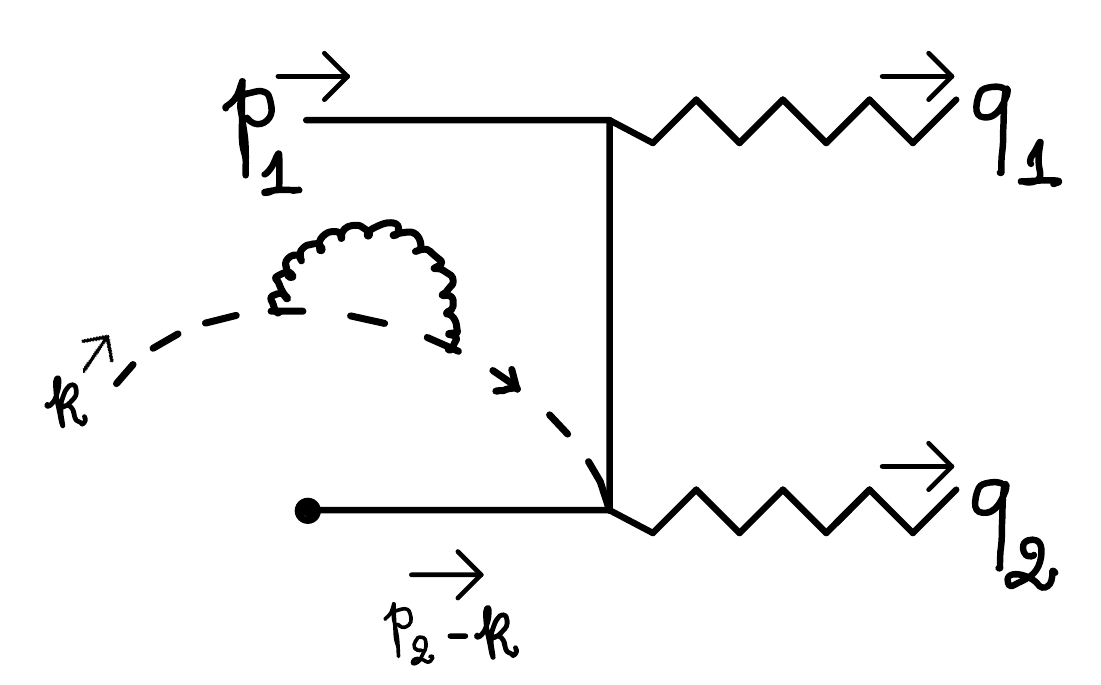} \, .
                                   \nonumber \\
  &&
\end{eqnarray}
In Eq.~\eqref{eq:ghost_diphoton_E}, the remaining terms are either finite in the $k \parallel p_2$ limit  (first two diagrams in the right-hand side) or are factorized (two last diagrams).   The factorized singularities are cancelled against the analogous contribution to the form factor IR counterterm.   
This is the result we set out to show, here in the case of the production of two off-shell photons.  The arguments, however, go over directly to the full set of processes for electroweak production we are considering.   A higher number of
electroweak vertices simply increases the number of times we need to apply the tree-level Ward identity (\ref{eq:Feyn-iden}), shown graphically in (\ref{eq:Feyn_iden_picture}).
Thus, as anticipated, no further infrared counterterms are required to produce our local integrand in the subtracted amplitude of Eq.\ (\ref{eq:cal-M-finite-12}).   To complete our
construction of a numerically-computable subtracted amplitude, we much render the infrared-subtracted amplitudes ultraviolet-convergent.  This is the subject of the next section.


\section{Ultraviolet subtractions}
\label{sec:UV}

The diagrams of Fig.\ \ref{fig:two-gluons-ew},  the set for which  the
Ward identity factorizes single-collinear gluon contributions, require
QCD and electroweak renormalization in general.  As in Ref.\ \cite{Anastasiou:2020sdt}, we
seek to add infrared finite counterterms that render the diagrams
convergent in the ultraviolet.   These counterterms can be computed
separately, and tailored to be equivalent to any desired
renormalization scheme.

Our counterterms remove the ultraviolet
singularities of one loop amplitudes for electroweak production, as  
\begin{equation}
  \label{eq:Ampl1R}
{\cal M}^{\left( 1\right)}_{\rm UV \; finite}(k) ={\cal M}^{\left(
    1\right)}(k) + R_{k \to \infty} {\cal M}^{\left(
    1\right)}(k),  
\end{equation}
where the term $R_{k \to \infty} {\cal M}^{\left(
    1\right)}(k)$ subtracts the singularity by furnishing an infrared finite approximation of the
amplitude in the ultraviolet multiplied  by $-1$.  The symbol $R_{k \to \infty}$ can be thought
of as a linear operation which acts on all one-loop Feynman graphs
that are singular in the ultraviolet and
substitutes them with the negative of a suitable ultraviolet approximation.

At two loops, the amplitude is rendered finite in the ultraviolet by 
\begin{eqnarray}
  \label{eq:Ampl2R}
{\cal M}^{\left( 2\right)}_{\rm UV \; finite }(k,l) &=& {\cal M}^{\left(
    2\right)}(k,l) + R_{k \to \infty} {\cal M}^{\left(
    2\right)}(k,l) + R_{l \to \infty} {\cal M}^{\left(
                                                           2\right)}(k,l)
                                                           \nonumber
  \\[2mm]
                                                       && \hspace{-3cm}
                                                          +R_{k,l \to \infty }\left\{
 {\cal M}^{\left(
    2\right)}(k,l) + R_{k \to \infty} {\cal M}^{\left(
    2\right)}(k,l) + R_{l \to \infty} {\cal M}^{\left(
                                                           2\right)}(k,l)
                                                          \right\}\, .
\end{eqnarray} 
In addition to one-loop counterterms $R_{k \to \infty}$ and $R_{l \to \infty}$  (in the first line) for ultraviolet
divergences in one loop subgraphs,  $R_{k.l \to \infty}$ (in the second line) 
introduces local counterterms that remove ultraviolet singularities in
the limit where both loop momenta become infinite simultaneously.
To construct the counterterms, we
proceed as in Ref.\ \cite{Anastasiou:2020sdt}, Taylor expanding in the
ratios of external momenta to loop momenta and screening infrared
singularities by replacing massless propagators with massive. For
example, 
\begin{eqnarray}
R_{l,k \to \infty}: \quad  \frac{1}{(l+k+r)^2} &\to& - \frac{1}{(l+k)^2} + \frac{2 (l+k) \cdot
  r}{\left((l+k)^2\right)^2} + \ldots  \nonumber \\
&\to &
-\frac{1}{(l+k)^2-M^2} + \frac{2 (l+k) \cdot
  r}{\left((l+k)^2-M^2\right)^2} 
\, . 
\end{eqnarray}
The two-loop superficial singularities in $R_{k,l \to \infty}$ are logarithmic and we construct
straightforward counterterms regarding substitutions such as the ones
above and keeping the leading term in the ultraviolet expansion. This
procedure is described in detail in
Ref.~\cite{Anastasiou:2020sdt}. The treatment of ultraviolet
singularities in one-loop subgraphs requires a further discussion due
to its interplay with the factorization of  collinear singularities. 

At the order we consider (one loop in the hard part) there are two
sets of QCD UV counterterms in $R_{l \to \infty}$, those for  fermion self energies and
those for vertex corrections.  Their construction needs to be
performed carefully, as we now explain, in what concerns ambiguities
in non-leading (finite)
contributions in the ultraviolet expansion. 
As a concrete example of such an ambiguity, consider a possible
ultraviolet counterterm of the integrand expression that is
proportional to the abelian part of the one-loop quark-gluon vertex,
\begin{eqnarray}
  R_{l \to \infty}: \quad \frac{\slashed{l} \gamma^\mu \slashed{l} }{l^2 (l-r_a)^2 (l+r_b)^2}
\to  - \, \frac{\slashed{l} \gamma^\mu \slashed{l} }{\left(l^2-M^2\right)^3}\, ,
\end{eqnarray}
where the fixed momenta $r_{a,b}$ may include loop $k$.
Alternatively, the same expression can be also approximated by a
 counterterm with the form
\begin{eqnarray}
  R_{l \to \infty}: \quad \frac{\slashed{l} \gamma^\mu \slashed{l} }{l^2 (l-r_a)^2 (l+r_b)^2}
\to  - \, \frac{2 l^\mu \slashed{l} }{\left(l^2-M^2\right)^3}
+ \frac{\gamma^\mu}{\left(l^2-M^2\right)^2}\, .
\end{eqnarray}
These two example counterterms both match the UV singularity of the
original  term but differ by finite contributions in the
ultraviolet. However, our choice is constrained to be consistent for
all graphs that enter the Ward identity that guarantees the
factorization of infrared singularities when the second loop momentum,
$k$ in this example, becomes collinear to an initial state quark or antiquark. 

We have already seen that all the diagrams of Fig.\
\ref{fig:two-gluons-ew} satisfy the Ward identities necessary for
factorization in region $(1_k,H_l)$ locally in loop momenta $k$ and
$l$.  If we carry out the expansions that define UV counterterms
consistently, we naturally expect that the sets of integrals that define counterterms satisfy the same Ward identities locally.    Let us see how this comes about for the QCD loops that require renormalization in Fig.\ \ref{fig:two-gluons-ew}.   Again, these are either self-energies or vertex corrections, corresponding to diagrams where the set of momenta $\{q_b\}$ in Fig.\ \ref{fig:two-gluons-ew}a,d is empty, or vertex corrections, also corresponding to no $q_b$ emissions in Fig.\ \ref{fig:two-gluons-ew}b,c.       
 
The self energy diagrams for Fig.\ \ref{fig:two-gluons-ew}d are of the usual form, and we define their integrands by
\bea
\eqs[0.20]{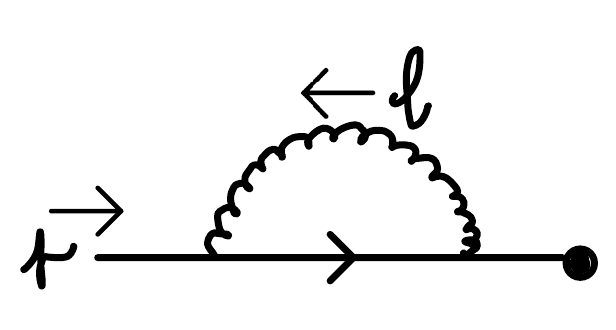} &\equiv& \Pi_{qq}(r, l) 
\nn\\[2mm]
&=&\, - \, C_F \, 
g_s^2\, \frac{1}{l^2 (r+l)^2}\, \gamma^\alpha (\rsla  + \lsla ) \gamma_\alpha\, .
\label{eq:Pi-10d}
\eea
The counterterm integrand corresponding to $\Pi_{qq}(r, l)$ is chosen as
\bea
R_{l \to \infty} \Pi_{qq}(r, l)
&=& g_s^2 \, \, C_F \,  2 \, \left(1-\epsilon \right)
\nonumber \\ && \hspace{-01.5cm}                                    
                                  \times  \Bigg\{
 \frac{2  l \cdot r  \, \slashed{l} }{\left( l^2 - M^2\right)^3}
                                        -\frac{
                                          \slashed{r}+\slashed{l}}{
                                          \left(l^2-M^2 \right)^2}
                                        \Bigg\}
\label{eq:R-Pi-10d}
\eea
where $M$ is an arbitrary mass parameter.

For the self energy diagrams of Fig.\ \ref{fig:two-gluons-ew}a ($b=0$ in ${\cal M}_{2+b}$), we
include the shift subtractions of Fig.\ \ref{fig:planars} with
nonabelian color factors. Shift subtractions add a contribution to the
quark self-energy integrand that is given by 
\bea
&& \Pi_{qq}^{\rm shift}(r, k, l)\ =
\frac{C_A}{2 \, C_F} \left\{ 
  \eqs[0.2]{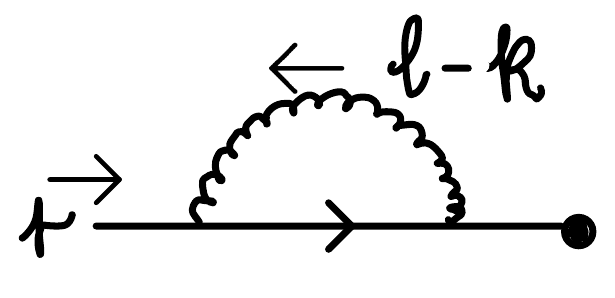} 
  -
   \eqs[0.2]{Pqq.pdf} 
\right\}= - \, \frac{C_A}{2} \, 
 g_s^2\,
\nonumber \\ &&
\hspace{1cm}
\times 
\left[ 
 \frac{1}{(l-k)^2 (r+l-k)^2}\, \gamma^\alpha (\rsla  + \lsla -\slashed{k}) \gamma_\alpha\,
                - \frac{1}{l^2 (r+l)^2}\, \gamma^\alpha (\rsla  + \lsla ) \gamma_\alpha\, 
\right]\, .
\label{eq:PqqShift}
\eea
The above expression is singular in the ultraviolet limit $l \to
\infty$. We subtract the singularity with a counterterm 
\begin{eqnarray}
R_{l \to \infty}  \Pi_{qq}^{\rm shift}(r, k, l)
  &=& g_s^2\,
                                                     C_A 
                                                     \,  
                                                     \left( 1- \epsilon
                                                                                   \right)
                                                                                   \nonumber
  \\ &&  \hspace{-2cm}
        \times \Bigg\{     
                                                   \frac{\slashed{k}
                                                   }{\left( l^2 -M^2\right)^2
                                                   }
                                                     -\frac{4 l \cdot k
                                                     \,
                                                   \slashed{l}}{\left(
                                                     l^2-M^2 \right)^3 }
                                                   \Bigg\}\, .
                                                   \label{eq:R-Pi-prime-10a}
\end{eqnarray}
As we have already remarked, infrared shift counterterms, such as the one of
Eq.~(\ref{eq:PqqShift}), vanish upon integration. This property also holds for the
corresponding ultraviolet counterterm of Eq.~(\ref{eq:R-Pi-prime-10a}),
\begin{equation}
\int d^Dl  \, R_{l \to \infty} \Pi^{\rm shift}_{qq}\left( k, l\right)  =0\, .
\end{equation}  

We will verify that the counterterms defined in Eqs.\
(\ref{eq:R-Pi-10d}) and (\ref{eq:R-Pi-prime-10a}) satisfy the Ward
identity relating the vertex corrections to self energies, locally in
momentum space.   At the integrand level, the relevant vertex
corrections are given by the sum of the QED and QCD vertex diagrams,
\begin{equation}
  \label{eq:GammaQCDQED}
  \Gamma^{\mu, c}_{qqg}\left( r, k, l \right) \equiv
  \eqs[0.2]{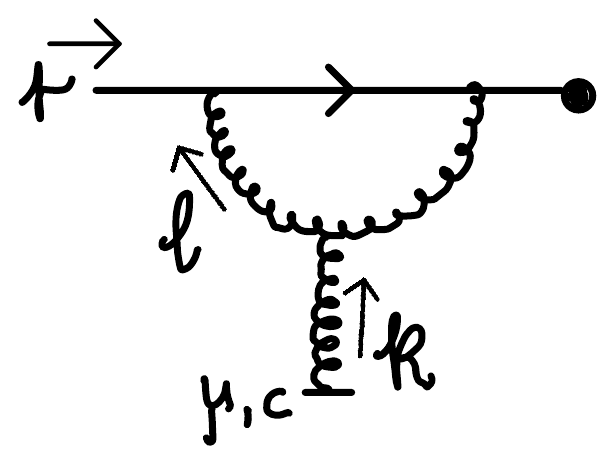}
  + \eqs[0.2]{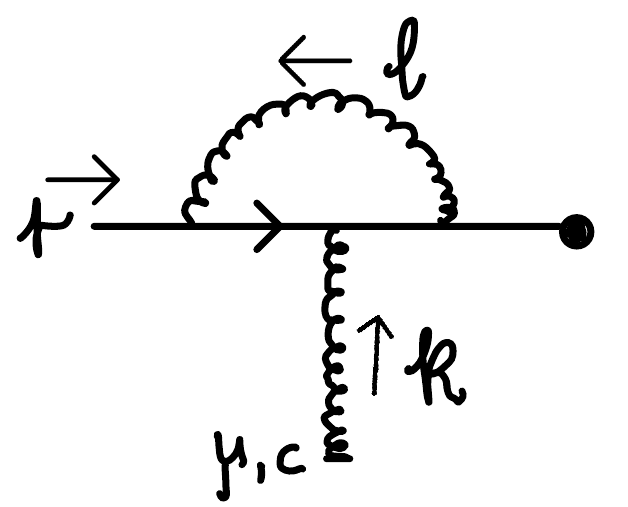}\, .
\end{equation}
The expression corresponding to the right-hand side of the above
equation is obtained with a direct application
of Feynman rules assuming the depicted loop momentum routing.
The diagrams of Eq.~(\ref{eq:GammaQCDQED}) are divergent in the limit $l
\to \infty$. 
For the corresponding counterterm, we pick
\begin{eqnarray}
R_{l \to \infty}   \Gamma^{\mu, c}_{qqg}\left( r, k, l \right)                               =
                                                                               -g_s^3T_c^{(q)}
                                                                               \nonumber
  \\ &&\hspace{-5.0cm} \times \Bigg\{   \left( C_F - C_A \right) \left( 1- \epsilon \right)
    \frac{ 4 \slashed l l^\mu }{\left(
                                                  l^2-M^2 \right)^3}
-
\left[
\epsilon \, C_A + 2 \left(1-\epsilon \right) \, C_F
  \right]
    \frac{\gamma^\mu}{\left( l^2- M^2 \right)^2}
        \Bigg\}\, .
        \label{eq:GammaCT}
\end{eqnarray}
It is now straightforward to verify the 
Ward identity 
\bea
 k_\mu   R_{l \to \infty}   \Gamma^{\mu, c}_{qqg}\left( r, k, l \right)   &=&  g_s \,  T_c^{(q)}
 \, \Bigg[
R_{l \to \infty}   \Pi_{qq}\left( r_a, l\right) - R_{l \to \infty}  \Pi_{qq}\left( r_a+k, l\right)    
                                                          \nonumber \\
                                                          &&
\hspace{-0.7cm}                                                          
-R_{l \to \infty} 
                                                          \Pi_{qq}^{\rm
                                                          shift}\left(r_a,
                                                          k,l \right)
                                                      +g_s^2 \, C_A \,   
    \frac{ \slashed {k} }{\left( l^2- M^2\right)^2} \Bigg].
\label{eq:identity-fig-10}
\eea
On the right hand side, the first, second and fourth terms in square brackets are in the ``standard" form of the vertex
function Ward identity, two self-energy terms and a term associated with ghosts.  
Compared to these standard terms, we have an additional
term, $\Pi_{qq}^{\rm shift}$, which will compensate for the ultraviolet behavior the shift term appropriate for
all the uncrossed ladder diagrams of Fig.\ \ref{fig:two-gluons-ew}a.   The
self energies $\Pi_{qq}\left( r_a, l\right)$ and $ \Pi_{qq}\left( r_a+k, l\right)$, match self energies in Fig.\ \ref{fig:two-gluons-ew}d.   

It is also worth noting that the shift counterterm preserves the Ward identities associated with the electroweak vertices as well.  This is
because both the self energies and the QCD vertex corrections to
electroweak vertices are both QCD uncrossed ladders in Fig.\
\ref{fig:two-gluons-ew}a. In particular, a generic one-loop QCD correction
to an electroweak vertex has the functional form,
\begin{eqnarray}
&& \Gamma_{qq \boxed{\bullet}}(r_a, r_b, l) \equiv \eqs[0.15]{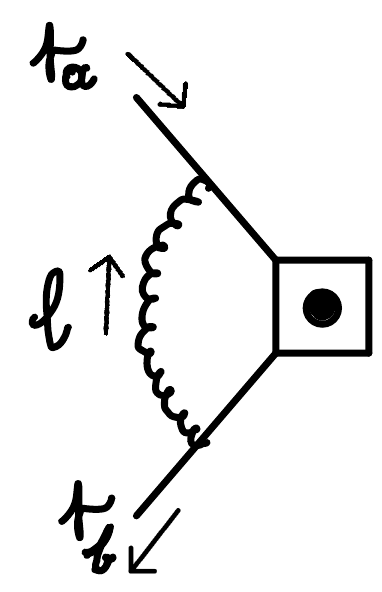} = -i g^2 C_F \frac{\gamma^\nu \left(
     \slashed{l}+\slashed{r}_a \right)
                           \, \boxed{\bullet} \, \left(
                             \slashed{l}+\slashed{r}_b \right)
                           \gamma_\nu }{l^2 \, \left(l+r_a\right)^2 \,
                                 \left(l-q_r\right)^2  }\, .
\end{eqnarray}
In the above, the tree-level electroweak vertex is 
  denoted with the symbol $\boxed{\bullet}$. To cancel the ultraviolet
  singularity we pick the following counterterm, 
 \begin{eqnarray}
  R_{l \to \infty} \Gamma_{qq \boxed{\bullet}}(r_a, r_b, l) &=& i g_s^2 C_F \frac{\gamma^\nu \slashed{l}
                           \, \boxed{\bullet} \, \slashed{l}  \gamma_\nu }{\left(l^2-M^2\right)^3}\, .
 \end{eqnarray}

We will now focus on one-loop corrections for QCD vertices and
propagators in S-type and V-type diagrams, which we treated with
special rules in order to ensure factorization of collinear
singularities.   Correspondingly, special counterterms specific to these
diagrams are required for the cancellation of ultraviolet
singularities.

In Eq.~(\ref{eq:Jbar})  we have given an alternative, yet equivalent upon integration, 
form for one-loop corrections  of vertices adjacent to an incoming leg,
compared to the expression described by Eq.~(\ref{eq:GammaQCDQED}).  
To account for this difference in the
ultraviolet, in addition to applying the counterterm of
Eq.~(\ref{eq:GammaCT}), we also renormalize the expressions of
Eq.~(\ref{eq:FRuleLPquark}) and Eq.~(\ref{eq:FRuleLPquarkTrans}), 
\begin{eqnarray}
  R_{l \to \infty} \Delta_{1} {\cal J}_c^\mu(k,l) &\equiv&
  g_s^3 \, T_c^{(q)} \, \left( C_F-C_A \right) \,
  2 \left( 1-\epsilon \right) \frac{\slashed{\eta}_1}{2 p_1
  \cdot \eta_1} \nonumber \\&& \hspace{1cm} \times \, 
  \left[
    -\frac{p_1^\mu}{\left( l^2-M^2\right)^2}
+\frac{4 l \cdot p_1 \, l^\mu}{\left( l^2-M^2\right)^3}
  \right]\, ,
\end{eqnarray}
\begin{equation}
 R_{l \to \infty} \Delta_{1\perp} {\cal J}_c^\mu(k,l) \equiv
  g_s^3 \, T_c^{(q)} \, \left( C_F-C_A \right) \,  
2 \left( 1-\epsilon \right) \frac{\slashed{l}_\perp \left( l^\mu + {\tilde l}^\mu \right) }{\left( l^2-M^2\right)^3}\, .
\end{equation}
Upon integration over the $l$ momentum, the above counterterms, as 
expected, vanish
\begin{equation}
  \int d^d l \, R_{l \to \infty} \Delta_{1} {\cal J}_c^\mu(k,l) =0, \quad
  \int d^d l \, R_{l \to \infty} \Delta_{1\perp} {\cal J}_c^\mu(k,l) =0\, .
\end{equation}

For S-type diagrams, self-energy corrections on quark propagators
adjacent to an incoming quark or antiquark are evaluated according to
Eq.~(\ref{eq:Nsquark})  and Eq.~(\ref{eq:Nsantiquark}). To remove the
ultraviolet singularity in $N_{S-q}(k, l)$ and $N_{S-{\bar q}}(k, l)$
we add a counterterm which reads
 \begin{equation}
  R_{l \to \infty} N_{S-q}(k,l) =R_{l \to \infty}  N_{S-{\bar q}}(k,l)
   \equiv  - \left(1-\epsilon\right)
   \frac 1 {\left( l^2 - M^2 \right)^2}\, .
\end{equation}
This counterterm  differs locally from the self-energy counterterm of
Eq.~(\ref{eq:R-Pi-10d}) used for  internal
lines.  However, the two counterterms are equivalent after
integration of the loop-momentum $l$. Indeed, with a simple tensor
reduction to master integrals we can show  that
\begin{equation}
 \int d^dl  \, R_{l \to \infty} \Pi_{qq}(r, l) = \slashed{ r} \int d^d
 l \,  R_{l \to \infty} N_{S-q}(k, l)\, .
\end{equation}

We now have all the ingredients to check that our ultraviolet
counterterms respect the Ward identity that cancels the singularity
in the ${\rm \left( H_l, 2_k\right)}$ region. We find, 
\begin{eqnarray}
  &&R_{l \to \infty} \left[
  {\cal J}^\mu_{c,{\rm canonical}}(k,l)
  +{\tilde \Delta}_{1} {\cal J}_c^\mu(k,l)
     \right] \, u(p_1)
\nonumber \\ &=&
     R_{l \to \infty} \left[
  {\cal J}^\mu_{c,{\rm canonical}}(k,l)
  +\Delta_{1} {\cal J}_c^\mu(k,l)
+\Delta_{1\perp} {\cal J}_c^\mu(k,l)
  \right] \, u(p_1)    
  \nonumber \\
  &=& \quad
      g_s^3 \, T_c^{(q)} \,\Bigg\{
      \left( C_F-C_A \right) \,
   \left( 1-\epsilon \right)  \, \left[
 \frac{\gamma^\mu_{\perp\left(p_1, p_2 \right)}}{\left(l^2
     -M^2 \right)^2}
     -\frac{
 4 \slashed{l}_{\perp\left( p_1, p_2\right)} l_{\perp\left( p_1, p_2\right)}^\mu
     }{\left(l^2
     -M^2 \right)^3}
      \right]
      \nonumber  \\ && \hspace{2.5cm} 
      + \, C_A \,\frac{\gamma^\mu }{\left( l^2 -M^2 \right)^2}
\Bigg\} \, 
      u(p_1)\, .
\end{eqnarray}
Upon contracting with $k_\mu$, as occurs in the $k \parallel p_2$
collinear limit,  we see that the terms in the square bracket above vanish due to $p_2 \cdot
\gamma_{\perp\left( p_1, p_2\right)} =p_2 \cdot
l_{\perp\left( p_1, p_2\right)} = 0$.    The last term,
proportional to $C_A \, \gamma^\mu$, is associated with ghosts. When
combined with the analogous terms of Eq.~(\ref{eq:identity-fig-10}), 
from one-loop vertex corrections that are not adjacent to the
incoming legs, it factorizes. 

In summary, we have constructed one-loop $R_{l \to \infty }$ ultraviolet counterterms that respect
the same Ward identities locally as the amplitude that we constructed
in the previous sections.  Combined with two-loop $R_{k,l \to \infty
}$ counterterms, we are now in a position to subtract
all singularities, ultraviolet and infrared, simultaneously, using the
scheme of Eq.~(\ref{eq:cal-M-finite-12}). In the next section, we
detail the steps that we follow for an example two-loop amplitude.


\section{Numerical check}
\label{sec:check}

In the previous sections,  we presented a systematic method to remove
the ultraviolet and infrared singularities of amplitudes for
generic electroweak production through two loops. To check our method,
we apply it  to the $q(p_1) +  \bar q(p_2) \to \gamma^*(q_1) + \gamma^*(q_2)$  QCD
amplitude. 

We first generate~\cite{Nogueira:1991ex} the integrand for the Feynman diagrams, applying Feynman
rules in the Feynman gauge  and assigning appropriate momentum flows
following the rules as in Fig.~\ref{fig:loopmomenta}. The tree amplitude is 
\begin{equation}
{\cal M}_{\gamma^* \gamma^*}^{\left( 0 \right)} =
\eqs[0.15]{./Diags/A1} + \left( \gamma^*(q_1)
  \leftrightarrow       \gamma^*(q_2) \right) \, ,
\end{equation}
where the second contributing diagram not explicitly shown is
obtained by Bose symmetry, exchanging the momenta and polarizations of
the external photons.
The one-loop amplitude integrand is given by
\begin{equation}
  {\cal M}_{\gamma^* \gamma^*}^{\left( 1 \right)} (k)= \sum_{i=1}^4 {\cal
    D}_i^{\left( 1 \right)}\left( k \right) 
+ \left( \gamma^*(q_1)
  \leftrightarrow       \gamma^*(q_2) \right) \, ,
\end{equation}
 \begin{table}[ht]
  \begin{center}
\begin{tabular}{|c|c|c|c|}
  \hline
  &&& \\
  $\eqs[0.20]{./Diags/B1}$ & $\eqs[0.20]{./Diags/B2}$ &
                                                        $\eqs[0.20]{./Diags/B3}$
                           & $\eqs[0.20]{./Diags/B4}$
   \\
   ${\cal D}^{(1)}_1$ &${\cal D}^{(1)}_2$&${\cal D}^{(1)}_3$&${\cal D}^{(1)}_4$ \\ 
  \hline
\end{tabular}
\end{center}
\caption{
\label{tab:Ampl1}
  One-loop diagrams for $q \bar q \to \gamma^* \gamma^*$}
\end{table}
where the integrands ${\cal
  D}_i^{\left( 1 \right)}\left( k \right)$
of the one-loop Feynman diagrams are derived by a
direct application of Feynman rules in Feynman gauge on the graphs of
Table~\ref{tab:Ampl1}. In all graphs the momentum $k$
of the gluon flows out of the quark-gluon vertex which is nearest
within the fermion line to the antiquark $\bar q(p_2)$. 

The part of the two-loop amplitude integrand which is discussed in
this publication is constructed as
\begin{eqnarray}
  \label{eq:Ampl2}
  {\cal M}_{\gamma^* \gamma^*}^{\left( 2 \right)} \left(k,l \right)
  &=&
 \sum_{i=1}^{4} {\overline{{\cal D}}}^{(2)}_{i}\left(k, l\right)                       
                                                          +  \sum_{i=5}^{38}
                                                           {\cal
                                                            D}_i^{\left(
                                                            2
                                                            \right)}\left(
                                                            k, l
                                                            \right)
                                                            \nonumber  \\ && \hspace{-2cm}
                                                            +\sum_{i=1}^{4} {\cal D}^{(2)}_{{\rm LP},i}\left(k, l\right)
                                                           +  \frac{C_A}{2 C_F} \, \sum_{i=13}^{20}
  \left[
    {\cal D}_i^{\left( 2 \right)}\left(  k, l-k \right)
    -{\cal D}_i^{\left( 2 \right)}\left( k , l\right)
                                                                             \right]
            \nonumber \\ && 
                                   \nonumber \\ &&
     \hspace{1cm}
     + \left( \gamma^*(q_1)
  \leftrightarrow       \gamma^*(q_2) \right) \, .
\end{eqnarray}
Equation (\ref{eq:Ampl2}) does not include two-loop diagrams with vacuum
polarization corrections to a gluon propagator or two-loop diagrams
with fermion loops. Up to straightforward multiplications with colour
factors, these diagrams also appear in the QED process of $e^+ e^- \to
\gamma^* \gamma^*$. We can subtract their
singularities locally with the procedures developed in
Section 4 of Ref.\ \cite{Anastasiou:2020sdt} and we will not discuss
them in this publication any further.  
  
\begin{table}[ht]
  \begin{center}
\begin{tabular}{|c|c|c|c|}
  \hline
  &&& \\
  $\eqs[0.20]{./Diags/C1}$ & $\eqs[0.20]{./Diags/C2}$ &
                                                        $\eqs[0.20]{./Diags/C3}$
                           & $\eqs[0.20]{./Diags/C4}$
   \\
   ${\cal D}^{(2)}_1$ &${\cal D}^{(2)}_2$&${\cal D}^{(2)}_3$&${\cal D}^{(2)}_4$ \\ 
  \hline
\end{tabular}
\end{center}
\caption{\label{tab:A2Stype} Two-loop S-type diagrams for $q \bar q \to \gamma^* \gamma^*$}
\end{table}
\begin{table}[ht]
  \begin{center}
\begin{tabular}{|c|c|c|c|}
  \hline
  &&& \\
  $\eqs[0.20]{./Diags/C5}$ & $\eqs[0.20]{./Diags/C6}$ &
                                                        $\eqs[0.20]{./Diags/C7}$
                           & $\eqs[0.20]{./Diags/C8}$
   \\
   ${\cal D}^{(2)}_5$ &${\cal D}^{(2)}_6$&${\cal D}^{(2)}_7$&${\cal D}^{(2)}_8$ \\ 
  \hline
  &&& \\
  $\eqs[0.20]{./Diags/C9}$ & $\eqs[0.20]{./Diags/C10}$ &
                                                        $\eqs[0.20]{./Diags/C11}$
                           & $\eqs[0.20]{./Diags/C12}$
   \\
   ${\cal D}^{(2)}_9$ &${\cal D}^{(2)}_{10}$&${\cal D}^{(2)}_{11}$&${\cal D}^{(2)}_{12}$ \\ 
  \hline
\end{tabular}
\end{center}
\caption{\label{tab:A2Vtype} Two-loop V-type diagrams for $q \bar q \to \gamma^* \gamma^*$}
\end{table}
\begin{table}[ht]
  \begin{center}
\begin{tabular}{|c|c|c|c|}
  \hline
  &&& \\
  $\eqs[0.20]{./Diags/C13}$ & $\eqs[0.20]{./Diags/C14}$ &
                                                        $\eqs[0.20]{./Diags/C15}$
                           & $\eqs[0.20]{./Diags/C16}$
   \\
   ${\cal D}^{(2)}_{13}$ &${\cal D}^{(2)}_{14}$&${\cal D}^{(2)}_{15}$&${\cal D}^{(2)}_{16}$ \\ 
  \hline
  &&& \\
  $\eqs[0.20]{./Diags/C17}$ & $\eqs[0.20]{./Diags/C18}$ &
                                                        $\eqs[0.20]{./Diags/C19}$
                           & $\eqs[0.20]{./Diags/C20}$
   \\
   ${\cal D}^{(2)}_{17}$ &${\cal D}^{(2)}_{18}$&${\cal D}^{(2)}_{19}$&${\cal D}^{(2)}_{20}$ \\ 
  \hline
\end{tabular}
\end{center}
\caption{\label{tab:A2Planar} Two-loop ``abelian'' planar diagrams (excluding S-type) for $q \bar q \to
  \gamma^* \gamma^*$ with a collinear singularity in $k \parallel p_1$
or $k \parallel p_2$}
\end{table}
\begin{table}[ht]
  \begin{center}
\begin{tabular}{|c|c|c|c|}
  \hline
  &&& \\
  $\eqs[0.15]{./Diags/C21}$ & $\eqs[0.15]{./Diags/C22}$ &
                                                        $\eqs[0.15]{./Diags/C23}$
                           & $\eqs[0.15]{./Diags/C24}$
   \\
   ${\cal D}^{(2)}_{21}$ &${\cal D}^{(2)}_{22}$&${\cal D}^{(2)}_{23}$&${\cal D}^{(2)}_{24}$ \\ 
  \hline
  &&& \\
  $\eqs[0.15]{./Diags/C25}$ & $\eqs[0.15]{./Diags/C26}$ &
                                                        $\eqs[0.15]{./Diags/C27}$
                           & $\eqs[0.15]{./Diags/C28}$
   \\
   ${\cal D}^{(2)}_{25}$ &${\cal D}^{(2)}_{26}$&${\cal D}^{(2)}_{27}$&${\cal D}^{(2)}_{28}$ \\ 
  \hline
    &&& \\
  $\eqs[0.15]{./Diags/C29}$ & $\eqs[0.15]{./Diags/C30}$ &
                                                        $\eqs[0.15]{./Diags/C31}$
                           & $\eqs[0.15]{./Diags/C32}$
   \\
   ${\cal D}^{(2)}_{29}$ &${\cal D}^{(2)}_{30}$&${\cal D}^{(2)}_{31}$&${\cal D}^{(2)}_{32}$ \\ 
  \hline
      &&& \\
  $\eqs[0.15]{./Diags/C33}$ & $\eqs[0.15]{./Diags/C34}$ &
                                                        $\eqs[0.15]{./Diags/C35}$
                           & $\eqs[0.15]{./Diags/C36}$
   \\
   ${\cal D}^{(2)}_{33}$ &${\cal D}^{(2)}_{34}$&${\cal D}^{(2)}_{35}$&${\cal D}^{(2)}_{36}$ \\ 
  \hline
      &&& \\
   & $\eqs[0.15]{./Diags/C37}$ &
                                                        $\eqs[0.15]{./Diags/C38}$
                           &
   \\
          &${\cal D}^{(2)}_{37}$&${\cal D}^{(2)}_{38}$& \\ 
  \hline
\end{tabular}
\end{center}
\caption{\label{tab:A2Rest} The remaining two-loop diagrams  for $q \bar q \to
  \gamma^* \gamma^*$ ( excluding the ones with fermion loops and vacuum
polarization corrections to the gluon propagator which have been
treated in Ref.~\cite{Anastasiou:2020sdt}).  }
\end{table}
\begin{table}
  \begin{center}
\begin{tabular}{|c|c|c|c|}
  \hline
  &&& \\
  $\eqs[0.22]{./Diags/Clp1}$ & $\eqs[0.22]{./Diags/Clp2}$ &
                                                        $\eqs[0.22]{./Diags/Clp3}$
                           & $\eqs[0.22]{./Diags/Clp4}$
   \\
   ${\cal D}^{(2)}_{{\rm LP}, 1}$ &${\cal D}^{(2)}_{{\rm LP}, 2}$&${\cal D}^{(2)}_{{\rm LP}, 3}$&${\cal D}^{(2)}_{{\rm LP}, 4}$ \\ 
  \hline
\end{tabular}
\end{center}
\caption{\label{tab:A2LPs} Two-loop diagrams of null integrated value
  that eliminate loop polarizations in the  $q \bar q \to \gamma^*
  \gamma^*$ amplitude integrand.}
\end{table}
The integrands of the
Feynman diagrams ${\cal D}_i^{\left( 2 \right)}(k, l)$ and $\overline{{\cal D}}_i^{\left( 2 \right)}(k, l)$ which are
included  in Eq.~(\ref{eq:Ampl2}) are derived from the graphs depicted
in Tables~\ref{tab:A2Stype},\ref{tab:A2Vtype},\ref{tab:A2Planar},\ref{tab:A2Rest}. 
We now describe the explicit steps we take for the construction of
each term  in Eq.~(\ref{eq:Ampl2}).
\begin{enumerate}
\item We assign loop momentum flows to all diagrams
$D_i^{(2)}\left(k,l\right)$ as depicted in Fig.~\ref{fig:loopmomenta}, according to the following rules. 
\begin{itemize}
  \item In diagrams with a triple-gluon vertex $\left( {\cal
      D}^{(2)}_{9}-{\cal D}^{(2)}_{12}, {\cal D}^{(2)}_{21}, {\cal D}^{(2)}_{23},{\cal
D}^{(2)}_{32}, {\cal D}^{(2)}_{34}\right)$, we assign momentum
    labels $k, l-k, -l$ to the three gluons,  in the order that they get
    attached to the fermion line starting from the external antiquark
    ${\bar q}(p_2)$ and with an outgoing direction to the
    corresponding quark gluon vertices. 
    \item In the remaining diagrams, without a triple-gluon vertex, we assign
      momentum labels $k,l$ to the two virtual gluons, in the order that they get
    attached to the fermion line starting from the external antiquark
    ${\bar q}(p_2)$ and with an outgoing direction to the
    corresponding quark gluon vertices.
  \end{itemize}
   \item Diagrams $D_1^{(2)}\left(k, l\right), \ldots,
    D_4^{(2)}\left(k, l\right) $ (in Table~\ref{tab:A2Stype}) belong
    to the S type. We derive a dual integrand  $\overline{{\cal
        D}}^{(2)}_i$ for them by treating the quark
    self-energy according to the rules of
    Eqs~(\ref{eq:Nsquark}),(\ref{eq:Nsantiquark}) and the remaining parts
    of these graphs with standard Feynman rules.
\item     
     Feynman diagrams  $D_5^{(2)}\left(k, l\right), \ldots,
    D_{36}^{(2)}\left(k, l\right) $  are computed  with only a
    conventional application of Feynman rules (always in the Feynman gauge).   
  \item As we have discussed in Section~\ref{sec:VandS},  diagrams $D_5^{(2)}\left(k, l\right), \ldots,
    D_{12}^{(2)}\left(k, l\right) $ (in Table~\ref{tab:A2Vtype}) belong
    to the V type and give rise to  loop polarizations.  We add
    to the right-hand side of Eq.~\ref{eq:Ampl2} four diagrams ${\cal
      D}_{{\rm LP}, i}^{(2)}(k, l)$, depicted in
    Table~\ref{tab:A2LPs}, which are engineered to eliminate loop
    polarizations.     The exotic vertices correspond to the currents
    introduced in Section~\ref{sec:VandS}, which are $\Delta_1$ of
    Eq.~(\ref{eq:FRuleLPquark}) in ${\cal
      D}_{{\rm LP}, 1}^{(2)}(k, l)$, $\Delta_2$ of
    Eq.~(\ref{eq:FRuleLPqbar}) in ${\cal
      D}_{{\rm LP}, 3}^{(2)}(k, l)$,
    ${\tilde \Delta}_1$ of
    Eq.~(\ref{eq:FRuleLPquarkComb}) in ${\cal
      D}_{{\rm LP}, 2}^{(2)}(k, l)$ and ${\tilde \Delta}_2$ of
    Eq.~(\ref{eq:FRuleLPqbarComb}) in ${\cal
      D}_{{\rm LP}, 4}^{(2)}(k, l)$.
    All other parts of these graphs are computed with conventional
    Feynman rules.  We emphasize that the novel diagrams,  ${\cal
      D}_{{\rm LP}, i}^{(2)}(k, l)$, are important locally to
    guarantee factorization of collinear singularities, but they do
    not change the value of the amplitude as they integrate to zero.
    
    \item In the last term of the second line of Eq.~(\ref{eq:Ampl2}) we use the
      integrands of planar diagrams with collinear singularities in
      Table~\ref{tab:A2Planar} to construct shift counterterms as
      described in Section~\ref{sec:construction}. 
    \end{enumerate}
\begin{table}[ht]
  \begin{center}
\begin{tabular}{|c|c|c|c|}
  \hline
  &&& \\
    $\eqs[0.18]{./Diags/FA1}$&
    $\eqs[0.18]{./Diags/FB1}$&
    $\eqs[0.18]{./Diags/FC1}$&
    $\eqs[0.18]{./Diags/FC2}$
  \\  
  ${\cal F}^{(0)}_{1}$&
  ${\cal F}^{(1)}_{1}$&
  ${\cal F}^{(2)}_{1}$&
  ${\cal F}^{(2)}_{3}$
  \\
  \hline
&&& \\
    $\eqs[0.18]{./Diags/FC3}$&
    $\eqs[0.18]{./Diags/FC4}$&
    $\eqs[0.18]{./Diags/FC5}$&
    $\eqs[0.18]{./Diags/FC6}$
  \\  
  ${\cal F}^{(2)}_{3}$&
  ${\cal F}^{(2)}_{4}$&
  ${\cal F}^{(2)}_{5}$&
  ${\cal F}^{(2)}_{6}$
  \\
  \hline
  &&& \\
    $\eqs[0.18]{./Diags/FC7}$&
    $\eqs[0.18]{./Diags/FC8}$&
    $\eqs[0.18]{./Diags/FClp1}$&
    $\eqs[0.18]{./Diags/FClp2}$
  \\  
  ${\cal F}^{(2)}_{7}$&
  ${\cal F}^{(2)}_{8}$&
  ${\cal F}^{(2)}_{{\rm LP}, 1}$&
  ${\cal F}^{(2)}_{{\rm LP}, 2}$
\\
                                  \hline
\end{tabular}
\end{center}
\caption{\label{tab:FF} Diagrams for the form factor
  amplitude $q(p_1) + \bar q(p_2) \to {\cal H}$ through two loops.}
\end{table}

We now remove the ultraviolet
divergences of the one and two loop amplitudes at the integrand level, applying the
formulas of Eq.~(\ref{eq:Ampl1R}) and Eq.~(\ref{eq:Ampl2R}) in
Section~\ref{sec:UV}. This yields an ultraviolet finite one and
two-loop amplitude,
\begin{equation}
  {\cal M}_{\gamma^* \gamma^*}^{\left( 1 \right)} \left(k,l \right)
  \to   {\cal M}_{\gamma^* \gamma^*}^{\left( 1 \right),  {\rm R}} \left(k,l \right)
\; , 
  \label{eq:Ampl1Rdiphoton}
\end{equation}
\begin{equation}
  {\cal M}_{\gamma^* \gamma^*}^{\left( 2 \right)} \left(k,l \right)
  \to   {\cal M}_{\gamma^* \gamma^*}^{\left( 2
    \right),  {\rm R}} \left(k,l \right)
\; . 
  \label{eq:Ampl2Rdiphoton}
\end{equation}
Eqs~(\ref{eq:Ampl1R})  and Eqs~(\ref{eq:Ampl2R}) can be implemented by
introducing counterterm Feynman rules, as one often does in UV
renormalization of integrated amplitudes.  Alternatively, one could
implement an algorithm that applies to each graph and subgraph,
identifying the singular limit and regulating it  with the counterterms
described in Sec.~\ref{sec:UV}. In the numerical check of this
section, we have chosen the latter approach.  

The integrands constructed in Eq.~(\ref{eq:Ampl1Rdiphoton}) and
Eq.~(\ref{eq:Ampl2Rdiphoton}) are still singular in infrared limits. However, in their
 construction, we have achieved that all collinear singularities
factorize after we symmetrize over the loop momenta $k \leftrightarrow
l$ and sum over diagrams. We remove the remaining infrared singularities with counterterms derived from
a form-factor amplitude for a generic $2 \to 1$ process.  Through
two loops, the form factor integrand is
generated from the diagrams of Table~\ref{tab:FF} as
\begin{eqnarray}
  \label{eq:Ftree}
  {\cal F}^{(0)}\left[ {\cal H}\right] &=& {\cal F}^{(0)}_1 \left[
                                           {\cal H}\right] \, , \\
  \label{eq:F1}
  {\cal F}^{(1)}\left[ {\cal H}\right]\left( k\right) &=& {\cal
                                                          F}^{(1)}_1
                                                          \left[ {\cal
                                                          H}\right]\left(
                                                          k\right) \, ,
                                                          \\
  \label{eq:F2}
  {\cal F}^{(2)}\left[ {\cal H}\right]\left( k, l \right) &=&
                                                              \sum_{i=1}^2
                                                             {\overline{ {\cal
                                                              F}}}^{(2)}_i
                                                              \left[
                                                              {\cal
                                                              H}\right]\left(
                                                              k,
                                                              l\right)
                                                              + \sum_{i=3}^8
                                                             {{ {\cal
                                                              F}}}^{(2)}_i
                                                              \left[
                                                              {\cal
                                                              H}\right]\left(
                                                              k,
                                                              l\right)\, ,
                                                              \nonumber
  \\
  && \hspace{-2cm}  + \sum_{i=1}^2
                                                             {{ {\cal
                                                              F}}}^{(2)}_{{\rm
     LP}, i}
                                                              \left[
                                                              {\cal
                                                              H}\right]\left(
                                                              k,
                                                              l\right)
     + \frac{C_A}{2 C_F} \left[
 {{ {\cal
                                                              F}}}^{(2)}_{5}
                                                              \left[
                                                              {\cal
                                                              H}\right]\left(
                                                              k,
     l-k \right)
     -
     {{ {\cal
                                                              F}}}^{(2)}_{5}
                                                              \left[
                                                              {\cal
                                                              H}\right]\left(
                                                              k,
     l \right)
     \right]\, .
\end{eqnarray} 
The tree, one-loop and two-loop form factor integrands above are
generated for a generic vertex ${\cal H}$, which is color-diagonal and
takes the form of an arbitrary matrix in spinor space. The two-loop
integrand ${\cal F}^{(2)}\left[ {\cal H}\right]\left( k, l \right)$
is constructed applying identical rules for treating S type diagrams
$\overline{\cal F}^{(2)}_1, \overline{\cal F}^{(2)}_2 $, cancelling
loop polarizations with $\overline{\cal F}^{(2)}_{{\rm LP}, 1},
\overline{\cal F}^{(2)}_{{\rm LP}, 2} $  and shift counterterms, all of which
we have included  in the construction of the right-hand side of
Eq.~(\ref{eq:Ampl2}). We note that the second line
of Eq.~(\ref{eq:F2}) integrates to zero. The integration of the first
line in Eq.~(\ref{eq:F2})  can be performed with standard methods and
relevant results are available to even higher orders than the
two loops~\cite{Moch:2005tm,Chakraborty:2022yan,Lee:2021lkc,Blumlein:2019oas,vonManteuffel:2019gpr,Ahmed:2016qjf,Ahmed:2016wiv,vonManteuffel:2015gxa,Ahmed:2015qpa,Ahmed:2015qia}
that we require here.    To remove ultraviolet singularities from the
one and two-loop form factor amplitudes, we construct 
${\cal F}^{\left( 1\right) , {\rm R}}\left[ {\cal H} \right](k)$
and ${\cal F}^{\left( 2\right),  {\rm R}}\left[ {\cal H} \right](k,l)$ using,
again,  the subtraction of Eq.~(\ref{eq:Ampl1R}) and Eq.~(\ref{eq:Ampl2R}).  

Before we present an expression for the infrared finite remainder of
the ${\cal M}_{\gamma^* \gamma^*}$ amplitude, we introduce a class of objects and
a corresponding notation that are required for the construction of the
spin-matrix ${\cal H}$.  Following the notation of Eq.\ (\ref{eq:cal-M-finite-12}), every diagram ${\cal D} \in {\cal D}^{(j)}_i,
{\cal F}^{(j)}_i $ takes the form
\begin{equation}
{\cal D}  = \bar v\left( p_2\right) \; \widetilde {\cal D} \; u(p_1).
\end{equation}
We define a dual matrix in spin-space of a diagram ${\cal D}$ (or sum
of diagrams) 
by replacing the spinor $\bar{v}(p_2)$ and $u(p_1)$ factors with the projector of Eq.~(\ref{eq:projector}),
\begin{equation}
  {\cal D} \to {\cal S} \left( {\cal D}\right) \equiv   \mathbf P_1
\; \widetilde {\cal D} \;
\mathbf P_1.
\end{equation}
We remark that if we insert the matrix ${\cal S}\left( \cal D \right)$
as a vertex into the zeroth-order form factor, we obtain back ${\cal D}$,
\begin{equation}
 {\cal S} \left( {\cal D}\right) \to  {\cal D} = {\cal
   F}^{\left(0\right)}\left[
{\cal S} \left( {\cal D}\right)
   \right].
 \end{equation}
 We now have all ingredients for constructing integrands for the
 one and two loop amplitudes that are free of all infrared
 singularities. This is a realization of Eq.\ (\ref{eq:cal-M-finite-12}).  We have
 \begin{equation}
   \label{eq:Hard1}
  {\cal H}_{\gamma^* \gamma^*}^{(1)}\left( k \right) = 
   {\cal M}_{\gamma^* \gamma^*}^{(1) , {\rm R}}\left( k \right) 
- {\cal F}^{(1) , {\rm R}}\left[ {\cal S} \left( {\cal M}_{\gamma^* \gamma^*}^{(0)} \right)  \right]\left( k \right)\, .
 \end{equation}
 and
 \begin{eqnarray}
   \label{eq:Hard2}
 && 2 {\cal H}_{\gamma^* \gamma^*}^{(2)}\left( k,l \right) =
    \nonumber \\
   && \hspace{-1cm}
   {\cal M}_{\gamma^* \gamma^*}^{(2),  {\rm R}}\left( k, l \right)
-  {\cal F}^{(2) , {\rm R}}\left[ {\cal S} \left( {\cal M}_{\gamma^*
                                                            \gamma^*}^{(0)} \right)  \right]\left( k, l \right)
-  {\cal F}^{(1) , {\rm R}}\left[ {\cal S} \left( {\cal H}_{\gamma^*
      \gamma^*}^{(1)}(k) \right)  \right]\left( l \right)
                                                            \nonumber \\              
&& \hspace{1cm} + \left( k \leftrightarrow l \right) \, .
 \end{eqnarray}

We now check explicitly with a semi-numerical calculation that our
construction in Eq.~(\ref{eq:Hard1}) and Eq.~(\ref{eq:Hard2}) leads to
expressions in their right-hand sides that are free of infrared
singularities and ultraviolet singularities.  Our check is carried out with the same method
as in the analogous numerical study of the
abelian contribution to the amplitude of Ref.~\cite{Anastasiou:2020sdt}.    
At one loop, our construction of Eq.~(\ref{eq:Hard1}) is almost identical, up to colour
factors and the functional form of the chosen ultraviolet
counterterms for the one-loop quark propagator and one-loop quark
gluon vertex, to the analogous QED process of
Ref.~\cite{Anastasiou:2020sdt}, and we readily reproduce that the
one-loop   remainder ${\cal H}_{\gamma^* \gamma^*}^{(1)}\left( k
\right)$ is finite in all soft, collinear and ultraviolet limits.
The construction of ${\cal H}_{\gamma^* \gamma^*}^{(2)}\left( k
\right)$ has required several novel ingredients with respect to the QED process
of  Ref.~\cite{Anastasiou:2020sdt} that we test and verify with the calculations of this section. 

For ease of comparison, we fix external momenta, spinors
(in the Weyl representation) and polarization
vectors to the same values to the ones of 
Ref.~\cite{Anastasiou:2020sdt}. These values are
\begin{equation}
  p_1 = 
\left(\begin{array}{c}
1 
\\
 0 
\\
 0 
\\
 1 
      \end{array}\right), \,
    p_2 =
    \left(\begin{array}{c}
1 
\\
 0 
\\
 0 
\\
 -1 
          \end{array}\right), \,
        q_1 =
        \left(\begin{array}{c}
1 
\\
 0 
\\
 {1}/{3} 
\\
 {1}/{7} 
              \end{array}\right), \,
            q_2 =
            \left(\begin{array}{c}
1 
\\
 0 
\\
 -{1}/{3} 
\\
 -{1}/{7} 
\end{array}\right)\, ,
\end{equation}
and
\begin{equation}
\epsilon(q_1) = \left(\begin{array}{c}
4 
\\
 -4 
\\
 9 
\\
 7 
\end{array}\right), \, 
\epsilon(q_2) =
\left(\begin{array}{c}
2 
\\
 -1 
\\
 -3 
\\
 -7 
      \end{array}\right), \,
u(p_1) = 
\left[\begin{array}{c}
0 
\\
 -7 
\\
 3 
\\
 0 
\end{array}\right], \, 
\overline{v}(p_2)^T = \left[\begin{array}{c}
0 
\\
 13 
\\
 -9 
\\
 0 
\end{array}\right]\, .
\end{equation}
We also construct the loop momenta from  vectors $K_+, L_+$ parallel to  $p_1$,
vectors $K_-, L_-$ parallel to  $p_2$ and vectors $K_\perp, L_\perp$
perpendicular to both  $p_1, p_2$, with values   
\begin{equation}
K_+ = \left(\begin{array}{c}
\frac{33}{17} 
\\
 0 
\\
 0 
\\
 \frac{33}{17} 
\end{array}\right), \, 
 K_- = \left(\begin{array}{c}
-\frac{48}{19} 
\\
 0 
\\
 0 
\\
 \frac{48}{19} 
             \end{array}\right), \,
           K_\perp =\left(\begin{array}{c}
0 
\\
 \frac{21}{23} 
\\
 \frac{21}{41} 
\\
 0 
\end{array}\right)  \, ,         
\end{equation}
\begin{equation}
L_+ =\left(\begin{array}{c}
\frac{47}{23} 
\\
 0 
\\
 0 
\\
 \frac{47}{23} 
\end{array}\right), \, 
 L_- =\left(\begin{array}{c}
-\frac{7}{61} 
\\
 0 
\\
 0 
\\
 \frac{7}{61} 
\end{array}\right), \,
           L_\perp =  \left(\begin{array}{c}
0 
\\
 -\frac{37}{73} 
\\
 -\frac{39}{67} 
\\
 0 
\end{array}\right)\, .
\end{equation}
We evaluate the IR- and UV-subtracted finite two-loop amplitude
integrand ${\cal H}_2(k,
l)$ for loop momentum values
\begin{equation}
 \label{eq:takelimits}
 k = \delta^{n_+} K_+ +\delta^{n_-} K_- + \delta^{n_\perp} K_\perp,
 \quad 
 l = \delta^{m_+} L_+ +\delta^{m_-} L_- + \delta^{m_\perp} L_\perp, 
\end{equation} 
where $\delta$ is kept as an analytic parameter. The exponents  $n_+,
n_-, n_\perp, m_+, m_-, m_\perp$ serve to study the behaviour in all
singular limits~\cite{Libby:1978qf} by setting their values to
appropriate integers. The sets of values and the corresponding infrared
or ultraviolet limits are listed in the Table~\ref{tab:singularities}. 
\begin{table}[ht]
\begin{center}
\begin{tabular}{|c|c||c|c|c||c|c|c|}
  \hline
 Singularity type & Limit & $n+ $ & $n_- $ & $n_\perp$ & $m_+$ & $m_-$& $m_\perp$  \\
  \hline
  double soft &$k,l \to 0 $ & 1 & 1 & 1 & 1 & 1& 1  \\
  \hline 
 &$k \to 0, l \parallel p_1$  & 1 & 1 & 1 & 0 & 2& 1  \\
  soft/collinear &$l \to 0, k \parallel p_1 $ & 0 & 2 & 1 & 1 & 1& 1  \\
  &$k \to 0, l \parallel p_2$  & 1 & 1 & 1 & 2 & 0& 1  \\
      &$l \to 0, k \parallel p_2 $ & 2 & 0 & 1 & 1 & 1& 1  \\
  \hline
two-loop collinear  &$k \parallel p_1, l \parallel p_1 $ & 0 & 2 & 1 & 0 & 2& 1  \\
  &$k \parallel p_2, l \parallel p_2 $ & 2 & 0 & 1 & 2 & 0& 1  \\
 \hline 
 collinear pairs     &$k \parallel p_1, l \parallel p_2 $ & 0 & 2 & 1 & 2 & 0& 1  \\
      &$k \parallel p_2, l \parallel p_1 $ & 2 & 0 & 1 & 0 & 2& 1  \\
  \hline
 single soft     &$k \to 0 $ & 1 & 1 & 1 & 0 & 0& 0  \\
      &$l \to 0 $ & 0 & 0 & 0 & 1 & 1 & 1  \\
  \hline
 &$k \parallel  p_1 $ & 0 & 2 & 1 & 0 & 0& 0  \\
 single collinear &$k \parallel  p_2 $ & 2 & 0 & 1 & 0 & 0& 0  \\
  &$l \parallel  p_1 $ & 0 & 0 & 0 & 0 & 2& 1  \\
  &$l \parallel  p_2 $ & 0 & 0 & 0 & 2 & 0 & 1  \\
\hline
  single UV    &$k \to \infty $ & -1 & -1 & -1 & 0 & 0& 0  \\
      &$l \to \infty $ & 0 & 0 & 0 & -1 & -1 & -1  \\
  \hline
 double UV  & $k,l \to \infty $ & -1 & -1 & -1 & -1 & -1& -1 \\
  \hline
\end{tabular}
\end{center}
\caption{
\label{tab:singularities}
  List of singular limits for  the $q \bar q \to \gamma^*
  \gamma^*$ amplitude.}
\end{table}
After we substitute Eq.~(\ref{eq:takelimits}) in the expression of
${\cal H}_2(k, l)$, we perform the Dirac algebra in conventional
dimensional regularization. Specifically, we encounter Dirac
structures with contracted gamma matrices, 
\begin{equation}
{\bar v}(p_2)\ldots \gamma^\mu \ldots \gamma_\mu \ldots  u(p_1)\, ,
 \end{equation} 
where in the ellipses represent products of other gamma matrices
contracted, similarly, with each other or with momenta or polarization
vectors. We use the Clifford algebra of $\gamma$ matrices to bring the
mutually contracted matrices to adjacent position where we can substitute
\begin{equation}
\gamma^\mu \gamma_\mu = (4 - 2 \epsilon) \, {\bf 1}.
\end{equation}
After this manipulation, which extracts explicitly the dependence of
the integrand on the spacetime dimensionality, and after collecting powers of
the parameter $\delta$,  spinor and scalar products are
computed numerically with exact arithmetic. 
We maintain explicit dependence on the dimensional regulator $\epsilon$ and the parameter
$\delta$. This permits us to perform a Taylor expansion in $\delta$ for each of the loop momentum configurations listed in Table~\ref{tab:singularities}.
In all cases, we find that
\begin{equation}
  \lim_{k,l \, \in \,  {\rm Table}~\mbox{\tiny \ref{tab:singularities}}}
d^dk \, d^dl \, {\cal H}^{(2)}_{\gamma^* \gamma^*}\left(k, l \right) = {\cal O}\left( \delta \right),
\end{equation}
which confirms that our counterterms have removed all singularities. 


\section{Conclusions and outlook}
\label{sec:conclusions}

In this paper we have shown how to  implement the factorization of infrared singularities at a local level in momentum space for a large class of two-loop QCD amplitudes for electroweak production initiated by the annihilation of quark pairs.    In the result, all dependence on the masses and momenta of the produced electroweak bosons is locally infrared finite and amenable to numerical evaluation.   In particular, the local subtractions implemented in this method do not introduce exotic denominators, and the resulting integrals can be deformed in their complex planes as necessary to avoid threshold and related singularities.   The number of infrared counterterms necessary to achieve these results is not large; in fact smaller than the number of diagrams.   Once these counterterms are combined algebraically, the factorization of collinear singularities is ``automatic".

We believe that the developments in this paper are a significant step toward a very general method for the construction of numerically-computable short distance functions in factorized cross sections, at two loops in QCD and perhaps beyond.   We anticipate that a similar, although not identical, procedure will allow an extension to gluon-initiated amplitudes for the same electroweak processes.   Beyond that, we are hopeful that this approach can be extended to final states involving both mixed electroweak production and QCD radiation, and eventually pure QCD processes.   

In the context of generalizations of this work, we note that our local factorization is in the same spirit as the exploitation of the local cancellation
\cite{Sterman:1978bj,Capatti:2022tit} of infrared singularities in suitably-defined sums over final states.   Both 
can alleviate the need for infrared regulation and/or the segmentation of phase space.   In fact, final-state cancellations rely on only the hermiticity of the interaction Hamiltonian and are in this sense ``automatic" once appropriate integrands are combined consistently.   The analogous principle underlying infrared factorization is causality, the application of which is complicated by the unphysical modes of perturbative gauge theories.  These are the origins of both the treatment of loop polarizations and the shift subtractions described in this paper.   We conjecture that our two-loop solutions to these problems should be extendable to higher order.    More ambitiously, we hope that by combining the local factorization developed here with the local cancellation of final-state singularities, we may make possible a new, numerically-based approach to precision in factorizable collider processes.



\section*{Acknowledgements}
We thank Julia Karlen, Matilde Vicini and Mao Zeng for useful discussions. 
This research was supported in part by the National Science Foundation
under Grants PHY-1915093 and PHY-2210533, by the Swiss National Science Foundation
under contract SNF200021\_179016 and by the European Commission through the ERC grant “pertQCD”.


\providecommand{\href}[2]{#2}\begingroup\raggedright\endgroup

\end{document}